\documentclass[12pt]{article}
\usepackage{epsfig,amssymb,amsmath,psfrag,
rotate}

\allowdisplaybreaks

\def \be  {\begin{equation}}
\def \ee  {\end{equation}}
\def \ba  {\begin{eqnarray}}
\def \ea  {\end{eqnarray}}
\def \baa {\begin{eqnarray*}}
\def \eaa {\end{eqnarray*}}
\def \lab #1 {\label{#1}}

\newcommand\re[1]{(\ref{#1})}

\def \matrix #1 {\left(\begin{array}{cc} #1 \end{array}\right)}

\def \tr {\mathop{\rm tr}\nolimits}

\newcommand{\bit}[1]{\mbox{\boldmath$#1$}}
\newcommand{\Q}{{\cal Q}}
\newcommand{\xB}{x_{\rm B}}
\newcommand{\tK}{\widetilde{K}}

\newcommand{\ft}[2]{{\textstyle\frac{#1}{#2}}}
\begin{document}

\begin{titlepage}

\thispagestyle{empty}

\vspace*{1cm}

\centerline{\large \bf Compton scattering: from deeply virtual to quasi-real}

\vspace{15mm}

\centerline{\sc A.V. Belitsky$^a$, D. M\"uller$^b$, Y. Ji$^a$}

\vspace{15mm}

\centerline{\it $^a$Department of Physics, Arizona State University}
\centerline{\it Tempe, AZ 85287-1504, USA}

\vspace{5mm}

\centerline{\it $^b$Institut f\"ur Theoretische Physik II, Ruhr-Universit\"at Bochum}
\centerline{\it D-44780 Bochum, Germany}

\vspace{2.5cm}

\centerline{\bf Abstract}

\vspace{5mm}
\noindent
We address the question of interpolation of the virtual Compton scattering process off a polarized nucleon target between the deeply virtual regime
for the initial-state photon and its near on-shell kinematics making use of the photon helicity-dependent Compton Form Factors (CFFs) as a main
ingredient of the formalism. The five-fold differential cross section\footnote{Cross section formulae are available in a  {\sc Mathematica} code upon
request, contact dieter.mueller@tp2.rub.de.} for the reaction with all possible polarization options for the lepton and nucleon spins is evaluated in terms
of CFFs in the rest reference frame of the initial-state nucleon. We suggest a rather simple parametrization of the Compton hadronic tensor in terms
of CFFs which are free from kinematical singularities and are directly related, at large photon virtualities, to Generalized Parton Distributions. We also
provide a relation of our basis spanned by a minimal number of Dirac bilinears to the one introduced by Tarrach for the parametrization of the virtual
Compton tensor and utilize the former to establish a set of equalities among our CFFs and Generalized Polarizabilities. As a complementary result,
we express Compton scattering in the Born approximation in terms of CFFs as well.

\vspace{3cm}

\end{titlepage}

\newpage
\pagestyle{empty}
\tableofcontents
\setcounter{footnote} 0

\newpage

\pagestyle{plain}
\setcounter{page} 1

\newpage
\pagestyle{plain}
\setcounter{page} 1

\section{Introduction 
}
\label{Sec-AziAngDep}

\begin{figure}[t!]
\begin{center}
\includegraphics[width=12cm]{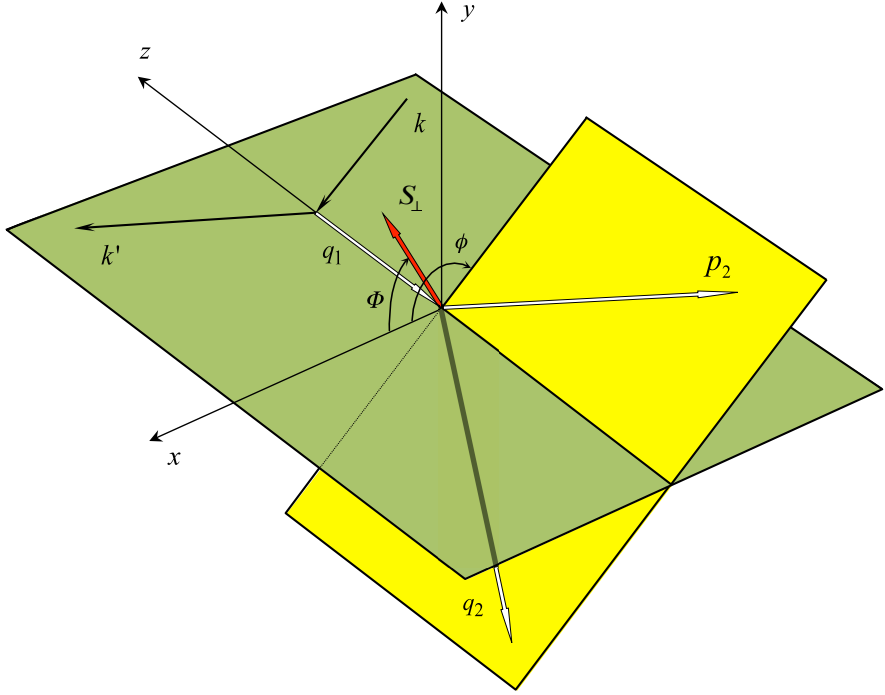}
\end{center}
\caption{\small The target rest frame, used in this work, is the same as adopted in our previous consideration \cite{Belitsky:2001ns}. The $z$-axis is
directed counter-along the photon three-momentum $\bit{q}_1$, the $x$-component of the incoming electron momentum $\bit{k}$ is chosen to be
positive. The angles parametrizing the five-fold cross section (\ref{WQ}) are defined as follows: $\phi$ is the azimuthal angle between the lepton
plane and the recoiled proton momentum,  while the difference $\varphi \equiv \Phi-\phi$ for fixed $\phi$ is determined by the direction of the
transverse nucleon polarization vector component $\bit{S}_\perp=(\cos\Phi, \sin\Phi)$. }
\label{fig:frame}
\end{figure}
Virtual Compton scattering on a nucleon, $\gamma^\ast (q_1) N(p_1) \to \gamma (q_2) N(p_2)$,
plays a distinguished role in the quest to access its internal content and unravel the
mysteries of strong interactions. The reason for this is multifold.  Experimentally, the
scattering process off a proton can be measured in a straightforward fashion,
free of complications of composite probes, via scattering of leptons on a hydrogen target.
The five-fold differential cross section for the emission of an on-shell photon
to the final state, $\ell (k) N (p_1) \to \ell (k^\prime) N (p_2) \gamma (q_2)$, reads
\be
\label{WQ}
d \sigma
=
\frac{\alpha_{\rm em}^3  \xB y^2 } {16 \, \pi^2 \,  {\cal Q}^4 \sqrt{1 + \epsilon^2}}
\left| \frac{\cal T}{e^3} \right|^2
d \xB d {\cal Q}^2 d |t| d \phi d \varphi
\, ,
\ee
in the approximation that neglects the mass of the lepton. The phase space is parameterized by the Bjorken variable $\xB = \mathcal{Q}^2/(2 p_1 \cdot q_1)$,
which is in turn determined by the momentum $q_1 = k - k'$ of the initial-state photon of virtuality $\mathcal{Q}^2 = - q_1^2$, the square of the $t$-channel
momentum $t = (p_2 - p_1)^2$, the azimuthal angle $\phi$ of the recoiled nucleon, and for a transversally polarized target yet another (relative) angle $\varphi$,
where the latter two are defined in the rest frame of the target as depicted in Fig.~\ref{fig:frame}. Finally, we introduce the variable $y = p_1\cdot q_1/p_1\cdot k$
for the lepton energy loss and a shorthand nation for $\epsilon = 2 \xB M/\mathcal{Q}$ that incorporates nonvanishing target mass effects. In the above five-fold
cross section, the leptoproduction amplitude $\mathcal{T}$ is a linear superposition of the Bethe-Heitler (BH) and virtual Compton scattering (VCS) amplitudes,
depending on whether the real photon is emitted off the lepton or nucleon, respectively.  In the scattering amplitude
\begin{equation}
{\cal T}
= {\cal T}^{\rm BH} + {\cal T}^{\rm VCS}
\, ,
\end{equation}
the former is determined in terms of the nucleon matrix element of the quark electromagnetic current $j_\mu$
\begin{align}
\label{quarkEMcurrent}
J_\mu = \langle p_2 | j_\mu (0) | p_1 \rangle
\, ,
\end{align}
while the hadronic Compton tensor,
\begin{align}
\label{ComptonAmplitudeT}
T_{\mu\nu} =i\int d^4 z \, {\rm e}^{\ft{i}{2} (q_1 + q_2) \cdot z}
\langle p_2 | T \left\{ j_\mu (z/2) j_\nu (- z/2) \right\} | p_1 \rangle
\, ,
\end{align}
encodes information on more intricate long-distance dynamics. This is the main observable for our subsequent analysis.

The variation of the virtuality $\mathcal{Q}^2$ of the initial state photon allows one to probe a wide range of distance scales, interpolating
between short- and long-wave structures of the nucleon. A number of observables  are available to achieve this goal, all representing different
facets of the same reaction. For real to slightly virtual initial-state photon, produced as a bremsstrahlung off the lepton beam, and low energy
$\omega^\prime = q_2^0$ of the outgoing photon, the Compton amplitude admits conventional multipole expansion with leading contributions
defining the electric $\alpha$ and magnetic $\beta$ polarizabilities of the nucleon, see, e.g., Ref.~\cite{Drechsel:2002ar} for a review. The latter
characterize the linear response of the nucleon to the electric and magnetic fields
of the incoming photon which slightly distorts the hadron and, as a consequence, induces (in the quasi-static approximation) nontrivial electric
$\bit{d} = \alpha \bit{E}_{\rm in}$ and magnetic $\bit{\mu} = \beta \bit{B}_{\rm in}$ dipole moments. The latter then interact with the electromagnetic
fields of the outgoing photon through multipole couplings $\bit{d} \cdot \bit{E}_{\rm out} + \bit{\mu} \cdot \bit{B}_{\rm out}$. The experimental values
for the coefficients $\alpha$ and $\beta$ are very small indicating that the nucleon is a very rigid object allowing only for a very small deformation.
Understanding of their magnitude within effective field theories comes about as a result of a subtle cancelation of the pion cloud and quark-core
effects. For an off-shell initial-state photon with virtuality $\mathcal{Q}^2$ that scatters on a polarized spin one-half target, one can introduce ten
\cite{Guichon:1995pu} generalized, --- referring to the functional dependence on $\mathcal{Q}^2$ rather than being mere numbers,
--- polarizabilities which reduce to six, once one imposes charge conjugation and crossing symmetry constraints \cite{Drechsel:1997xv}.

Increasing the momentum transfer in the $t-$channel results in large-angle scattering of the emitted real photon in the final state. As a consequence,
one enters the domain of the wide-angle Compton scattering. In this kinematics the process receives quantitative description within QCD
factorization approach with the leading asymptotic behavior driven by the hard gluon exchanges between nucleon's constituents \cite{Lepage:1980fj}
and by the Feynman soft mechanism at moderate $t$ with the amplitude arguably described by a hand-bag diagram \cite{Radyushkin:1998rt,Diehl:1998kh}.
The real Compton Form Factors emerging in the latter framework are actually moments of more general functions encoding the partonic degrees of
freedom in the nucleon.

In the deeply virtual regime of large Euclidean $\mathcal{Q}^2$ and fixed $t$, the probe resolves individual nucleon's constituents and the process
admits a full-fledged description in terms of the Generalized Parton Distributions (GPDs) \cite{Mueller:1998fv,Radyushkin:1996nd,Ji:1996nm}. However,
the Compton amplitude itself is only determined by an integral of GPDs accompanied by a perturbatively computable coefficient function. These convolutions
are known as Compton Form Factors (CFFs) \cite{Belitsky:2001ns}. Making use of the gauge invariance, discrete symmetries and crossing, one can establish
that there are twelve independent CFFs when the outgoing photon is real. They describe information about the hadron for all possible polarization settings of
the nucleon and the resolving photons. Since this decomposition is general, the CFFs define the amplitude in all kinematical regimes, interpolating between
the aforementioned polarizabilities at low energies, and thus describing the response of the nucleon as a whole to the external probes, all the way up to probing
partonic degrees of freedom at high energies.

The goal of the present study is to elaborate on our previous analysis
\cite{Belitsky:2010jw} and provide a complete set of exact results for helicity amplitudes
describing the virtual Compton scattering, on the one hand, filling the gap for transversely polarized
target as well as contributions of the double helicity-flip effects that were not entirely worked
out before, as well as deliver a set of relations between CFFs and polarizabilities introduced in earlier studies \cite{Drechsel:1997xv}
making use of the Tarrach's decomposition of the Compton tensor \cite{Tarrach_1975tu}, on the other. Thus we establish a useful dictionary that can be used to
re-express the results of experimental measurements in terms of the same observables, Compton Form Factors.

Our subsequent presentation is organized as follows. In the next section, we review the formalism of helicity amplitudes, used previously by us in the
deeply virtual kinematics, and provide a set of exact concise formulas for all polarization settings, unpolarized, longitudinal and transverse, of the nucleon
target. In Sect.~\ref{ComptonParametrization}, we address the question of gauge-invariant decomposition of the hadronic tensor. We start there with an
exactly solvable toy example of a point particle and then suggest a rather simple parameterization of the Compton tensor in terms of CFFs that are free of
kinematical singularities.   The connection to the structure functions defined by Tarrach and the form of CFFs in the Born approximation is established by
means of helicity amplitudes. In Sect.~\ref{GPs}, we develop, based on our findings, the low-energy expansion of the CFFs and provide a complete set of
relations to the generalized polarizabilities of Guichon et al., introduced in Ref.~\cite{Guichon:1995pu}.  Finally, we summarize and point out further
applications of our unified framework. A number of appendices serve to explain different reference frames used to determine the kinematics in our
computations, as well as a number of explicit results that prove to be too long to fit in the main body of the paper. We add that all results reported here are
also available as a {\sc Mathematica} code for easy computer implementation by practitioners.

\section{Cross section in terms of helicity amplitudes}
\label{HelicityAmplitudes}

In recent investigations \cite{Belitsky:2010jw,Belitsky:2008bz}, we demonstrated that the deviation between the data on hard electroproduction of photons
and theoretical estimates for corresponding observables within the approximation scheme of Ref.~\cite{Belitsky:2001ns} could be reconciled by calculating
kinematical corrections in the hard scale exactly while ignoring dynamical high-twist contributions altogether. The neglect of the latter was motivated by the
hierarchy of low-energy scales associated with hadronic matrix elements of high-twist operators which are smaller than soft kinematical scales encountered
in the problem, i.e., the nucleon mass and the $t$-channel momentum transfer. Incorporation of the kinematical power-suppressed effects was achieved by
separating them between the leptonic and hadronic parts independently and by evaluating photon helicity amplitudes utilizing the polarization vectors for
the incoming and outgoing photons in the target rest frame. In addition to providing an efficient computational scheme, it has another advantage of localizing
the azimuthal angular dependence in the lepton helicity amplitudes for the choice of the reference frame with the $z$-axis counter-aligned with the incoming
photon three-momentum, as shown in Fig.\ \ref{fig:frame}. It also allows one for a straightforward reduction to the harmonic expansion introduced in Refs.\
\cite{Belitsky:2001ns,DieGouPirRal97}.

\subsection{Form factor parameterization of hadronic helicity amplitudes}

Let us start with the hadronic component of the leptoproduction amplitude of a real photon. We define the nucleon helicity amplitudes for the
(deeply virtual or quasi-real) Compton scattering as
\be
\label{DVCS2helicity}
{\cal T}^{\rm VCS}_{ab}(\phi) = (-1)^{a-1}
\varepsilon^{\mu\ast}_2(b) T_{\mu \nu} \varepsilon^{\nu}_1(a) \, ,
\ee
by contracting the VCS tensor (\ref{ComptonAmplitudeT}) with the photon polarization vectors. Here, the overall phase $(-1)^{a-1}$ accounts for the signature
factor in the completeness relation for the photon polarization vectors. The $a$ and $b$ indices take the following values $a \in \{0, \pm 1\}$ and $b = \pm 1$.
The $\varepsilon$-vectors for the virtual photon are given in our reference frame by
\begin{eqnarray}
\label{Polarization1}
\varepsilon_1^\mu(\pm)
\!\!\!&=& \!\!\!
\frac{e^{\mp i\phi}}{\sqrt{2}}(0,1, \pm i,0)\,, \qquad
\varepsilon_1^\mu(0) = \frac{1}{\epsilon}(-\sqrt{1+\epsilon^2},0, 0,1),
\end{eqnarray}
while for the real photon they are
\begin{eqnarray}
\label{Polarization2}
\varepsilon_2^{\mu\ast}(\pm)
\!\!\!&=&\!\!\!
\frac{1}{\sqrt{2}}
\left(
0,
\frac{1+\frac{ \epsilon^2}{2} \frac{{\Q}^2 + t}{\Q^2 + \xB t}}{ \sqrt{1+\epsilon^2}} \cos\phi
\pm i \sin\phi,
\mp i \cos\phi
+
\frac{1+\frac{\epsilon^2}{2} \frac{{\Q}^2 + t}{\Q^2 + \xB t}}{\sqrt{1+\epsilon^2}}
\sin\phi,
\frac{-\epsilon \Q \widetilde{K}/\sqrt{1+\epsilon^2}}{\Q^2+\xB t}
\right)
\, . \nonumber\\
\end{eqnarray}
Here, we introduced for later convenience  a kinematical factor with mass dimension one
\begin{eqnarray}
\label{K-tilde}
{\widetilde K}
= \sqrt{(1-\xB) \xB+\frac{\epsilon^2}{4}}\; \sqrt{\frac{(t_{\rm min}-t)(t-t_{\rm max})}{\Q^2}}\,,
\end{eqnarray}
which vanishes at the minimally (maximally) allowed value of the $t$-channel momentum transfer $-t=-t_{\rm min}$  ($-t=-t_{\rm max}$)
with
\begin{align}
t_{\rm min} = - \mathcal{Q}^2 \frac{2 (1 - \xB) \left( 1 - \sqrt{1 + \epsilon^2}\right)+ \epsilon^2}{4 \xB (1 - \xB) + \epsilon^2}\,,
\qquad
t_{\rm max} = - \mathcal{Q}^2 \frac{2 (1 - \xB) \left( 1  + \sqrt{1 + \epsilon^2}\right)+ \epsilon^2}{4 \xB (1 - \xB) + \epsilon^2}
\, .
\end{align}
In turn, ${\widetilde K}$ vanishes if $\xB$ reaches for fixed $-t$ and $\Q^2$ the maximal allowed value
\begin{align}
x_{\rm B\, max}=
1-\frac{\Q^2+t}{ \Q^2+t+ \left(\!\sqrt{-t\left(4 M^2-t\right)}-t\!\right)\frac{\Q^2}{2M^2} }.
\end{align}
Consequently, this factor encodes the
phase space boundary in hadronic variables. In the explicit computation of Eq.\ (\ref{DVCS2helicity}), we use the Lorentz covariant decomposition
for the $\varepsilon$-vectors in terms of momentum four-vectors defining the process, which we often write for convenience in terms of the
$t$-channel momentum transfer, the sum of nucleons' momenta,  the averaged momentum of the photons, and a vector orthogonal to the previous three%
\footnote{As in our previous work, we also adopt here to the conventions of Itzykson and Zuber \cite{ItzZub80}, i.e., for the Levi-Civita tensor we choose the
normalization $\epsilon^{0123}=+1$.}:
\begin{align}
\label{AveagedMomenta}
\Delta^\mu= p_2^\mu-p_1^\mu\,, \quad
p^\mu = p_1^\mu + p_2^\mu\,, \quad
\quad q^\mu = \ft12 (q_1^\mu + q_2^\mu)\, , \quad
\epsilon^\mu_{\phantom{\mu} pq\Delta}\equiv \epsilon^\mu_{\phantom{\mu}\alpha\beta\gamma} p^\alpha q^\beta \Delta^\gamma\,.
\end{align}
The coefficients in such an expansion are given in terms of the kinematical invariants introduced above.  A complete set of relations is
deferred to the Appendix \ref{KinematicalDecomposition}.

The computations of the cross section (\ref{WQ}) by means of the hadron helicity amplitudes (\ref{DVCS2helicity}), presented in the following two
sections, require an explicit tensor decomposition of the Compton amplitude. Unfortunately,  no consensus exists on the form of parametrization of
such a tensor even for DVCS kinematics.  In the latter case, the partonic interpretation of the $T_{\mu\nu}$ arises from the application of the Operator
Product Expansion (OPE) techniques which valid to a given accuracy in the $1/\mathcal{Q}$-expansion and this leaves a substantial ambiguity in the
parametrization of the hadronic amplitude depending on the fashion that one restores its gauge invariance broken by the leading order approximation.

To go around this problem, we first parameterize directly the photon helicity amplitudes (\ref{DVCS2helicity}). Thereby, we  describe the nucleon-to-nucleon
transition for given photon helicities in terms of two even parity and two odd parity bilinear Dirac spinor covariants, analogously to the manner they
appear in the standard form factor parameterization of the vector and axial-vector currents. Moreover, we take into account that the above helicity
amplitudes for opposite pairs of helicities are not independent of each other and are rather related by the parity conservation,  generically written as
\begin{eqnarray}
\label{helTsympro}
 {\cal T}^{\rm VCS}_{--} ({\cal F})  &\!\!\!=\!\!\!&  {\cal T}^{\rm VCS}_{++}({\cal F})
\Big|_{{\cal F}^{P=\pm 1}\to \pm {\cal F}^{P=\pm 1}}
\, , \nonumber\\
{\cal T}^{\rm VCS}_{0-} ({\cal F})  &\!\!\!=\!\!\!&  {\cal T}^{\rm VCS}_{0+}({\cal F})
\Big|_{{\cal F}^{P=\pm 1}\to \pm {\cal F}^{P=\pm 1}}
\, ,\\
 {\cal T}^{\rm VCS}_{-+}({\cal F})  &\!\!\!=\!\!\!&   {\cal T}^{\rm VCS}_{+-}({\cal F})
\Big|_{{\cal F}^{P=\pm 1}\to \pm {\cal F}^{P=\pm 1}}
\, , \nonumber
\end{eqnarray}
where ${\cal F}^{P}$ stays for CFFs with definite parity $P=\pm 1$ (even parity $P=1$ refers to vector case while odd one $P=-1$ refers to axial-vector case).
As a consequence, we have a set ${\cal F}$ of three times four independent CFFs and our helicity amplitudes can be expressed in terms of six linear functions,
three depending on two even (or two odd) CFFs.  Furthermore, we can summarize diverse formulation in a single parametrization since the representation of
the photon polarization vectors in terms of kinematical variables allows us to use the Dirac equation for the free nucleon spinors. Consequently, we write the
helicity amplitudes (\ref{DVCS2helicity}) in following form
\begin{align}
\label{cal-Tab}
{\cal T}^{\rm VCS}_{ab}
=
 {\cal V}({\cal F}_{ab}) - b\, {\cal A}({\cal F}_{ab}) \quad\mbox{for}\quad b\in\{+,-\}
\end{align}
in terms of the vector and axial-vector form factor parametrization,
\begin{eqnarray}
\label{cal-TabV}
{\cal V}({\cal F}_{ab}) \!\!\!&=&\!\!\!
\bar{u}_2 \left( \, {\not\!\! m} {\cal H}_{ab} + i \sigma_{\alpha \beta} \frac{m^\alpha \Delta^\beta}{2 M}\, {\cal E}_{ab}\right) u_1
\\
\label{cal-TabA}
{\cal A}({\cal F}_{ab})\!\!\!&=&\!\!\!
\bar{u}_2 \left( \, {\not\!\! m} \gamma_5 \,  \widetilde{\cal H}_{ab} + \gamma_5 \frac{m\cdot\Delta}{2 M}\, \widetilde{\cal E}_{ab}\right) u_1\,,
\end{eqnarray}
with a convention-dependent vector $m^\mu$ and bispinors $u_i \equiv u (p_i, S_i)$, normalized as $\bar{u} (p,S) u (p, S) = 2M$.
Such a uniform functional form for all photon helicity options, which closely matches the GPD notation, is very convenient for the evaluation of the
cross section.  However, one has to take special care tracing potential kinematical singularities. Emphasizing the simplicity of the underlying
analysis, in Sect.~\ref{ComptonParametrization} we will state how CFFs can be defined in a singularity free manner.

A few other comments are in order. Our conventions imply the relations
\begin{equation}
{\cal F}_{--} = {\cal F}_{++}\,, \quad  {\cal F}_{0-} = {\cal F}_{0+}
 \,, \;\;  \mbox{and} \;\;\; {\cal F}_{-+} = {\cal F}_{+-}\quad \mbox{for}\quad  {\cal F} \in \{{\cal H},{\cal E},\widetilde{\cal H},\widetilde{\cal E}\}\,.
\end{equation}
The above vector $m^\mu$ in the GPD framework is often equated to a fixed light-like vector and reflects, loosely
speaking, also the accuracy in restoring gauge invariance lost within the twist-two accuracy. A couple of fixed light-like vector choices were
explored in the literature, see, e.g., discussion in Ref.~\cite{Belitsky:2005qn}. Going beyond the leading twist approximation, the choice $m=q^\mu/p\cdot q$
is physically motivated and guarantees a proper behavior under Lorentz transformations as well as allows one for a simple implementation of the Bose
symmetry \cite{Belitsky:2001ns}.  Another choice $m^\mu = q_1^\mu/p_1\cdot q_1$ can be advocated by the fact that in our reference frame this vector
contains only longitudinal degrees of freedom. Finally, $m^\mu \propto q_2^\mu$  can also be taken as a light-like vector in the GPD framework
\cite{Braun:2011dgBraun:2011zr,Braun:2012bgBraun:2012hq}. Using the free Dirac equation for the nucleon spinors it becomes obvious that the
parameterization (\ref{cal-Tab})--(\ref{cal-TabA}) in terms of spinor bilinears is complete and, hence, different choices of $m^\mu$ correspond to a linear
transformation in the space of CFFs. As in previous work, we will use the following vector $$m^\mu=q^\mu/p\cdot q$$ throughout our current analysis.

The above helicity CFFs can be expressed in terms of the ones emerging in the GPD framework. However, since the latter relies on a truncation of the
$1/\Q$-expansion, the resulting relations will depend on a particular parametrization of the Compton tensor and identification of CFFs as a convolution of
GPDs and perturbative coefficient functions valid only to a very low accuracy in $1/\mathcal{Q}$-expansion. In Sect.~\ref{ComptonParametrization} an
exact set of  CFF relations will be given, see below Eqs.\ (\ref{ExactHelicityFpb})--(\ref{ExactHelicityF0p}), here we quote the leading contribution to the
helicity form factors from the twist-two  $\mathcal{F}\, (\equiv \mathcal{F}^{{\rm tw}-2})$, the effective twist-three
$\mathcal{F}^{\rm eff}$ and gluon transversity $\mathcal{F}_T$ CFFs,
\begin{eqnarray}
\label{cff2-cff++}
{\cal F}_{++} &\!\!\!=\!\!\!& {\cal F} + {\cal O}\left(1/{\cal Q}^2\right)
\\
\label{cff2-cff0+}
{\cal F}_{0+} &\!\!\!=\!\!\!& \frac{\sqrt{2}\widetilde{K}}{\sqrt{1+\epsilon^2} \Q \left(2-\xB+\frac{\xB t}{\Q^2}\right)} {\cal F}^{\rm eff}+ {\cal O}\left(1/{\cal Q}^2\right)
+{\cal O}(\alpha_s)\,,
\\
\label{cff2-cff+-}
{\cal F}_{+-} &\!\!\!=\!\!\!&  \frac{{\widetilde K}^2}{2 M^2\left(2-\xB+\frac{\xB t}{\Q^2}\right)^2}  {\cal F}_{T}  + {\cal O}\left(1/{\cal Q}^2\right)
\, ,
\end{eqnarray}
where some typical kinematical factors were treated here exactly preparing the stage for full-fledged formulas. In the amplitude $\mathcal{F}_{0+}$, we
used an effective GPD inspired CFF
\begin{eqnarray}
{\cal F}^{\rm eff} =  - 2 \xi \left(
\frac{1}{1 + \xi} {\cal F} + {\cal F}^{\rm{tw}-3}_+ - {\cal F}^{\rm{tw}-3}_-\right) + {\cal O}\left(1/{\cal Q}^2\right) + {\cal O}\left(\alpha_s/{\cal Q}\right)
\, .
\end{eqnarray}
These contain a twist-two induced part and twist-three quantities ${\cal F}^3_\pm$ that are given in Ref.~\cite{Belitsky:2001ns}, see Eqs.~(84)--(87)
there. Note, however, that these effective CFFs are also affected by the twist-two gluon transversity, formally suppressed by $\alpha_s$, and also
high-twist contributions. As it has been discussed in Ref.~\cite{Belitsky:2008bz} for a scalar target, the transversity admixture to the longitudinal helicity-flip
amplitudes presently is not under theoretical control and its clarification requires a twist-three analysis at NLO accuracy. Above we made use of the
generalized Bjorken variable $\xi$ that is expressed via $\xB$ as follows $\xi \simeq \xB/(2 - \xB)$.  Further insights on the interplay between the current
conservation, the choice of the partonic scaling variables and, respectively, the choice of the auxiliary light-like vectors and kinematical effects  can be found
in Ref.~\cite{Belitsky:2008bz} and below in Sect.~\ref{sec:tensor}.

Having fixed the parametrization of the hadronic helicity amplitudes (\ref{cal-Tab})--(\ref{cal-TabA}) by the choice $m^\mu \equiv q^\mu/p\cdot q$,
we will turn now in the next two sections to how they are incorporated into the squared of the VCS amplitude as well as its interference with the
Bethe-Heitler process.  We emphasize once more that the uncertainties from kinematical and dynamical higher twist contributions, appearing in the
relation of hadronic and partonic quantities for the deeply virtual kinematics,  are entirely encoded in the relations of helicity dependent CFFs ${\cal F}_{ab}$
to the set of CFFs that one adopts for the evaluation of the hadronic tensor. Thus, the results that follow are exact, free of any approximations.

\subsection{Squared Compton scattering amplitude}
\label{sect-TDVCS}

Let us now calculate the square of the (D)VCS amplitude, entering the cross section (\ref{WQ}), where the lepton mass is set to zero and the polarization
of the final state lepton remains unobserved.
Using the completeness relations for the photon polarization vectors, we can rewrite this square as
\begin{eqnarray}
\label{T2^DVCS}
|{\cal T}^{\rm VCS}|^2
=
\frac{1}{{\cal Q}^2}
\sum_{a = {\scriptscriptstyle -}, 0, {\scriptscriptstyle +}}
\sum_{b = {\scriptscriptstyle -}, 0, {\scriptscriptstyle +}}
{\cal L}_{ab} (\lambda, \phi) {\cal W}_{ab}
\, , \quad
\end{eqnarray}
in terms of the hadronic,
\be
\label{W_ab}
{\cal W}_{ab}
=
{\cal T}^{\rm VCS}_{a +} \left({\cal T}^{\rm VCS}_{b +}\right)^\ast
+
{\cal T}^{\rm VCS}_{a -} \left({\cal T}^{\rm VCS}_{b -}\right)^\ast
\,  ,
\ee
and leptonic,
\be
{\cal L}_{ab}(\lambda,\phi) =
\varepsilon^{\mu\ast}_1(a) L_{\mu \nu}(\lambda)\varepsilon^{\nu}_1(b)\,,
\ee
squared amplitudes, labeled by the helicity states of the initial and final photons. Here, the familiar leptonic tensor for the initial-state lepton with
helicity $\lambda=\pm 1$ reads
\begin{align}
\label{L-tensor}
L_{\mu\nu} = 2 \mathcal{Q}^{-2}
\left( k_\mu k^\prime_\nu + k_\nu k^\prime_\mu - k \cdot k^\prime g_{\mu\nu} + i \lambda \varepsilon_{\mu\nu k k^\prime}  \right)
\, .
\end{align}
Note that $|\lambda|\le 1$ can be also regarded as the polarizability of the lepton beam.
More explicitly, one finds for the squared VCS amplitude (\ref{T2^DVCS})
\begin{eqnarray}
\label{HelAmpDVCS}
{\cal Q}^2|{\cal T}^{\rm VCS}|^2 \!\!\!&=&\!\!\!
{\cal L}_{++}(\lambda) {\cal W}_{++}  +{\cal L}_{++}(-\lambda){\cal W}_{--}  +{\cal L}_{00} {\cal W}_{00}
\\
&+&\!\!\!
{\cal L}_{0+}(\lambda,\phi) {\cal W}_{0+}
+ {\cal L}_{0+}(-\lambda,-\phi) {\cal W}_{0-}
+{\cal L}_{0+}(-\lambda,\phi) {\cal W}_{+ 0}
+ {\cal L}_{0+}(\lambda,-\phi){\cal W}_{- 0}
\nonumber\\
&+&\!\!\!
{\cal L}_{+-}(\phi)
{\cal W}_{+-}  +{\cal L}_{+-}(-\phi) {\cal W}_{-+}
\,.
\nonumber
\end{eqnarray}

The squared leptonic helicity amplitudes can be calculated exactly yielding known results, e.g., in the form already presented in
Ref.~\cite{Belitsky:2008bz}:
\begin{eqnarray}
\label{cL^DVCS_++}
{\cal L}_{++} (\lambda)
\!\!\!&=&\!\!\!
\frac{1}{ y^2(1 + \epsilon^2)} \left(2 - 2 y + y^2 + \frac{\epsilon ^2}{2} y^2 \right)
-
\frac{2-y}{\sqrt{1+\epsilon^2} y} \lambda
\, , \\
\label{cL^DVCS_00}
{\cal L}_{00}
\!\!\!&=&\!\!\!
\frac{4}{y^2(1+\epsilon^2)} \left(1-y - \frac{\epsilon ^2}{4} y^2\right)
\, , \\
\label{cL^DVCS_0+}
{\cal L}_{0+}(\lambda,\phi)
\!\!\!&=&\!\!\!
\frac{2 - y - \lambda y \sqrt{1 + \epsilon^2}}{y^2 (1 + \epsilon^2)}
\sqrt{2} \sqrt{1 - y - \frac{\epsilon ^2}{4} y^2} \, e^{-i \phi}
\, , \\
\label{cL^DVCS_-+}
{\cal L}_{+-}(\phi)
\!\!\!&=&\!\!\!
\frac{2}{y^2 (1 + \epsilon^2)} \left( 1-y - \frac{\epsilon ^2}{4} y^2 \right) e^{i 2 \phi}
\,.
\end{eqnarray}
The remaining squared amplitudes are related to the above by parity- and time-reversal invariance,
\begin{eqnarray}
\begin{array}{ll}
{\cal L}_{0 -}(\lambda,\phi)
=
{\cal L}_{0+}(-\lambda,-\phi) \, ,
&\qquad
{\cal L}_{\pm,0}(\lambda,\phi)
=
{\cal L}_{0,\pm}(-\lambda,\phi)
\, , \qquad
\\[1mm]
{\cal L}_{--}(\lambda)
=
{\cal L}_{++}(-\lambda) \, ,
&\qquad
{\cal L}_{-+} (\phi)
=
{\cal L}_{+-}(-\phi)
\, .
\end{array}
\end{eqnarray}

The squared helicity amplitudes of the hadronic tensor (\ref{W_ab}) take the following form in the spinor representation (\ref{cal-Tab})--(\ref{cal-TabA}),
\begin{eqnarray}
{\cal W}_{ab}\!\!\!&=&\!\!\!
\sum_{S^\prime}\sum_{c=\pm1}\Big[
{\cal V} ({\cal F}_{ac}) - c\, {\cal A}({\cal F}_{ac})\Big]\Big[{\cal V}^\dagger ({\cal F}^\ast_{bc}) - c\, {\cal A}^\dagger ({\cal F}^\ast_{bc})\Big]\,,
\nonumber
\end{eqnarray}
and will be evaluated exactly for given nucleon polarizations.  We decompose the  polarization vector of the initial nucleon in its transverse and
longitudinal  components,
\begin{align}
\label{S-vector}
S^\mu(\Phi,\theta) =  \sin\theta \, S^\mu_{\rm T}(\Phi)  + \cos\theta \, S^\mu_{\rm L}\,,
\end{align}
where the angle is $\Phi=\varphi+\phi$ is introduced in Fig.\ \ref{fig:frame}, while the individual vectors
\begin{align}
S^\mu_{\rm T}(\varphi+\phi) = \left(0, \cos(\varphi+\phi),\sin(\varphi+\phi),0\right)\,, \qquad
S^\mu_{\rm L} =\left(0,0,0,1\right)\,,
\end{align}
can be expressed in the basis of momenta (\ref{AveagedMomenta}), see Appendix \ref{KinematicalDecomposition}. The outgoing nucleon will be treated
in our considerations as unpolarized, since we are not aware of any plans to perform recoil polarization measurements in experiments for a rather
challenging virtual Compton scattering reaction. If needed, our work can be generalized along these lines.

The Fourier coefficients, given by the square of the VCS helicity amplitudes, can be re-expressed as bilinear combinations of CFFs with their functional
dependence reflecting the nucleon polarization states. Consequently, we decompose the square of the VCS amplitude into four terms exhibiting the
spin of the target as follows,
\begin{eqnarray}
\label{calC}
\sum_{S^\prime}
\Big[ {\cal V}({\cal F}) + {\cal A}({\cal F}) \Big]
\Big[ {\cal V}^\dagger({\cal F}^\ast) + {\cal A}^\dagger({\cal F}^\ast) \Big]
& \!\!\! =\!\!\! &
\Bigg[{\cal C}^{\rm VCS}_{\rm unp} +
\Lambda\cos(\theta) \frac{1}{\sqrt{1+\epsilon^2}}\, {\cal C}^{\rm VCS}_{\rm LP}
\\
&&\hspace{-4cm}+
\Lambda\sin(\theta) \sin(\varphi) \frac{i\widetilde K}{2M}{\cal C}^{\rm VCS}_{{\rm TP}-} +
\Lambda\sin(\theta) \cos(\varphi) \frac{\widetilde K}{2M\sqrt{1+\epsilon^2}}{\cal C}^{\rm VCS}_{{\rm TP}+}\Bigg]({\cal F},{\cal F}^\ast)\,,
\nonumber
\end{eqnarray}
where we made the polarizability $\Lambda$ of the nucleon target explicit. The naming of different ${\cal C}_{\cdots}^{\rm VCS}({\cal F},{\cal F}^\ast)$
functions is self-explanatory. These arise as bilinear combinations of CFFs making use of the definition (\ref{cal-Tab}). Their form will be given below.
Moreover, we may consider ${\cal F}$ and ${\cal F}^\ast$ as independent variables so that a uniform functional form can be employed in the evaluation of
all initial-to-final photon-helicity state transitions: spanning the range between conserved to longitudinal-to-transverse, and transverse-to-transverse
helicity-flip contributions.

Let us clearly spell out some of the changes in our definitions compared to the ones we used in our earlier studies. Note that in comparison to
Ref.~\cite{Belitsky:2010jw}, we redefined the combination ${\cal C}^{\rm VCS}_{\rm LP}$ by pulling out an overall factor of $1/\sqrt{1+\epsilon^2}$. Moreover,
with respect to the approximate expressions of Ref.~\cite{Belitsky:2001ns} we also changed here the overall normalization of the transversity contributions.
Note also that in the relations (\ref{cff2-cff0+}) and (\ref{cff2-cff+-}) between  longitudinal and transverse helicity flip CFFs and GPD inspired CFFs there
appears a factor $\tK$ and $\tK^2$, respectively. Compared to Ref.~\cite{Belitsky:2001ns}, such kinematical factors are  now stripped off if we express first
and second order harmonics in terms of helicity-dependent CFFs. Another modification is that the leptonic part for exact kinematics can be simply obtained
by a set of substitution rules from our previous DVCS results that has been already discussed in Ref.~\cite{Belitsky:2010jw} and will not be repeated here.
Finally, we also note that some of the remaining corrections in the hadronic part can be considered as a reparametrization of the scaling variable, i.e.,
$$
\xi \simeq \frac{\xB}{2-\xB} \quad \rightarrow \quad \xi = \frac{\xB}{2-\xB + \frac{\xB t}{\Q^2}}\,.
$$

Now we are able to cast our findings into the form suggested in Ref.~\cite{Belitsky:2001ns}. Namely, from the squared VCS amplitude (\ref{HelAmpDVCS}),
the computed leptonic helicity amplitudes (\ref{cL^DVCS_-+}) and the definition of the hadronic coefficients ${\cal C}^{\rm DVCS}$ as functions of the
helicity-dependent CFFs, one can immediately read off the harmonic expansion, which we write here by analogy to Ref.~\cite{Belitsky:2001ns} as
\be
\label{AmplitudesSquared}
 |{\cal T}^{\rm VCS}(\phi,\varphi)|^2
=
\frac{e^6}{y^2 {\cal Q}^2}\left\{ c^{\rm VCS}_0(\varphi) + \sum_{n=1}^2
\left[ c^{\rm VCS}_n(\varphi)\, \cos(n\phi) + s^{\rm VCS}_n(\varphi)\, \sin(n \phi)
\right] \right\} \, .
\ee
The evaluation of the Fourier harmonics in Eq.\ (\ref{AmplitudesSquared}) is straightforward and provides  for the coefficients in the decomposition
\begin{eqnarray}
c_{n}^{\rm VCS}(\varphi) = c_{n,\rm unp}^{\rm VCS}
+  \cos\theta\, c_{n,\rm LP}^{\rm VCS} + \sin\theta\, c_{n,\rm TP}^{\rm VCS}(\varphi)
\\
s_{n}^{\rm VCS}(\varphi) = s_{n,\rm unp}^{\rm VCS}
+  \cos\theta\, s_{n,\rm LP}^{\rm VCS} + \sin\theta\, s_{n,\rm TP}^{\rm VCS}(\varphi)
\end{eqnarray}
the following results:
\begin{itemize}
\item Unpolarized target
\end{itemize}\vspace{-0.5cm}
\begin{eqnarray}
\label{Res-Mom-DVCS-0-imp} c_{0,\rm unp}^{\rm VCS}
\!\!\!&=&\!\!\!
2 \frac{2-2 y + y^2+\frac{\epsilon ^2}{2} y^2}{1+\epsilon^2}\,
{\cal C}_{\rm unp}^{\rm VCS} ({\cal F}_{++},{\cal F}_{++}^\ast\big|{\cal F}_{-+},{\cal F}_{-+}^\ast  )
+ 8 \frac{1-y - \frac{\epsilon ^2}{4} y^2
}{
 1+\epsilon^2
}\, {\cal C}_{\rm unp}^{\rm VCS}({\cal F}_{0+}, {\cal F}_{0+}^\ast) \, ,
\nonumber\\ {\ }\\
\label{Res-Mom-DVCS-1-imp}
\left\{ {c^{\rm VCS}_{1,\rm unp} \atop s^{\rm VCS}_{1,\rm unp}} \right\} \!\!\!&=&\!\!\!
\frac{4 \sqrt{2} \sqrt{1-y - \frac{\epsilon^2}{4} y^2}}{1+\epsilon^2}\,
\left\{ { 2 - y  \atop - \lambda y\sqrt{1 + \epsilon^2}} \right\}
\left\{ {\Re{\rm e} \atop \Im{\rm m}} \right\} \,
{\cal C}^{\rm VCS}_{\rm unp} \left( {\cal F}_{0+}\big|{\cal F}_{++}^\ast,{\cal F}_{-+}^\ast\right),
\\ {\ }\nonumber\\
c^{\rm VCS}_{2,{\rm unp}}
\!\!\!&=&\!\!\!
8\frac{1-y - \frac{\epsilon^2}{4} y^2}{1+\epsilon^2} \,
\Re{\rm e}\,
{\cal C}^{\rm VCS}_{{\rm unp}} \left( {\cal F}_{-+}, {\cal F}_{++}^\ast \right).
\end{eqnarray}
\vspace{0.0cm}
\begin{itemize}
\item Longitudinally polarized target
\end{itemize}
\begin{eqnarray}
\label{Res-Mom-DVCS-0-impLP} c_{0,{\rm LP}}^{\rm VCS}
\!\!\!&=&\!\!\! \frac{2 \lambda\Lambda\,   y
(2-y)}{1+\epsilon^2}\, {\cal C}^{\rm VCS}_{\rm LP} ({\cal F}_{++},{\cal F}_{++}^\ast\big|{\cal F}_{-+},{\cal F}_{-+}^\ast)
\, ,
\\ {\ } \nonumber \\
\label{Res-Mom-DVCS-1-impLP}
\left\{ { c^{\rm VCS}_{1,{\rm LP}} \atop s^{\rm VCS}_{1,{\rm LP}} } \right\} \!\!\!&=&\!\!\!
-4\sqrt{2}\Lambda  \frac{\sqrt{1-y - \frac{\epsilon^2}{4} y^2} }{(1+\epsilon^2)^{3/2}}
\left\{{ - \lambda y \sqrt{1 + \epsilon^2}  \atop  2 - y} \right\}
\left\{ {\Re{\rm e} \atop \Im{\rm m}} \right\}
{\cal C}^{\rm VCS}_{\rm LP}\left( {\cal F}_{0+}\big|{\cal F}_{++}^\ast,{\cal F}^\ast_{-+}\right),
\\ {\ }\nonumber\\
s^{\rm VCS}_{2,{\rm LP}}
\!\!\!&=&\!\!\! -8\Lambda  \frac{1-y - \frac{\epsilon ^2}{4} y^2}{\left(1+\epsilon^2\right)^{3/2}}
\,
\Im{\rm m}\, {\cal C}^{\rm VCS}_{{\rm LP}}
\left({\cal F}_{-+},{\cal F}_{++}^\ast\right).
\end{eqnarray}
\begin{itemize}
\item Transversally polarized target
\end{itemize}
\begin{eqnarray}
\label{Res-Mom-DVCS-0-impTP} c_{0,{\rm TP}}^{\rm VCS}
\!\!\!&=&\!\!\!
- 4 \frac{1-y - \frac{\epsilon ^2}{4} y^2
}{
 1+\epsilon^2
}\, \frac{\widetilde K}{M}\, \Lambda\sin(\varphi) \Im{\rm m} \, {\cal C}_{{\rm TP}_-}^{\rm VCS}({\cal F}_{0+}, {\cal F}_{0+}^\ast)
\\
\!\!\!&+&\!\!\!
\frac{2-y}{1+\epsilon^2}\frac{{\widetilde K}}{M} \Bigg[\lambda\Lambda \cos(\varphi)\, y\,
{\cal C}^{\rm VCS}_{{\rm TP}+}
\nonumber\\
&&\hspace{14mm}
-
\Lambda\sin(\varphi)  \frac{2 - 2 y + y^2 + \ft12 \epsilon^2 y^2}{2-y}
\Im{\rm m} \, {\cal C}^{\rm VCS}_{{\rm TP}-}
\Bigg]({\cal F}_{++},{\cal F}_{++}^\ast\big|{\cal F}_{-+},{\cal F}^\ast_{-+})
,
\nonumber \\
{\ }  \nonumber \\
\label{Res-Mom-DVCS-1-impTP} \left\{ { c^{\rm VCS}_{1,{\rm TP}}
\atop s^{\rm VCS}_{1,{\rm TP}} } \right\}\!\!\!&=&\!\!\!-2\sqrt{2}
\frac{\sqrt{1-y - \frac{\epsilon^2}{4} y^2} }{1+\epsilon^2} \frac{{\widetilde K}}{M}
\Bigg[
\frac{\Lambda\cos(\varphi)}{\sqrt{1+\epsilon^2}} \left\{{
-\lambda y  \sqrt{1+\epsilon^2} \atop  2-y } \right\} \left\{ {\Re{\rm
e} \atop \Im{\rm m}} \right\}  \, {\cal C}^{\rm VCS}_{{\rm TP}+}
\\
&&\phantom{-\frac{\widetilde K}{M(1+\epsilon^2)} \frac{K}{(2-\xB)}}+
\Lambda\sin(\varphi) \left\{{2-y   \atop  \lambda y\sqrt{1+\epsilon^2}} \right\} \left\{ {\Im{\rm m}} \atop   {\Re{\rm e}}  \right\}  \, {\cal C}^{\rm VCS}_{{\rm TP}-}
\Bigg]\left( {\cal F}_{0+}\big|{\cal F}_{++}^\ast,{\cal F}^\ast_{-+}\right),
 \nonumber\\{\ } \nonumber\\
\left\{
{ c^{\rm VCS}_{2,{\rm TP}} \atop s^{\rm VCS}_{2,{\rm TP}} }
\right\}
\!\!\!&=&\!\!\!  -4 \frac{1-y - \frac{\epsilon ^2}{4} y^2 }{(1+\epsilon^2)^{3/2}}
\frac{{\widetilde K}}{M}\,
\Im{\rm m}
\left\{
{\sqrt{1+\epsilon^2}\, \Lambda\sin(\varphi)\, {\cal C}^{\rm VCS}_{{\rm TP}-}
\atop
\phantom{\sqrt{1+\epsilon^2}\,}\Lambda\cos(\varphi)\, {\cal C}^{\rm VCS}_{{\rm TP}+ } }
\right\}
\left( {\cal F}_{-+},{\cal F}_{++}^\ast \right).
\end{eqnarray}
Here we introduced incoherent sums of  transverse helicity-flip and non-flip CFFs:
\begin{eqnarray}
\label{Ccal-1}
{\cal C}_{\rm S}^{\rm VCS} ({\cal F}_{++},{\cal F}_{++}^\ast\big|{\cal F}_{-+},{\cal F}_{-+}^\ast  )
\!\!\!&=&\!\!\!
{\cal C}_{\rm S}^{\rm VCS} ({\cal F}_{++},{\cal F}_{++}^\ast)
\pm {\cal C}_{\rm S}^{\rm VCS}({\cal F}_{-+},{\cal F}_{-+}^\ast )\,,
\\
{\ }\nonumber\\
\label{Ccal-2}
{\cal C}_{\rm S}^{\rm VCS} ({\cal F}_{0+}\big|{\cal F}_{++}^\ast,{\cal F}_{-+}^\ast  )
\!\!\!&=&\!\!\!
{\cal C}_{\rm S}^{\rm VCS} ({\cal F}_{0+},{\cal F}_{++}^\ast)
\pm
{\cal C}_{\rm S}^{\rm VCS}({\cal F}_{0+},{\cal F}_{-+}^\ast )\,,
\end{eqnarray}
where the $+$  and $-$ signs apply for ${\rm S}\in\{{\rm unp},{\rm TP}-\}$ and ${\rm S}\in\{{\rm LP},{\rm TP}+\}$ cases, respectively.

By means of Eq.~(\ref{HelAmpDVCS}), we find the following exact results for the bilinear CFF combinations that enter the VCS squared term:
\begin{itemize}
\item Unpolarized target
\end{itemize}
\vspace{-4mm}
\begin{eqnarray}
\label{Def-CDVCSunp}
{\cal C}_{\rm unp}^{\rm VCS} &\!\!\!=\!\!\!&
\frac{4 (1-\xB)\left(1+\frac{\xB t}{\Q^2}\right)}{\left(2-\xB + \frac{\xB t}{\Q^2}\right)^2 }
\big[{\cal H} {\cal H}^\ast + \widetilde{\cal H} \widetilde{\cal H}^\ast\big]+
\,\frac{\left(2+\frac{t}{\Q^2} \right)\epsilon^2 }{\left(2-\xB + \frac{\xB t}{\Q^2}\right)^2 } \widetilde{\cal H} \widetilde{\cal H}^\ast
- \frac{t}{4 M^2} {\cal E} {\cal E}^\ast
\\
&& -\frac{\xB^2}{\left(2-\xB + \frac{\xB t}{\Q^2}\right)^2 }
 \left\{
 \left(\!1+\frac{t}{\Q^2}\!\right)^2 \big[ {\cal H} {\cal E}^\ast + {\cal E} {\cal H}^\ast + {\cal E} {\cal E}^\ast \big] +
 \widetilde{\cal H} \widetilde{\cal E}^\ast + \widetilde{\cal E} \widetilde{\cal H}^\ast + \frac{t}{4 M^2} \widetilde{\cal E} \widetilde{\cal E}^\ast
\right\}, \nonumber
\end{eqnarray}\\ 
\begin{itemize}
\item Longitudinally polarized  target
\end{itemize}
\vspace{-4mm}
\begin{eqnarray}
\label{Def-CDVCSLP}
{\cal C}_{\rm LP}^{\rm VCS} &\!\!\!=\!\!\!&
\frac{4(1-\xB)\left(1+\frac{\xB t}{\Q^2}\right)+2\left(1-\xB+ \frac{\Q^2+t}{2\Q^2}\right)\epsilon^2}{\left(2-\xB + \frac{\xB t}{\Q^2}\right)^2 }
\big[ {\cal H} \widetilde{\cal H}^\ast + \widetilde{\cal H} {\cal H}^\ast \big]
\\
&&-\frac{\xB^2 \left(1 + \frac{\xB t}{\Q^2}- (1-\xB) \frac{t}{\Q^2}\right)}{\left(2-\xB + \frac{\xB t}{\Q^2}\right)^2 }\,
\big[ {\cal H} \widetilde{\cal E}^\ast + \widetilde{\cal E} {\cal H}^\ast +
\widetilde{\cal H} {\cal E}^\ast + {\cal E} \widetilde{\cal H}^\ast \big]
\nonumber\\
&&-\frac{4\xB (1-\xB) \left(1+ \frac{\xB t}{\Q^2}\right) \frac{t}{\Q^2}   +
\xB \left(1+\frac{t}{\Q^2}\right)^2 \epsilon^2 }{2\left(2-\xB + \frac{\xB t}{\Q^2}\right)^2}\,
\big[ \widetilde{\cal H} {\cal E}^\ast + {\cal E} \widetilde{\cal H}^\ast \big]
\nonumber\\
&& - \frac{\xB}{2-\xB + \frac{\xB t}{\Q^2}}
\left(\frac{\xB^2\left(1+\frac{t}{\Q^2}\right)^2 }{2\left(2-\xB + \frac{\xB t}{\Q^2}\right)}+\frac{t}{4 M^2}\right)\, \big[ {\cal E}
\widetilde{\cal E}^\ast + \widetilde{\cal E} {\cal E}^\ast \big] ,
\nonumber
\end{eqnarray}\\
\begin{itemize}
\item  Transversally polarized target
\end{itemize}
\vspace{-4mm}
\begin{eqnarray}
\label{Def-CDVCSTP+}
{\cal C}_{{\rm TP}+}^{\rm VCS} &\!\!\!=\!\!\!&
\frac{2}{\left(2-\xB+ \frac{\xB t}{ \Q^2}\right)^2
}
\Bigg\{
\xB\big[ {\cal H} \widetilde{\cal E}^\ast + \widetilde{\cal E} {\cal H}^\ast \big]
+
\frac{ 4  \xB (1-2\xB)M^2}{\Q^2} \big[ {\cal H}
\widetilde{\cal H}^\ast + \widetilde{\cal H} {\cal H}^\ast \big]
\\
&&
-
\left(2-\xB + \frac{\xB  t}{\Q^2} + \left(3+\frac{t}{\Q^2}\right)\frac{\epsilon^2}{2}\right)
\big[ \widetilde{\cal H} {\cal E}^\ast + {\cal E} \widetilde{\cal H}^\ast \big]
+
\frac{\xB^2}{2}  \left(1- \frac{t}{\Q^2}\right)  \big[{\cal E} \widetilde{\cal E}^\ast + \widetilde{\cal E} {\cal E}^\ast \big]
\!
\Bigg\} ,
\nonumber\\
{\ }\nonumber\\
\label{Def-CDVCSTP-}
{\cal C}_{{\rm TP}-}^{\rm VCS} &\!\!\!=\!\!\!&
\frac{2}{2-\xB+ \frac{\xB t}{ \Q^2}} \big[{\cal H} {\cal E}^\ast - {\cal E} {\cal H}^\ast\big]
 - \frac{2\xB}{\left(2-\xB+ \frac{\xB t}{ \Q^2}\right)^2} \big[ \widetilde{\cal H} \widetilde{\cal E}^\ast - \widetilde{\cal E} \widetilde{\cal H}^\ast
\big]
 \, .
\end{eqnarray}

Let us point out at this moment that the transverse double-flip CFFs, given in the approximation (\ref{cff2-cff+-}), can be expressed by the gluon transversity
CFFs  which were introduced in Ref.~\cite{Diehl:2001pm}  via the following linear map (cf.\ (\ref{Tamp}) and (\ref{FT2FT-1})--(\ref{FT2FT-4}) below)
\begin{eqnarray}
\label{FT2FTVCS-1}
{\cal H}_{T}  \!\!\!&=&\!\!\!
{\cal H}^{\scriptsize\cite{Diehl:2001pm}}_{T} +  {\cal E}^{\scriptsize\cite{Diehl:2001pm}}_{T}+ 2 \widetilde{\cal H}^{\scriptsize\cite{Diehl:2001pm}}_{T}
\\
&+&
\frac{t}{\widetilde{K}^2}\left[
\left(1-\xB+\frac{\xB t}{2\Q^2}\right)
\left(1+\frac{\xB t}{2\Q^2}\right){\cal H}^{\scriptsize\cite{Diehl:2001pm}}_{T}-\frac{\xB^2}{4}{\cal E}^{\scriptsize\cite{Diehl:2001pm}}_{T}+
\frac{\xB}{4}\left(2-\xB+\frac{\xB t}{\Q^2}\right)\widetilde{\cal E}^{\scriptsize\cite{Diehl:2001pm}}_{T}
\right]
\nonumber\\[3mm]
{\cal E}_{T}  \!\!\!&=&\!\!\! -2\widetilde{\cal H}^{\scriptsize\cite{Diehl:2001pm}}_{T}
\\
&-&
\frac{4M^2}{\widetilde{K}^2}\left[
\left(1-\xB+\frac{\xB t}{2\Q^2}\right)\left(1+\frac{\xB t}{2\Q^2}\right){\cal H}_{T}^{\scriptsize\cite{Diehl:2001pm}}-
\frac{\xB^2}{4}{\cal E}^{\scriptsize\cite{Diehl:2001pm}}_{T}+\frac{\xB}{4}\left(2-\xB+\frac{\xB t}{\Q^2}\right)\widetilde{\cal E}^{\scriptsize\cite{Diehl:2001pm}}_{T}
\right]
\nonumber\\[3mm]
\widetilde{\cal H}_{T}  \!\!\!&=&\!\!\!
\frac{M^2\left(2-\xB+\frac{\xB t}{\Q^2}\right)}{\widetilde{K}^2}\left[
\xB{\cal H}^{\scriptsize\cite{Diehl:2001pm}}_{T}+\frac{\xB t}{4M^2} {\cal E}^{\scriptsize\cite{Diehl:2001pm}}_{T}
-\left(2-\xB+\frac{\xB t}{\Q^2}\right) \frac{t}{4M^2}\widetilde{\cal E}^{\scriptsize\cite{Diehl:2001pm}}_{T}
\right],
\\[3mm]
\label{FT2FTVCS-4}
\widetilde{\cal E}_{T}  \!\!\!&=&\!\!\!
\frac{M^2\left(2-\xB+\frac{\xB t}{\Q^2}\right)}{\widetilde{K}^2}\left[
\left(2-\xB+\frac{\xB t}{\Q^2}\right)\widetilde{\cal E}^{\scriptsize\cite{Diehl:2001pm}}_{T}-
\frac{4M^2\xB^2+4\widetilde{K}^2}{\xB t}{\cal H}_{T}^{\scriptsize\cite{Diehl:2001pm}}
-\xB{\cal E}_{T}^{\scriptsize\cite{Diehl:2001pm}}\right]
\, .
\end{eqnarray}
These obviously suffer from kinematical $1/\tK^2$ singularities. In the case of our unaltered twist-three CFF definitions such kinematical singularities
cancel each other while for gluon transversity contributions we observe a partial cancelation in all four expressions for $\cal C$-functions (\ref{Def-CDVCSunp})--(\ref{Def-CDVCSTP-}):
$$
 {\cal C}_{\rm S}^{\rm VCS}({\cal F}_{T}, {\cal F}^\ast_{T}) \propto \tK^{-2}  \quad\mbox{for}\quad S\in  \{ {\rm unp},{\rm LP},{\rm TP}+,{\rm TP}-\}
$$
and
$$ {\cal C}_{\rm S}^{\rm VCS}({\cal F}_{T}, {\cal F}^\ast) \propto \left\{  \tK^{0} \atop \tK^{-2} \right\}
\quad\mbox{for}\quad S\in \left\{ { {\rm unp},{\rm LP} \atop  {\rm TP}+,{\rm TP}- }  \right\}.
$$
If we neglect power suppressed contributions, we
retrieve for the $\cal C$-functions the same functional form that was already found in Ref.~\cite{Belitsky:2001ns}. Moreover, the behavior of helicity flip CFFs, indicated by the additional $\tK$ and
$\tK^2$ factors in the relations (\ref{cff2-cff0+}), (\ref{cff2-cff+-}), ensure that all first (second) and second (first) order harmonics for
${\rm unp}$ and  ${\rm LP}$ (${\rm TP}+$ and ${\rm TP}-$) cases vanish in the limit $t\to t_{\rm min}$  as $\sqrt{t_{\rm min}-t}$ and  $t_{\rm min}-t$, respectively. Finally, we add that our results are consistent with the expanded ones of Ref.~\cite{Belitsky:2001ns} and that they have been numerically cross checked by means of the leptonic tensor (\ref{L-tensor}) and a hadronic Compton scattering tensor, given below in Eq.~(\ref{BornAmplitude}).

\subsection{Interference term}
\label{sect-INT}

Let us now turn to the interference term. Inserting the completeness condition for the initial
and final photon polarization states, one finds the interference term ${\cal I}$ as a linear superposition
\begin{eqnarray}
\label{Def-Sqa-Int-Hel}
{\cal I} =  \frac{\pm e^6}{t \, {\cal P}_1(\phi) {\cal P}_2(\phi)}
\sum_{a = {\scriptscriptstyle -}, 0, {\scriptscriptstyle +}}
\sum_{b = {\scriptscriptstyle -}, {\scriptscriptstyle +}}
\sum_{S^\prime}
\left\{
{\cal L}^\rho_{a b}(\lambda,\phi)  {\cal T}_{a b}J^\dagger_\rho
+
\left({\cal L}^\rho_{a b}(\lambda,\phi) {\cal T}_{a b}J^\dagger_\rho\right)^\ast \right\}
\, ,
\end{eqnarray}
of the products of hadronic and leptonic helicity amplitudes. The former were defined earlier in Eq.\
(\ref{cal-Tab}) and the matrix element of the quark electromagnetic current (\ref{quarkEMcurrent})
\begin{eqnarray}
\label{EMcurrentFFs}
J_\rho = \bar{u}_2 \Gamma_\rho (\Delta) u_1
\qquad
\mbox{with}
\qquad
\Gamma_\rho (\Delta)
=
\gamma_\rho\,  F_1(t) + i \sigma_{\rho \sigma} \frac{\Delta^\sigma}{2 M}\, F_2(t)
\end{eqnarray}
is determined by the Dirac and Pauli form factors $F_1$ and $F_2$. Moreover, $1/{\cal P}_1 (\phi) {\cal P}_2 (\phi)$ stands for the product of rescaled propagators of the Bethe-Heitler amplitude, specified in Eqs.~(28)--(31) of Ref.~\cite{Belitsky:2001ns}.

First we consider the hadronic part ${\cal T}_{a b}J^\dagger_\rho$ of the interference term (\ref{Def-Sqa-Int-Hel}) which, similarly to the leptonic part, has one open Lorentz index.
The former is given by the VCS helicity amplitudes (\ref{cal-Tab})--(\ref{cal-TabA}) and the electromagnetic current (\ref{EMcurrentFFs}). The
resulting (axial-) vector amplitudes will be decomposed in the basis (\ref{AveagedMomenta}) of the physical momenta. Due to the electromagnetic
current conservation, we can neglect terms proportional to $\Delta_\rho$, which vanish upon contraction with the leptonic part ${\cal L}_{ab}^\rho$.
The summation over the final nucleon polarization states yields the following expression in the vector sector
\begin{eqnarray}
\label{Def-Intsum-V}
\sum_{S^\prime} {\cal V}({\cal F}) J^\dagger_\rho
\!\!\!&=&\!\!\!
p_{\rho} \left[ {\cal C}^{{\cal I}}_{\rm unp}({\cal F}) - {\cal C}^{{\cal I},A}_{\rm unp}  \right]\! ({\cal F})
+
2 q_{\rho}  \frac{t}{\Q^2}  {\cal C}^{{\cal I},V}_{\rm unp} ({\cal F})
\\
&&\!\!\!\!\!\!\!
-p_{\rho} \frac{\Lambda \sin(\theta)\sin(\varphi) M}{i \widetilde K} \left[ {\cal C}^{{\cal I}}_{{\rm TP}-} - {\cal C}^{{\cal I},A}_{{\rm TP}-}  \right]\! ({\cal F})\,
-
2 q_{\rho}  \frac{t}{\Q^2} \frac{\Lambda \sin(\theta)\sin(\varphi) M}{i \widetilde K} {\cal C}^{{\cal I},V}_{{\rm TP}-} ({\cal F})
\nonumber\\
&&\!\!\!\!\!\!\!+ \frac{2 i \varepsilon_{pq\Delta\rho}}{\Q^2} \left[
\frac{\Lambda \cos(\theta)}{\sqrt{1+\epsilon^2}} {\cal C}^{{\cal I},V}_{\rm LP}
 +
\frac{\Lambda\sin(\theta)\cos(\varphi) M}{\sqrt{1+\epsilon^2}\widetilde K} \,
{\cal C}^{{\cal I},V}_{{\rm TP}+}
\right]({\cal F})\,,
\nonumber
\end{eqnarray}
and analogously in the axial-vector case
\begin{eqnarray}
\label{Def-Intsum-A}
\sum_{S^\prime}  {\cal A}({\cal F}) J^\dagger_\rho
\!\!\!&=&\!\!\!
 p_{\rho} \frac{\Lambda\cos(\theta)}{\sqrt{1+\epsilon^2}}
\left[{\cal C}^{{\cal I}}_{\rm LP} - {\cal C}^{{\cal I},V}_{\rm LP}\right] \!({\cal F})
+
2 q_\rho \frac{t}{\Q^2}\,  \frac{\Lambda\cos(\theta)}{\sqrt{1+\epsilon^2}}\, {\cal C}^{{\cal I},A}_{\rm LP} ({\cal F})
\\
&&\!\!\!\!\!\!\!+
p_{\rho} \frac{\Lambda\sin(\theta)\cos(\varphi) M}{\sqrt{1+\epsilon^2}\widetilde K} \left[ {\cal C}^{{\cal I}}_{{\rm TP}+}
- {\cal C}^{{\cal I},V}_{{\rm TP}+}  \right]\! ({\cal F})+
2 q_{\rho}  \frac{t}{\Q^2} \frac{\Lambda\sin(\theta)\cos(\varphi) M}{\sqrt{1+\epsilon^2}\widetilde K} {\cal C}^{{\cal I},A}_{{\rm TP}+} ({\cal F})
\nonumber\\
&&\!\!\!\!\!\!\!+
 \frac{2 i \varepsilon_{pq\Delta\rho}}{\Q^2}\left[ {\cal C}_{\rm unp}^{{\cal I},A} - \frac{\Lambda\sin(\theta)\sin(\varphi) M}{i \widetilde K} \,
{\cal C}^{{\cal I},A}_{{\rm TP}-}  \right](\cal F)\, .
\nonumber
\end{eqnarray}
As it becomes obvious from these two equations, the result for the transversely polarized target can be obtained from the ones of unpolarized and
longitudinally polarized cases by the following substitutions
\begin{eqnarray}
{\cal C}_{\rm unp}^{{\cal I},\cdots}({\cal F})\quad &\Rightarrow&\quad - \Lambda\sin(\theta)\sin(\varphi)  \frac{M}{i \widetilde K}\,
{\cal C}_{{\rm TP}-}^{{\cal I},\cdots}({\cal F})\,,
\\
\frac{\Lambda \cos(\theta)}{\sqrt{1+\epsilon^2}}\,{\cal C}_{\rm LP}^{{\cal I},\cdots}({\cal F})\quad &\Rightarrow&\quad
\frac{ \Lambda\sin(\theta)\cos(\varphi) }{\sqrt{1+\epsilon^2}}\,\frac{M}{\widetilde K}\, {\cal C}_{{\rm TP}+}^{{\cal I},\cdots}({\cal F})\,.
\end{eqnarray}

Now we turn to the leptonic helicity amplitudes,
\begin{eqnarray}
{\cal L}^\rho_{a b}(\lambda,\phi)
=
\varepsilon_1^{\mu\ast}(a) L_{\mu\phantom{\rho}\nu}^{\phantom{\mu}\rho}  \varepsilon_2^\nu(b)
\, ,
\end{eqnarray}
where
\begin{align}
L_{\mu\rho\nu}
=
\mathcal{Q}^{-6} (k-q_2)^2 (k-\Delta)^2
\tr \ft12 (1 - \lambda \gamma_5)
\left[
\gamma_\rho ({\not\! k} - {\not\!\Delta})^{-1} \gamma_\nu
+
\gamma_\rho ({\not\! k}^]\prime + {\not\!\Delta})^{-1} \gamma_\nu
\right]
\gamma_\mu {\not\! k}
\, .
\end{align}
This amplitude contains the entire azimuthal angular dependence of the interference term. Its contraction with the Lorentz vectors entering the decomposition
of the hadronic amplitudes (\ref{Def-Intsum-V}) and (\ref{Def-Intsum-A}) introduces the coefficients for the lepton helicity-independent,
\begin{align}
\label{Cab&Sab}
\left\{
\begin{array}{c}
  C_{ab} \\
  C^{V}_{ab} \\
  C^{A}_{ab}
\end{array}
  \right\}
  = \Re{\rm e} \, {\cal L}^\rho_{a b}(\lambda=0,\phi)
\left\{
\begin{array}{c}
   p_{\rho} \\
  2 q_{\rho}  \frac{t}{\Q^2} \\
 \frac{2 i \varepsilon_{pq\Delta\rho}}{\Q^2}
\end{array}
\right\}
\, , \qquad
\left\{
\begin{array}{c}
   \delta S_{ab} \\
   \delta S^{V}_{ab} \\
   \delta S^{A}_{ab}
\end{array}
  \right\}
  = \Im{\rm m} \frac{{\cal L}^\rho_{a b}(\lambda=0,\phi)}{\sqrt{1+\epsilon^2}}
\left\{
\begin{array}{c}
   p_{\rho} \\
  2 q_{\rho}  \frac{t}{\Q^2} \\
 \frac{2 i \varepsilon_{pq\Delta\rho}}{\Q^2}
\end{array}
\right\}\,,
\end{align}
and helicity-dependent components
\begin{align}
\label{dCab&dSab}
\left\{
\begin{array}{c}
  S_{ab} \\
  S^{V}_{ab} \\
  S^{A}_{ab}
\end{array}
  \right\}
  = \Im{\rm m} \, \frac{\partial}{\partial\lambda} {\cal L}^\rho_{a b}(\lambda,\phi)
\left\{
\begin{array}{c}
   p_{\rho} \\
  2 q_{\rho}  \frac{t}{\Q^2} \\
 \frac{2 i \varepsilon_{pq\Delta\rho}}{\Q^2}
\end{array}
\right\}\, , \qquad
\left\{
\begin{array}{c}
  \delta C_{ab} \\
   \delta C^{V}_{ab} \\
   \delta C^{A}_{ab}
\end{array}
  \right\}
  = \Re{\rm e} \, \frac{\partial}{\partial\lambda} \frac{{\cal L}^\rho_{a b}(\lambda,\phi)}{\sqrt{1+\epsilon^2}}
\left\{
\begin{array}{c}
   p_{\rho} \\
  2 q_{\rho}  \frac{t}{\Q^2} \\
 \frac{2 i \varepsilon_{pq\Delta\rho}}{\Q^2}
\end{array}
\right\} \, ,
\end{align}
respectively. Notice that in the deeply virtual regime, the leptonic coefficients with $V$ and $A$ superscripts are power suppressed.
We also introduce the harmonic expansion of these coefficients
\begin{eqnarray}
\label{C-S-FourExpansion}
(\delta)C_{ab}(\phi) = \frac{1}{\xB y^3}\sum_{n=0}^3 \cos(n \phi)\, (\delta)C_{ab}(n)\,, \quad
(\delta)S_{ab}(\phi) = \frac{1}{\xB y^3} \sum_{n=1}^3 \sin(n \phi)\, (\delta)S_{ab}(n)\,,
\end{eqnarray}
where we include a conventional factor $1/(\xB y^3)$.

As for  the square of the virtual Compton scattering amplitude, listed in Sect.~\ref{sect-TDVCS}, we decompose the interference term
in a Fourier harmonic sum and label entering contributions $c^{\cal I}_{k,{\rm S}}$ with respect to the polarization of the incoming nucleon state
${\rm S}\in\{{\rm unp},{\rm LP},{{\rm TP}+},{{\rm TP}-}\}$,
\be
\label{InterferenceTerm}
{\cal I}(\phi,\varphi)
=
\frac{\pm e^6}{\xB y^3 t {\cal P}_1 (\phi) {\cal P}_2 (\phi)}
\left[
\sum_{n = 0}^3
c_{n,{\rm S}}^{\cal I}(\varphi)\, \cos(n \phi) +
\sum_{n = 1}^3  s_{n,{\rm S}}^{\cal I}(\varphi)\, \sin(n \phi)
\right]
\, .
\ee
The Fourier coefficients $c^{\cal I}_{n,{\rm S}}$ and $s^{\cal I}_{n,{\rm S}}$ are straightforwardly obtained from the definitions
given in this section and can be exactly expressed in terms of effective linear combinations of helicity dependent CFFs (\ref{cal-Tab}).

However, as it
follows from Eqs.~(\ref{Def-Intsum-V}), (\ref{Def-Intsum-A}) together with (\ref{Cab&Sab}), (\ref{dCab&dSab}), an exact calculation of the interference
term (\ref{Def-Sqa-Int-Hel}) yields a result that is given by a superposition of factorized leptonic and hadronic components. Hence, we may introduce ``effective"
hadronic linear combinations of CFFs that read for the unpolarized and transversally polarized ${\rm TP} -$ components as follows
\begin{eqnarray}
{\cal C}_{ab,{\rm S}}^{{\cal I}}\!\left(n|{\cal F}_{ab}\right)  &\!\!\!= \!\!\!&
{\cal C}^{{\cal I}}_{\rm S}\!\left({\cal F}_{ab}\right) + \frac{C^{V}_{ab} (n) }{C_{ab} (n)}\,
{\cal C}_{\rm S}^{{\cal I},V}\!\left({\cal F}_{ab}\right) +  \frac{C^{A}_{ab} (n) }{C_{ab} (n)}\,
{\cal C}_{\rm S}^{{\cal I},A}\!\left({\cal F}_{ab}\right)\,,
\nonumber
\\
\label{Cab&Sab-unp&TP-}
{\cal S}_{ab,{\rm S}}^{{\cal I}}\!\left(n|{\cal F}_{ab}\right)  &\!\!\!= \!\!\!&
{\cal C}^{{\cal I}}_{\rm S}\!\left({\cal F}_{ab}\right) + \frac{S^{V}_{ab} (n)}{S_{ab} (n)}\,
{\cal C}^{{\cal I},V}_{\rm S}\!\left({\cal F}_{ab}\right) +  \frac{S^{A}_{ab} (n) }{S_{ab} (n)}\,
{\cal C}^{{\cal I},A}_{\rm S}\!\left({\cal F}_{ab}\right)\,,
\end{eqnarray}
where ${\rm S} \in \{{\rm unp},{\rm TP}-\}$, and for the longitudinally and transversally polarized ${\rm TP}+$ parts as
\begin{eqnarray}
{\cal C}_{ab,{\rm S}}^{{\cal I}}\!\left(n|{\cal F}_{ab}\right)  &\!\!\!= \!\!\!&
{\cal C}^{{\cal I}}_{\rm S}\!\left({\cal F}_{ab}\right) + \frac{\delta C^{V}_{ab} (n) }{\delta C_{ab} (n)}\,
{\cal C}_{\rm S}^{{\cal I},V}\!\left({\cal F}_{ab}\right) +  \frac{\delta C^{A}_{ab} (n) }{\delta C_{ab} (n)}\,
{\cal C}_{\rm S}^{{\cal I},A}\!\left({\cal F}_{ab}\right)\,,
\nonumber\\
\label{Cab&Sab-LP&TP+}
{\cal S}_{ab,{\rm S}}^{{\cal I}}\!\left(n|{\cal F}_{ab}\right)  &\!\!\!= \!\!\!&
{\cal C}^{{\cal I}}_{\rm S}\!\left({\cal F}_{ab}\right) + \frac{\delta S^{V}_{ab} (n)}{\delta S_{ab} (n)}\,
{\cal C}^{{\cal I},V}_{\rm S}\!\left({\cal F}_{ab}\right) +  \frac{\delta S^{A}_{ab} (n) }{\delta S_{ab} (n)}\,
{\cal C}^{{\cal I},A}_{\rm S}\!\left({\cal F}_{ab}\right)\,,
\end{eqnarray}
with ${\rm S} \in \{{\rm LP},{\rm TP}+\}$.
The explicit results of the calculation of the leptonic coefficients $(\delta)C^{\cdots}_{ab} (n)$ and $(\delta)S^{\cdots}_{ab} (n)$, defined in Eqs.~(\ref{Cab&Sab})--(\ref{C-S-FourExpansion}), are listed in Appendix \ref{FourierHarmonics}. From what was said above it follows that the  dominant term in the deeply virtual
kinematics is given by the coefficients ${\cal C}^{\cal I}_{\rm S}$. Also it turns out that for a given harmonic all helicity amplitudes will contribute. However, in
the regime of large photon virtualities, the first harmonics are dominated by the helicity conserved CFFs ${\cal F}_{++}$ (of twist-two in power counting),
while the second ones receive leading contribution from the longitudinal-to-transverse CFFs ${\cal F}_{0+}$ (twist-three). The third harmonic is governed by the
transverse-to-transfers CFFs ${\cal F}_{-+}$, determined at twist-two level by the gluon transversity GPDs. The latter contribution yields a $\cos(3\phi)$ harmonic
for unpolarized scattering, given by the real part of CFFs, and $\sin(3\phi)$ harmonic for the longitudinally polarized part, this time expressed in terms of the
imaginary part. The transversally polarized part is determined by the imaginary part of CFF combinations, leading to a $\cos(\varphi) \sin(3\phi)$  and
$\sin(\varphi) \cos(3\phi)$ harmonics. There also appear constant terms that are relatively suppressed by $1/{\Q}$ in the amplitudes and are dominated by
twist-two operator matrix elements.

We now list the explicit expressions for the Fourier coefficients in terms of linear photon helicity-dependent CFF combinations, where the separate terms are
ordered with respect to their importance in the deeply virtual region:
\begin{itemize}
\item Unpolarized target
\end{itemize}
\vspace{-5mm}
\begin{eqnarray}
\label{c0unpI}
c^{\cal I}_{0,{\rm unp}}\!\!\!&=&\!\!\!   C_{++} (0)\, \Re{\rm e}\, {\cal C}_{++,{\rm unp}}^{\cal I} \left(0|{\cal F}_{++}\right)+
\{_{++}\,\to\, _{0+}\} + \{_{++}\,\to\, _{-+}\},
\nonumber\\
{\ }\\
\left\{{
c^{\cal I}_{1}
\atop
s^{\cal I}_{1}
}
\right\}_{\rm unp}
\!\!\!&=&\!\!\!
\left\{{
C_{++} (1)
\atop
\lambda\, S_{++} (1)
}
\right\}
\left\{{\Re{\rm e}
\atop
\Im{\rm m}
}
\right\}
\left\{{
 {\cal C}_{++}^{{\cal I}} \left(1|{\cal F}_{++}\right)
\atop
{\cal S}_{++}^{{\cal I}} \left(1|{\cal F}_{++}\right)
}
\right\}_{\rm unp}+ \{_{++}\,\to\, _{0+}\} + \{_{++}\,\to\, _{-+}\},
\nonumber\\
{\ }\\
\left\{{
c^{\cal I}_{2}
\atop
s^{\cal I}_{2}
}
\right\}_{\rm unp}
\!\!\!&=&\!\!\!
\left\{{
C_{0+}(2)
\atop
\lambda\, S_{0+} (2)
}
\right\}
\left\{{\Re{\rm e}
\atop
\Im{\rm m}
}
\right\}
\left\{{
{\cal C}_{0+}^{{\cal I}}\left(2| {\cal F}_{0+} \right)
\atop
{\cal S}_{0+}^{{\cal I}}\left(2| {\cal F}_{0+} \right)
}
\right\}_{\rm unp}+ \{_{0+}\,\to\, _{++}\} + \{_{0+}\,\to\, _{-+}\},
\nonumber\\
{\ }
\\
c^{\cal I}_{3,{\rm unp}}\!\!\!&=&\!\!\!
C_{-+} (3)\,  \Re{\rm e}\,  {\cal C}_{-+,{\rm unp}}^{{\cal I}} \left(3| {\cal F}_{-+}\right)
+ \{_{-+}\,\to\, _{++}\} + \{_{-+}\,\to\, _{0+}\}
,
\nonumber\\
{\ }
\end{eqnarray}
where the CFF combinations ${\cal C}_{ab,{\rm unp}}^{{\cal I}}$ and ${\cal S}_{ab,{\rm unp}}^{{\cal I}}$ are defined in Eqs.~(\ref{Cab&Sab-unp&TP-})
and (\ref{Cunp^I})--(\ref{Cunp^IA}) together with ${C}_{ab}^{{\cal I},\cdots}$ and ${S}_{ab}^{{\cal I},\cdots}$, listed in the Appendix \ref{C&S-leptonic}.

\begin{itemize}
\item Longitudinally polarized target [i.e., $\cos\theta$ proportional part]
\end{itemize}
\vspace{-5mm}
\begin{eqnarray}
c^{\cal I}_{0,{\rm LP}}\!\!\!&=&\!\!\!   \Lambda\, \lambda\,
\delta C_{++} (0)\, \Re{\rm e}\, {\cal C}_{++,{\rm LP}}^{\cal I} \left(0|{\cal F}_{++}\right)+ \{_{++}\,\to\, _{0+}\} + \{_{++}\,\to\, _{-+}\},
\nonumber\\
{\ }\\
\left\{{
c^{\cal I}_{1}
\atop
s^{\cal I}_{1}
}
\right\}_{\rm LP}
\!\!\!&=&\!\!\!
\Lambda \left\{{
 \lambda\, \delta C_{++} (1)
\atop
\delta S_{++} (1)
}
\right\}
\left\{{\Re{\rm e}
\atop
\Im{\rm m}
}
\right\}
\left\{{
 {\cal C}_{++}^{{\cal I}} \left(1|{\cal F}_{++}\right)
\atop
{\cal S}_{++}^{{\cal I}} \left(1|{\cal F}_{++}\right)
}
\right\}_{\rm LP}+ \{_{++}\,\to\, _{0+}\} + \{_{++}\,\to\, _{-+}\},
\nonumber\\
{\ }\\
\left\{{
c^{\cal I}_{2}
\atop
s^{\cal I}_{2}
}
\right\}_{\rm LP}
\!\!\!&=&\!\!\!
\Lambda \left\{{
\lambda\,\delta C_{0+}(2)
\atop
\delta S_{0+} (2)
}
\right\}
\left\{{\Re{\rm e}
\atop
\Im{\rm m}
}
\right\}
\left\{{
{\cal C}_{0+}^{{\cal I}}\left(2| {\cal F}_{0+} \right)
\atop
{\cal S}_{0+}^{{\cal I}}\left(2| {\cal F}_{0+} \right)
}
\right\}_{\rm LP}+ \{_{0+}\,\to\, _{++}\} + \{_{0+}\,\to\, _{-+}\},
\nonumber\\
{\ }
\\
s^{\cal I}_{3,{\rm LP}}\!\!\!&=&\!\!\!
\Lambda\, \delta S_{-+} (3)\,  \Im{\rm m}\,  {\cal C}_{-+,{\rm LP}}^{{\cal I}} \left(3| {\cal F}_{-+}\right)
+ \{_{-+}\,\to\, _{++}\} + \{_{-+}\,\to\, _{0+}\}
,
\nonumber\\
{\ }
\end{eqnarray}
where the CFF combinations ${\cal C}_{ab,{\rm LP}}^{{\cal I}}$ and ${\cal S}_{ab,{\rm LP}}^{{\cal I}}$ are defined in
Eqs.~(\ref{Cab&Sab-LP&TP+}) and (\ref{CLP^I})--(\ref{CLP^IA}) together with ${\delta C}_{ab}^{{\cal I},\cdots}$ and ${\delta S}_{ab}^{{\cal I},\cdots}$,
listed in the Appendix \ref{dC&dS-leptonic}.

\begin{itemize}
\item Transversally polarized target [i.e., $\sin\theta$ proportional part]
\end{itemize}
\vspace{-5mm}
\begin{eqnarray}
c^{\cal I}_{0,{\rm TP}}\!\!\!&=&\!\!\!
\lambda\,\Lambda \cos(\varphi)\, \frac{M}{\widetilde{K}}\, \delta C_{++} (0)\, \Re{\rm e}\, {\cal C}_{++,{\rm TP}+}^{\cal I} \left(0|{\cal F}_{++}\right)
\nonumber\\
&-&\!\!\!
\Lambda \sin(\varphi)\, \frac{M}{\widetilde{K}}\, C_{++} (0)\, \Im{\rm m}\, {\cal C}_{++,{\rm TP}-}^{\cal I} \left(0|{\cal F}_{++}\right)
+ \{_{++}\,\to\, _{0+}\} + \{_{++}\to _{-+}\},
\\
{\ }
\nonumber\\
\left\{{
c^{\cal I}_{1}
\atop
s^{\cal I}_{1}
}
\right\}_{\rm TP}
\!\!\!&=&\!\!\!
\Lambda\cos(\varphi)\, \frac{M}{\widetilde{K}}\, \left\{{
 \lambda\, \delta C_{++} (1)
\atop
\delta S_{++} (1)
}
\right\}
\left\{{\Re{\rm e}
\atop
\Im{\rm m}
}
\right\}
\left\{{
 {\cal C}_{++}^{{\cal I}} \left(1|{\cal F}_{++}\right)
\atop
{\cal S}_{++}^{{\cal I}} \left(1|{\cal F}_{++}\right)
}
\right\}_{{\rm TP}+}
\nonumber\\
&+& \Lambda \sin({\varphi}) \frac{M}{\widetilde{K}}\, \left\{{
- C_{++} (1)
\atop
 \lambda\, S_{++} (1)
}
\right\}
\left\{{\Im {\rm m}
\atop
\Re{\rm e}
}
\right\}
\left\{{
 {\cal C}_{++}^{{\cal I}} \left(1|{\cal F}_{++}\right)
\atop
{\cal S}_{++}^{{\cal I}} \left(1|{\cal F}_{++}\right)
}
\right\}_{{\rm TP}-}+ \{_{++}\,\to\, _{0+}\} + \{_{++}\,\to\, _{-+}\},
\nonumber\\
{\ }\\
\left\{{
c^{\cal I}_{2}
\atop
s^{\cal I}_{2}
}
\right\}_{\rm TP}
\!\!\!&=&\!\!\!
\Lambda \cos(\varphi)\, \frac{M}{\widetilde{K}}\,
\left\{{
\lambda\,\delta C_{0+}(2)
\atop
\delta S_{0+} (2)
}
\right\}
\left\{{\Re{\rm e}
\atop
\Im{\rm m}
}
\right\}
\left\{{
{\cal C}_{0+}^{{\cal I}}\left(2| {\cal F}_{0+} \right)
\atop
{\cal S}_{0+}^{{\cal I}}\left(2| {\cal F}_{0+} \right)
}
\right\}_{{\rm TP}+}
\nonumber\\
&+& \Lambda \sin({\varphi}) \frac{M}{\widetilde{K}}\, \left\{{
- C_{++} (2)
\atop
 \lambda\, S_{++} (2)
}
\right\}
\left\{{\Im {\rm m}
\atop
\Re{\rm e}
}
\right\}
\left\{{
 {\cal C}_{0+}^{{\cal I}} \left(2|{\cal F}_{0+}\right)
\atop
{\cal S}_{0+}^{{\cal I}} \left(2|{\cal F}_{0+}\right)
}
\right\}_{{\rm TP}-}+ \{_{0+}\,\to\, _{++}\} + \{_{0+}\,\to\, _{-+}\},
\nonumber\\
{\ }
\\
s^{\cal I}_{3,{\rm TP}}\!\!\!&=&\!\!\!
\Lambda \cos(\varphi)\, \frac{M}{\widetilde{K}}\, \delta S_{-+} (3)\,  \Im{\rm m}\,  {\cal C}_{-+,{\rm TP}+}^{{\cal I}} \left(3| {\cal F}_{-+}\right)
+ \{_{-+}\,\to\, _{++}\} + \{_{-+}\,\to\, _{0+}\}\,,
\\
c^{\cal I}_{3,{\rm TP}}\!\!\!&=&\!\!\!
-\Lambda \sin(\varphi)\, \frac{M}{\widetilde{K}}\, C_{-+} (3)\,  \Im{\rm m}\,  {\cal C}_{-+,{\rm TP}-}^{{\cal I}} \left(3| {\cal F}_{-+}\right)
+ \{_{-+}\,\to\, _{++}\} + \{_{-+}\,\to\, _{0+}\}\,,
\label{c3TPI}
\end{eqnarray}
where the CFF combinations ${\cal C}_{ab,{\rm TP}-}^{{\cal I}}$ and ${\cal S}_{ab,{\rm TP}-}^{{\cal I}}$
[${\cal C}_{ab,{\rm TP}+}^{{\cal I}}$ and ${\cal S}_{ab,{\rm TP}+}^{{\cal I}}$] are defined in
 Eqs.~(\ref{Cab&Sab-unp&TP-}) and (\ref{CTP-^I})--(\ref{CTP-^IA}) [Eqs.~(\ref{Cab&Sab-LP&TP+}) and (\ref{CTP+^I})--(\ref{CTP+^IA})]
together with ${C}_{ab}^{{\cal I},\cdots}$ and ${S}_{ab}^{{\cal I},\cdots}$
[${\delta C}_{ab}^{{\cal I},\cdots}$ and ${\delta  S}_{ab}^{{\cal I},\cdots}$], listed in Appendix \ref{C&S-leptonic} [Appendix \ref{dC&dS-leptonic}].

For the linear combinations of CFFs, evaluated from Eqs.~(\ref{Def-Intsum-V}) and
(\ref{Def-Intsum-A}), we find for the helicity dependent CFFs the following exact expressions:
\begin{itemize}
\item Unpolarized target
\end{itemize}
\vspace{-4mm}
\begin{eqnarray}
\label{Cunp^I}
{\cal C}_{\rm unp}^{{\cal I}} {(\cal F})
\!\!\!&=&\!\!\!
F_1 {\cal H} - \frac{t}{4 M^2}  F_2 {\cal E}
+
\frac{\xB}{2 - \xB +  \frac{\xB t}{\Q^2}}
(F_1+F_2)\widetilde{\cal H}  \,,
\\
\label{Cunp^IV}
{\cal C}^{{\cal I},V}_{\rm unp} ({\cal F})
\!\!\!&=&\!\!\!
\frac{\xB}{2 - \xB +  \frac{\xB t}{\Q^2}}   (F_1+F_2) ( {\cal H}+  {\cal E})  \, ,
\\
\label{Cunp^IA}
 {\cal C}^{{\cal I},A}_{\rm unp} ({\cal F})
\!\!\! &=&\!\!\! \frac{\xB}{2 - \xB +  \frac{\xB t}{\Q^2}}  (F_1+F_2)\widetilde{\cal H} \,,
\end{eqnarray}
\begin{itemize}
\item Longitudinally polarized target
\end{itemize}
\vspace{-4mm}
\begin{eqnarray}
\label{CLP^I}
{\cal C}^{{\cal I}}_{\rm LP} ({\cal F})
\!\!\!&=&\!\!\!
\frac{2}{2-\xB+\frac{\xB t}{\Q^2}}F_1\Bigg[\left\{(1-\xB)  \left(\! 1+\frac{\xB t}{\Q^2} \!\right)+\frac{\xB}{2} +
\frac{\xB^2 M^2}{\Q^2}\left(\!3+\frac{t}{\Q^2}\!\right)\right\} \widetilde{\cal H}
\\
&&\!\phantom{\frac{2}{2-\xB+\frac{\xB t}{\Q^2}}\Big\{(1-\xB)  \left(1+\frac{\xB t}{\Q^2} \right)} +\frac{\xB}{2}\left\{\frac{t}{4M^2}-\frac{\xB}{2}\left(1-\frac{t}{\Q^2}\right)\right\} \widetilde{\cal E} \Bigg]
\nonumber\\
&+&\!\!\!\frac{\xB}{2-\xB+\frac{\xB t}{\Q^2}}(F_1+F_2)\left[ {\cal H}+ \frac{\xB}{2}\left(\!1-\frac{t}{\Q^2}\!\right){\cal E}
-\frac{(1 -2\xB) t}{\Q^2} \widetilde{\cal H} -\frac{t}{4M^2} \widetilde{\cal E} \right],
\nonumber\\
\label{CLP^IV}
{\cal C}^{{\cal I},V}_{\rm LP} ({\cal F})
\!\!\!&=&\!\!\!
\frac{\xB}{2 - \xB +  \frac{\xB t}{\Q^2}}
(F_1+F_2)\left[{\cal H}+ \frac{\xB}{2} \left(1-\frac{t}{\Q^2}\right) {\cal E}\right],
\\
\label{CLP^IA}
{\cal C}^{{\cal I},A}_{\rm LP} ({\cal F})
\!\!\!&=&\!\!\!
\frac{\xB}{2 - \xB +  \frac{\xB t}{\Q^2}}  (F_1+F_2)
\left[ \widetilde{\cal H}  + 2\xB \frac{M^2}{\Q^2} \widetilde{\cal H}
+ \frac{\xB}{2} \widetilde{\cal E} \right]
\, .
\end{eqnarray}
\begin{itemize}
\item Transversally polarized target
\end{itemize}
\vspace{-4mm}
\begin{eqnarray}
\label{CTP-^I}
{\cal C}_{{\rm TP}-}^{{\cal I}} {(\cal F})
\!\!\!&=&\!\!\!
\frac{1}{2 - \xB +   \frac{\xB t}{\Q^2}}\Bigg[\frac{\widetilde{K}^2}{M^2}  (F_2 {\cal H}-F_1 {\cal E})
\\
&+&\!\!\!
\xB^2 (F_1 + F_2) \Bigg\{\left(1 + \frac{t}{\Q^2}\right)^2 \left({\cal H} +\frac{t}{4 M^2} {\cal E}\right)
-
\widetilde{\cal H} -\frac{t}{4 M^2} \widetilde{\cal E}\Bigg\}\Bigg] \,,
\nonumber\\
\label{CTP-^IV}
{\cal C}^{{\cal I},V}_{{\rm TP}-} {(\cal F})
\!\!\!&=&\!\!\!
\xB (F_1+F_2) \left[ {\cal H}+ \frac{t}{4 M^2}  {\cal E}\right]  \, ,
\\
\label{CTP-^IA}
 {\cal C}^{{\cal I},A}_{{\rm TP}-} ({\cal F})
\!\!\! &=&\!\!\!-\frac{\xB^2}{2 - \xB +  \frac{\xB t}{\Q^2}}  (F_1+F_2)\left[\widetilde{\cal H} +\frac{t}{4 M^2} \widetilde{\cal E}\right] \,,
\\
{\ }\nonumber\\
\label{CTP+^I}
{\cal C}^{{\cal I}}_{{\rm TP}+} ({\cal F})
\!\!\!&=&\!\!\!
\frac{\xB^2 \left(1-(1-2\xB)\frac{t}{\Q^2}\right)}{2 - \xB +  \frac{\xB t}{\Q^2}}
(F_1+F_2)\left[{\cal H}+ \frac{t}{4 M^2} {\cal E}-\widetilde{\cal H} -\frac{t}{4 M^2} \widetilde{\cal E}\right]
\\
&-&\!\!\!
\frac{1}{2 - \xB +  \frac{\xB t}{\Q^2}} \frac{\widetilde{K}^2}{M^2}
 \left[
   \frac{\xB}{2} F_1 \left({\cal E}-\widetilde{\cal E} -
      \frac{4 M^2}{\Q^2}  \widetilde{\cal H}\right) +  \frac{\xB}{2}  F_2 {\cal E} +
  F_2  \widetilde{\cal H} \right],
\nonumber  \\
{\ }\nonumber\\
\label{CTP+^IV}
{\cal C}^{{\cal I},V}_{{\rm TP}+} ({\cal F})
\!\!\!&=&\!\!\!
\frac{\xB}{2 - \xB +  \frac{\xB t}{\Q^2}}
(F_1+F_2)\left[\xB  \left(1 - \frac{t}{\Q^2} (1 - 2 \xB)\right) \left({\cal H} +\frac{t}{4 M^2} {\cal E}\right)-
\frac{{\widetilde K}^2}{2 M^2}  {\cal E}\right],
\nonumber\\
{\ }\\
\label{CTP+^IA}
{\cal C}^{{\cal I},A}_{{\rm TP}+} ({\cal F})
\!\!\!&=&\!\!\!
-\frac{\xB}{2 - \xB +  \frac{\xB t}{\Q^2}}  (F_1+F_2)
\Bigg[\left\{2 - \xB + 2   \frac{\xB t}{\Q^2}+
  \left(3 + \frac{t}{\Q^2} - \frac{t}{M^2}\right) \frac{\epsilon^2}{2}\right\}  \widetilde{\cal H}
\nonumber\\
&&- \frac{\xB}{2}\left\{\xB  \left(1 - \frac{t}{\Q^2}\right) - \frac{t}{2 M^2}\right\} \widetilde{\cal E} \Bigg]
\, .
\end{eqnarray}
This completes the full set of exact results for electroproduction cross section with exact account for the kinematical power corrections, where
the pure BH cross section is given in Ref.~\cite{Belitsky:2001ns}.
As in the case of  DVCS $\cal C$-coefficients, the kinematical singularities appearing for twist-three and transversity CFFs (partially) cancel also in
the interference term and the expected behavior of the harmonics in the $-t \to -t_{\rm min}$ limit can be established. Thereby, the CFF combinations
\begin{equation}
\label{singfreeF}
{\cal H}_{a+} + \frac{t}{4M^2} {\cal E}_{a+}\,, \quad \widetilde{\cal H}_{a+} + \frac{t}{4M^2} \widetilde{\cal E}_{a+}\,,
\quad
{\cal H}_{a+} + \frac{\xB \left(1+\frac{t}{\Q^2}\right)}{2-\xB + \frac{\xB t}{\Q^2}}\widetilde{\cal H}_{a+}\,,
\quad
{\cal E}_{a+} + \frac{\xB \left(1+\frac{t}{\Q^2}\right)}{2-\xB + \frac{\xB t}{\Q^2}}\widetilde{\cal E}_{a+}
\end{equation}
behave as $\tK^{1-a}$.
Furthermore, with the map (\ref{FT2FTVCS-1})--(\ref{FT2FTVCS-4}) the approximated results, given in Ref.~\cite{Belitsky:2001ns}, are restored.
Again, we performed a numerical cross check of our results.

Let us add that the harmonics (\ref{c0unpI})--(\ref{c3TPI}) can also be explicitly evaluated as function of the helicity dependent
CFFs. For instance, the exact results for the odd harmonics of a unpolarized target read in the fashion of \cite{Belitsky:2001ns} as follows,
\begin{eqnarray}
\label{s^{I}_{1,{unp}}}
s^{\cal I}_{1,{\rm unp}} &\!\!\!=\!\!\! & \frac{8 \tK \sqrt{1-y-\frac{y^2 \epsilon^2}{4}} \lambda(2-y) y }{\Q \left(1+\epsilon^2\right)^2}
\\
&\times&\!\!\!\!
\Im{\rm m}\bigg\{
{\cal C}_{\rm unp}^{\prime\,{\cal I}}\!\Bigg(\!{\textstyle
\bigg[\frac{1+\sqrt{1+\epsilon ^2}}{2\left(1+\frac{t}{\Q^2}\right)^{-1}}+\epsilon^2-\frac{\xB t}{\Q^2}\bigg] {\cal F}_{++}
+\bigg[\frac{1-\sqrt{1+\epsilon^2}}{2\left(1+\frac{t}{\Q^2}\right)^{-1}}+\epsilon^2-\frac{\xB t}{\Q^2}\bigg] {\cal F}_{-+}
+\frac{\sqrt{2} \tK}{\Q} {\cal F}_{0+}
}\!\!\Bigg)
\nonumber\\
&& \phantom{\Im}+\Delta{\cal S}_{1,{\rm unp}}^{{\cal I}}\!\Bigg(\!{\textstyle
\bigg[\frac{1-\sqrt{1+\epsilon^2}}{2\left(1+\frac{t}{\Q^2}\right)^{-1}}-\frac{(1-\xB) t}{\Q^2}\bigg] {\cal F}_{++}
+\bigg[\frac{1+\sqrt{1+\epsilon^2}}{2\left(1+\frac{t}{\Q^2}\right)^{-1}}-\frac{(1-\xB) t}{\Q^2}\bigg] {\cal F}_{-+}
-\frac{\sqrt{2} \tK}{\Q} {\cal F}_{0+}
}\!\!\Bigg)
\bigg\},
\nonumber\\
\label{s^{I}_{2,{unp}}}
s^{\cal I}_{2,{\rm unp}} &\!\!\!=\!\!\! &
\frac{ 16 \tK^2  \left(1-y-\frac{y^2 \epsilon^2}{4}\right)  \lambda y}{\Q^2 \left(1+\epsilon^2\right)^2}
\\
&\times&\!\!\!\!
\Im{\rm m}\bigg\{
{\cal C}_{\rm unp}^{\prime\,{\cal I}}\!\Bigg(\!{\textstyle
\frac{1+\frac{\xB t}{\Q^2}+\frac{1}{2}\epsilon^2\left(1+\frac{t}{\Q^2}\right)}{\sqrt{2}\tK \Q^{-1}}{\cal F}_{0+}
+ \frac{ \frac{1}{2}\left(\Q^2-t\right)\epsilon^2  -\xB t}{2 \tK^2}
}\!\!\sum_{a\in\{1,-1\}}{\textstyle \bigg[\frac{1-a \sqrt{1+\epsilon^2}}{2 \left(1+\frac{t}{\Q^2}\right)^{-1}}-\frac{(1-\xB)t}{\Q^2}\bigg] {\cal F}_{a+}
}
\!\Bigg)
\nonumber\\
&&\phantom{\Im}+\Delta{\cal S}_{2,{\rm unp}}^{{\cal I}}\!\Bigg(\!\!{\textstyle
\frac{\Q}{\sqrt{2} \tK }} {\cal F}_{0+}
-\frac{\Q^2}{2 \tK^2}\!\sum_{a\in\{1,-1\}}
{\textstyle \bigg[\frac{1-a \sqrt{1+\epsilon^2}}{2 \left(1+\frac{t}{\Q^2}\right)^{-1}} - \frac{(1-\xB)t}{\Q^2} \bigg]{\cal F}_{a+}
}\!\!\Bigg)
\bigg\},
\nonumber
\end{eqnarray}
where we introduced slightly different $\cal C$-coefficients and power suppressed addenda
$${\cal C}_{\rm unp}^{\prime\,{\cal I}}({\cal F}) = {\cal C}_{\rm unp}^{{\cal I}}({\cal F})  + \frac{t}{\Q^2}\, {\cal C}_{\rm unp}^{{\cal I},A}({\cal F}),
\quad
\Delta{\cal S}_{1,{\rm unp}}^{{\cal I}}({\cal F}) = \frac{2\xB t}{\Q^2} {\cal C}_{\rm unp}^{{\cal I},V}({\cal F}) + \frac{2 (1-\xB)t}{\Q^2}  {\cal C}_{\rm unp}^{{\cal I},A}({\cal F}),$$
and
$$\Delta{\cal S}_{2,{\rm unp}}^{{\cal I}}({\cal F}) =
-\left(\!1-\frac{t}{\Q^2}+\frac{2\xB t}{\Q^2}\!\right)\! \left[\frac{\xB t}{\Q^2}{\cal C}_{\rm unp}^{{\cal I},V}+
\frac{(1-\xB) t}{\Q^2}{\cal C}_{\rm unp}^{{\cal I},A}
\right]\!\!({\cal F})
- \frac{t \left(1+\epsilon^2\right)}{\Q^2} \left(\!1+\frac{t }{\Q^2}\!\right) {\cal C}_{\rm unp}^{{\cal I},A}({\cal F}). $$
The first harmonic (\ref{s^{I}_{1,{unp}}}), in particular the term $\tK {\cal F}_{0+}/\Q$, is free of kinematical singularities.
In the hadronic coefficients of the second harmonic (\ref{s^{I}_{2,{unp}}}) possible $1/\tK^2$ singularities in the hadronic coefficients cancel each other, too, see (\ref{singfreeF}).
Similar expression hold true for the even harmonics, however, the addenda will then also depend on the photon polarization parameter.

\section{Parametrization of the Compton tensor}
\label{ComptonParametrization}

Experimental studies of the Compton effect on the  nucleon target have a long history, with theoretical considerations preceding them.
The efforts of the past decade focused on the deep Euclidean kinematics giving access to partonic constituents of matter as we pointed
about in the previous few sections. Over the years various parametrizations of the hadronic tensor were devised tailored to the specific
needs of observables of interest. The story goes back to Prange \cite{PhysRev.110.240} who provided a decomposition originally given in
terms of bilinear combinations of Dirac spinors and often rewritten by means of two-dimensional Pauli spinors \cite{PhysRev.126.789}. Another widely used
Lorentz-invariant representation was introduced by Tarrach \cite{Tarrach_1975tu} and employed in recent years for consideration of quasi-real
kinematics \cite{Drechsel:1997xv} since this decomposition is free from kinematical singularities, on the one hand, and with all hadronic functions of
kinematical invariants admitting a well-defined dispersion representation that possess correct analytical properties \cite{Drechsel:2002ar}, on
the other. Finally, the developments of the last decade of the formalism of the deeply virtual Compton scattering were mimicking structures used in
the analysis of the forward deep-inelastic scattering and thus yet another parametrization was devised as a consequence. However, the emerging
Lorentz structures were recovered making use of the OPE for the correlation function (\ref{ComptonAmplitudeT}) of the quark electromagnetic
currents, demonstrating that electromagnetic gauge invariance, broken in the leading twist approximation, can be approximately restored by
accounting for twist-three effects. More recently this program was pushed beyond the first subleading corrections in Ref.\
\cite{Braun:2011dgBraun:2011zr,Braun:2012bgBraun:2012hq} by incorporating
dynamical effects in target mass and momentum transfer in the $t$-channel. Seeking a unified picture for observables used at high and low energies,
we will rely on the DVCS set-up for the Compton tensor and construct interpolation between different kinematical limits for CFFs. We will provide
a gauge-invariant decomposition of the Compton tensor starting from the analysis of the deeply virtual regime and then provide
a set of formulas connecting helicity CFFs (\ref{cal-Tab}) with the ordinary CFFs admitting partonic interpretation.

\subsection{A toy example}

As a pedagogical example, let us first consider a point particle with spin-1/2 as our target. In this case the electromagnetic current
(\ref{EMcurrentFFs}) reduces to $J_\mu = \overline{u}_2 \gamma_\mu u_1$. The Compton matrix element can be then obtained from familiar lowest-order
QED diagrams, i.e., the $s$- and $u$-channel hand-bag graphs, see Eq.\ (\ref{BornAmplitude}) below with $F_1 = 1$ and $F_2 = 0$. One can easily
verify by means of the Dirac equation that the resulting Compton scattering tensor exactly respects current conservation. To find a representation in
which gauge the symmetry is explicitly manifested, we employ the following trick
$$
1 = \frac{{\not\! p} {\not\! q}+ {\not\! q} {\not\! p}}{2\, p\cdot q}\,,
$$
the equations of motion
$$
{\not\! p}\, u_1 = 2 M u_1 + {\not\!\! \Delta}\, u_1\,, \qquad
\bar{u}_2\, {\not\! p}  = 2 M \bar{u}_2 - \bar{u}_2\, {\not\!\! \Delta} \, ,
$$
and subsequent use of the Dirac-matrix algebra.
This procedure yields a tensor containing non-flip ``transverse'' contributions only as a consequence of the leading-order approximation,
\begin{eqnarray}
\label{Ten-BM-point-like}
T^{\rm p.p.}_{\mu \nu}
\!\!\!&=&\!\!\!
-\left[g_{\mu\nu} -  \frac{q_{1\mu} p_{\nu}}{p \cdot q} -  \frac{q_{2\nu} p_{\mu}}{p \cdot q}
+ \frac{q_1\cdot q_2}{p \cdot q}\, \frac{p_{\mu} p_{\nu}}{p \cdot q}\right]\frac{q \cdot V_{\rm p.p.}}{p \cdot q}
\\
&&\!\!\! -\left[
 \varepsilon_{\mu\nu q \rho}
 +\varepsilon_{q \Delta \nu \rho} \left(\frac{q_{2\mu }}{q_1\cdot q_2}- \frac{p_\mu}{2 p\cdot q}\right)
 + \varepsilon_{q \Delta \mu \rho} \left(\frac{q_{1\nu }}{q_1\cdot q_2}- \frac{p_\nu}{2 p\cdot q}\right)
 \right]\frac{2 q_1\cdot q_2}{q_1^2 + q_2^2}\,\frac{A_{\rm p.p.}^\rho}{p\cdot q}\,.
\nonumber
\end{eqnarray}
Note that the kinematical pole  in the projection operators
$$
\frac{q_{1\nu }}{q_1\cdot q_2}- \frac{p_\nu}{2 p\cdot q}
\quad\mbox{and}\quad
\frac{q_{2\mu }}{q_1\cdot q_2}- \frac{p_\mu}{2 p\cdot q}\,,
$$
is removed by the overall factor of $q_1\cdot q_2$ in Eq.\ (\ref{Ten-BM-point-like}).  The  vector and axial-vector CFFs in the tensor (\ref{Ten-BM-point-like})
have a very simple  form and read for an on-shell final-state photon
\begin{eqnarray}
\label{V-pointlike}
V_{\rm p.p.}^{\rho}=\bar{u}_2 \gamma^\rho u_1 {\cal H}^{\rm p.p.}\,,
&&   {\cal H}^{\rm p.p.} = \frac{-(2-\xB)\Q^2-\xB t}{2 \Q^2}\,\left[\frac{1}{1 - \xB}+\frac{1}{1+ \frac{\xB t}{\Q^2}}\right],\qquad
\\
\label{A-pointlike}
A_{\rm p.p.}^{\rho}=\bar{u}_2 \gamma^\rho \gamma_5 u_1 \widetilde{\cal H}^{\rm p.p.}\,, &&
\widetilde{\cal H}^{\rm p.p.} = \frac{-(2-\xB)\Q^2-\xB t}{2 \Q^2\left(1+\frac{t}{\Q^2}\right)}\,\left[\frac{1}{1 - \xB}-\frac{1}{1+ \frac{\xB t}{\Q^2}}\right].
\end{eqnarray}
Here, the CFFs have only two physical poles at $s=M^2$  and $u=M^2$, showing up in our variables at $\xB=1$ and $\xB=-\Q^2/t $, respectively, and
they have the proper symmetry under $s\leftrightarrow u$ exchange.  Moreover, these CFFS are free of kinematical singularities and are related to each
other by a multiplicative factor
\begin{equation}
\label{Ht2H-pointlike}
\widetilde{\cal H}^{\rm p.p.} = \frac{\xB}{2-\xB +\frac{\xB t}{ \Q^2}} {\cal H}^{\rm p.p.} \, , \quad\mbox{or}\quad
\widetilde{\cal H}^{\rm p.p.} = \frac{\Q^2}{s-u} {\cal H}^{\rm p.p.} \,.
\end{equation}
We emphasize that even with the definitions (\ref{V-pointlike}) and (\ref{A-pointlike}), the form of the hadronic tensor is not uniquely fixed, rather
by means of the Dirac equation and the relation (\ref{Ht2H-pointlike}) one can find a different, however, equivalent forms of the Compton
scattering tensor. On the other hand, if a frame of reference is picked up, the helicity amplitudes (\ref{cal-Tab}) are independent of any parametrization
ambiguities. However, we remind that the relation between these helicity CFFs and ``partonic'' CFFs does heavily depend on the chosen tensor decomposition.

Let us employ our parametrization of the helicity-dependent CFFs (\ref{cal-TabV})--(\ref{cal-TabA}), used for evaluation of the differential cross sections in
the preceding section, where we adopted the projection of the (axial-) vector CFFs with the averaged photon momentum $m_\mu = q_\mu/q \cdot p$.
These CFFs can be  straightforwardly computed from the original tree diagrams. Alternatively, we may stick to the (axial-)vector CFFs (\ref{V-pointlike})
and (\ref{A-pointlike}) and compute the helicity amplitudes starting from Eq.\ (\ref{Ten-BM-point-like}). This is the route that we will choose in the realistic case
of composite target below. The result of this analysis can be summarized in the following set of helicity CFFs,
\begin{align}
\label{H_+b-pointlike}
{\cal H}^{\rm p.p.}_{+b}
&= \left[
\frac{1+ b\sqrt{1+\epsilon^2}}{2\sqrt{1+\epsilon^2}}+
\frac{
(1-\xB)\left(1+\frac{t}{\Q^2}\right) \epsilon^2 -
\left(2-\xB+\frac{\xB t}{\Q^2}\right) \frac{\xB^2 t}{\Q^2}
}{\sqrt{1+\epsilon^2} \left(2-\xB+\frac{\xB t}{\Q^2}\right)^2}
\right] {\cal H}^{\rm p.p.}
\,, \\
{\cal E}^{\rm p.p.}_{+ b} &=
\frac{\epsilon^2 \left(1+\frac{t}{\Q^2}-2(1-\xB)\frac{t}{\Q^2} \right)}{\sqrt{1+\epsilon^2}
\left(2-\xB+\frac{\xB t}{\Q^2}\right)^2}\, {\cal H}^{\rm p.p.}
\,,\\
\widetilde{\cal H}_{+b}^{\rm p.p.}
&=
\left[
\frac{1+b\sqrt{1+\epsilon^2}}{2\sqrt{1+\epsilon^2}}\left(\!1 -\frac{t}{\Q^2}\!\right)  +
\frac{1}{\sqrt{1+\epsilon^2}}\,\frac{\xB t}{\Q^2}
\right] \widetilde{\cal H}^{\rm p.p.}
\,, \\
\widetilde{\cal E}^{\rm p.p.}_{+b}
&=
\frac{4M^2}{\Q^2}\left[
\frac{1+b \sqrt{1+\epsilon^2}}{2\sqrt{1+\epsilon^2}} \left(\!3+\frac{t}{\Q^2}\!\right) -
\frac{1 + (1-\xB) \frac{t}{\Q^2} }{\sqrt{1+\epsilon^2}}
\right] \widetilde{\cal H}^{\rm p.p.}
\,,\\
\label{H_0+-pointlike}
{\cal H}^{\rm p.p.}_{0+}  &=
-\frac{\sqrt{2}\,\xB\,\widetilde{K}\left(2-\xB+\frac{\xB t}{\Q^2}-2 \epsilon^2\right)}{\Q \sqrt{1+\epsilon^2}\left(2-\xB+\frac{\xB t}{\Q^2}\right)^2}\,
{\cal H}^{\rm p.p.}
\,, \\
{\cal E}_{0+}^{\rm p.p.}
&=
\frac{2\sqrt{2}\,\epsilon^2\,\widetilde{K}}{\Q \sqrt{1+\epsilon^2}\left(2-\xB+\frac{\xB t}{\Q^2}\right)^2}\,
{\cal H}^{\rm p.p.}
\,,\\
\label{tH_0+-pointlike}
\widetilde{\cal H}_{0+}^{\rm p.p.}
&=
\frac{\sqrt{2}\widetilde{K}}{\Q \sqrt{1+\epsilon^2}}\, \widetilde{\cal H}^{\rm p.p.}
\,, \\
\label{tE_0+-pointlike}
\widetilde{\cal E}_{0+}^{\rm p.p.}
&=
\frac{4 M^2}{\Q^2}\,\frac{\sqrt{2}\widetilde{K}}{\Q \sqrt{1+\epsilon^2}}\,
 \widetilde{\cal H}^{\rm p.p.}
 \, .
\end{align}
Several comments are in order. As observed in GPD calculations in the twist-three sector, see Eqs.~(84)--(87) in Ref.~\cite{Belitsky:2001ns},  the CFFs
in the vector and axial-vector sector mix with each other, while in our analysis of a point particle, we eliminated these admixtures by utilizing the
relation (\ref{Ht2H-pointlike}). Notice that the longitudinal and transverse spin-flip CFFs are power suppressed in the DVCS kinematics. The
longitudinal-to-transverse helicity-flip CFFs (\ref{H_0+-pointlike}) and (\ref{tH_0+-pointlike}) have the anticipated kinematical $\widetilde{K}$-factor in
front of them. The transverse-to-transverse helicity-flip CFFs also do not possess any kinematical singularities but do not have the anticipated kinematical
$\widetilde{K}^2$ factor in front of them, i.e., they do also not vanish at the kinematical
boundary $t=t_{\rm min}$. Hence, if one chooses to switch to the definitions such as (\ref{cff2-cff+-}), spurious kinematical $1/\tK^2$ singularities appear
in expressions for transversity CFFs, see also the map (\ref{FT2FTVCS-1})--(\ref{FT2FTVCS-4}) of such CFFs.  Obviously, such  spurious kinematical  singularities
can be simply pulled out by a redefinition of transversity CFFs and then they trivially will not appear in cross section expressions. Plugging in our point-particle
results (\ref{H_+b-pointlike})--(\ref{tE_0+-pointlike}) in the expression for the higher harmonics in the Fourier expansion, given in Sect.~\ref{sect-TDVCS}
and \ref{sect-INT}, we realize that they vanish in the kinematical limit $t\to t_{\rm min}$ as expected.
The basis of helicity dependent CFFs in which this behavior is explicit will be given in the next paragraph.

We also computed the helicity dependent CFFs associated with the so-called Born term for the Compton scattering off the nucleon (\ref{BornAmplitude}),
see Eqs.\ (\ref{H_+a-Born})--(\ref{tE_+0-Born}) in the Appendix \ref{App:Born}. Again we observe that the unpolarized and transverse-to-transverse spin-flip
CFFs are free of kinematical singularities, however, certain longitudinal-to-transverse CFFs are suffering now from $1/\widetilde{K}$ poles in the
representation of Ref.~\cite{Belitsky:2001ns}. The absence of such spurious singularities can be made transparent by switching to the electric-like\footnote{By
analogy with the electric nucleon form factor.} combinations of CFFs
\begin{eqnarray}
\label{cff-electric}
{\mathcal G}_{0b}
=
{\mathcal H}_{0b}+\frac{t}{4M^2}{\mathcal E}_{0b} \quad\mbox{and}\quad
\widetilde{\mathcal G}_{0b}
=
\widetilde{\mathcal H}_{0b}+\frac{t}{4M^2}\widetilde{\mathcal E}_{0b}
\, ,
\end{eqnarray}
where ${\mathcal G}_{0b}$ and $\widetilde{\mathcal G}_{0b}$ are proportional to the desired kinematical factor $\tK$, see expressions in Appendix
\ref{App:Born}. For transverse helicity-flip CFFs we find that ${\mathcal G}_{-+}$ and $\widetilde{\mathcal G}_{-+}$ are now proportional to
$t-t_{\rm min}$ (not explicitly shown). The redefinitions (\ref{cff-electric})  reparametrize the (axial-)vector matrix elements (\ref{cal-TabV}), (\ref{cal-TabA})
as follows
\begin{align}
\label{cal-TabV-1}
{\cal V}({\cal F}_{ab})
&= \frac{1}{p\cdot q}
\bar{u}_2 \left( \, {\not\! q}\, {\cal G}_{ab} + \left[i \sigma_{\rho \sigma} \frac{q^\rho \Delta^\sigma}{2 M}- {\not\! q}\,\frac{t}{4M^2}\right] {\cal E}_{ab}\right) u_1
\\
\label{cal-TabA-1}
{\cal A}({\cal F}_{ab})
&= \frac{1}{p\cdot q}
\bar{u}_2 \left( \, {\not\! q}\, \gamma_5 \, \widetilde{\cal G}_{ab}
+
\left[\gamma_5 \frac{q\cdot\Delta}{2 M}-{\not\! q}\, \gamma_5\,\frac{t}{4M^2} \right] \widetilde{\cal E}_{ab}\right) u_1
\,.
\end{align}
It can be verified that the new nucleon helicity structures, proportional to ${\cal E}_{ab}$ or $\widetilde{\cal E}_{ab}$, will yield a kinematical factor $\tK^2$
in the hadronic $\cal C$ coefficients of both the squared VCS  and interference term. Hence, this  guarantees that higher harmonics vanish in the
limit $t\to t_{\rm min}$ as they should and, moreover,  we can reshuffle this factor that yields a redefinition
$$
\tK^2 {\cal E}_{0b} \quad \rightarrow  \quad  {\cal E}_{0b}  \quad\mbox{and}\quad   \tK^2 \widetilde{\cal E}_{0b} \quad \rightarrow  \quad \widetilde{\cal E}_{0b}
\, .
$$
This guarantees that the new longitudinal helicity-flip amplitudes have the anticipated $\tK$ factor as an overall factor, one the one hand, and that they
are free of kinematical singularities, on the other.  The corresponding modifications in Fourier coefficients $\cal C$ are straightforward and do not require
any further comments.

\subsection{Constructing Compton scattering tensor}
\label{sec:tensor}

To devise a general parametrization as we advertised above, let us start with the Compton scattering process off the nucleon for the case when both photons
possess large virtualities, such that the hard scale is set by the Euclidean virtuality $Q^2= - q^2$ with $q_\mu$ defined in Eq.\ (\ref{AveagedMomenta}).
As $Q^2 \to \infty$, the decoherence of the short- and long-range
interactions allows one to probe partonic content of the nucleon via collinear factorization. This approach naturally introduces a pair of
the light-cone vectors $n_\mu$ and $n^\ast_\mu$, such that $n\cdot n = n^\ast \cdot n^\ast =0$ and $n\cdot n^\ast =1$, since partons propagate
along the light cone. However, these cannot be fixed uniquely in terms of the external momenta of the process. Restricting to the
leading terms in the $1/Q$-expansion, i.e., the so-called twist-two approximation, the result for the Compton scattering tensor is cast to the
following form\footnote{Notice that the ${\rm T}$ label used here does not have anything to do with the $T$-subscript adopted earlier to label
the transversity CFFs in Eqs.\ (\ref{FT2FTVCS-1})--(\ref{FT2FTVCS-4}) as well as (\ref{Tamp}) below.}
\begin{align}
\label{T_{munu}-twist2}
T_{\mu\nu} = - g_{\mu\nu}^\perp\, n\cdot V_{\rm T}  - i \varepsilon^\perp_{\mu\nu}\, n\cdot A_{\rm T} +
\left(\frac{q_2^2}{n^\ast \cdot q}\, n^\ast_{\mu} -  q_{2\mu} \right)  \left(\frac{q_1^2 }{n^\ast\cdot q}\,n^\ast_{\nu} -q_{1\nu}\right) n\cdot V_{\rm L}
+\tau^\perp_{\mu\nu;\rho\sigma}\, \frac{\Delta^\rho T^{\sigma}}{M^2} \, .
\end{align}
Here the first two terms on the r.h.s.\ were computed in numerous papers (in particular for DVCS kinematics). The third term after the equality sign
contains a purely longitudinal part  and appears at next-to-leading order in QCD coupling \cite{Mankiewicz:1997bk} mimicking the violation of the
Callan-Gross relation in deep-inelastic scattering. The forth term stems from the double photon-helicity flip and is perturbatively induced at
one-loop by the gluon transversity GPDs \cite{BelMue00}. The projection on the leading-twist structures in Eq.\ (\ref{T_{munu}-twist2}) is
achieved by the means of the tensors
\begin{eqnarray}
\label{perp-projectors}
g_{\mu\nu}^\perp = g_{\mu\nu} -n_{\mu} n^\ast_{\nu}-n_{\nu} n^\ast_{\mu}\,,
\quad
\varepsilon^\perp_{\mu\nu} = \varepsilon_{\mu\nu-+}\,,
\quad
\tau^\perp_{\mu\nu;\rho\sigma} = \frac{1}{2}\left[
g_{\mu\rho}^\perp g_{\nu\sigma}^\perp +  g_{\mu\sigma}^\perp g_{\nu\rho}^\perp - g_{\mu\nu}^\perp g_{\rho\sigma}^\perp
\right]
\, .
\end{eqnarray}
The vector $V^\rho_{\rm T}$ ($V^\rho_{\rm L}$) and the axial-vector $A^\rho_{\rm T}$ CFFs describe transition amplitudes when the transverse
(longitudinal) photon helicity is (nearly) conserved while $T^{\rho}$ is associated with the aforementioned transverse photon helicity-flip contribution.
As a consequence of the leading-twist approximation, eight longitudinal--to--transverse and transverse--to--longitudinal photon helicity-flip amplitudes
are absent and, moreover, the Compton scattering tensor (\ref{T_{munu}-twist2}) respects current conservation only to leading order in $1/Q$.
However, once one goes beyond the twist-two approximation in the OPE analysis of the hadronic tensor, these missing amplitudes emerge and moreover,
making use of QCD equation-of-motions, the electromagnetic current conservation gets restored up to the same accuracy. Thus the ambiguity
in the construction of the Lorentz tensors is pushed up to the next order in the $1/Q^2$ expansion.

Since we would like to stay as close as possible to the VCS tensor decomposed in terms of Lorentz structures that have a simple limit in the deeply
virtual regime, below we propose a parametrization motivated by Eq.\ (\ref{T_{munu}-twist2}) that can also be used for quasi-real (or real) photons without
encountering kinematical singularities. Thereby, we impose the following natural requirements
\begin{itemize}
\item manifest current conservation and Bose symmetry;
\item a close match with conventions used in deeply virtual Compton kinematics;
\item singularity-free kinematical dependence.
\end{itemize}
Instead of the light-cone vectors we employ the external particle vectors in the construction of the tensors. To make the tensors dimensionless, we use
the scalar product $p \cdot q = (s-u)/2$, proportional to the positive energy variable $\nu$, in denominators. Equipped with the above conditions and
building blocks, the transverse metric tensor entering the leading-twist parametrization, which received corrections from the twist-three effects mentioned
above, gets promoted to the following expression
\begin{eqnarray}
\label{tg_{munu}}
g_{\mu\nu}^\perp\;\; \to\;\;   \widetilde{g}_{\mu \nu}
=g_{\mu\nu} -  \frac{q_{1\mu} p_{\nu}}{p \cdot q} -  \frac{q_{2\nu} p_{\mu}}{p \cdot q}
+ \frac{q_1\cdot q_2}{p \cdot q}\, \frac{p_{\mu} p_{\nu}}{p \cdot q}\,,
\end{eqnarray}
where the twist-four component $p_{\mu} p_{\nu}$ follows from the expansion of $g_{\mu\nu}^\perp$. This tensor already appeared in our toy example (\ref{Ten-BM-point-like}) and its gauge invariance is easily verified by making use of the relations
$$
p\cdot q_i= p\cdot q \quad \mbox{since}\quad q^\nu_1=q^\nu + \Delta^\nu/2\;\;\mbox{and}\;\; q^\mu_2= q^\mu - \Delta^\mu/2 \quad \mbox{with}\quad  p\cdot\Delta =0\,.
$$
Going to the Breit frame, where the transverse momentum is entirely carried by  the nucleons, i.e., $p_2^\perp=-p_1^\perp$ and $q_i^\perp=0$,
one realizes that the gauge invariant tensor (\ref{tg_{munu}}) projects onto photons with the same transverse helicity, having even parity in the $t$-channel.
The counterpart of this contribution, having $t$-channel  odd parity, is expressed in terms of the Levi-Civita tensor that generalizes the above $\varepsilon^\perp_{\mu\nu}$
beyond leading twist,
\begin{eqnarray}
\label{tepsilon_{munu}}
\varepsilon_{\mu\nu}^\perp\;\; \to\;\; \widetilde \varepsilon_{\mu \nu } =   \frac{1}{p\cdot q}\left[
\varepsilon_{\mu \nu  p q}  + \frac{p_{\mu}}{2\,p\cdot q} \varepsilon_{\Delta \nu p q } -
\varepsilon_{\mu \Delta  p q }\, \frac{p_{\nu}}{2\,p\cdot q} + \varepsilon_{\mu \nu \Delta q } \, \frac{p\cdot p}{2\,p\cdot q}
\right]\,.
\end{eqnarray}
Here, we added for later convenience a power suppressed $p\cdot p/ p\cdot q$-term, which respects current conservation by itself.
The coupling of longitudinal and transverse photon helicity states may be naturally encoded in terms of the following tensor structures
$$
\left(q_{2\mu} - \frac{q_2^2}{p\cdot q} p_\mu\right)\left(g_{\nu \rho}- \frac{p_{\nu}\, q_{1}^\rho}{p\cdot q}\right)\quad \mbox{and}\quad
\left(q_{1\nu} - \frac{q_1^2}{p\cdot q} p_\nu\right)\left(g_{\mu \rho}- \frac{p_{\mu}\, q_{2}^\rho}{p\cdot q}\right)\,,
$$
which are to be contracted with vector CFFs.  Here the projectors, satisfying the relations
\begin{eqnarray}
\label{P-trans}
\left(g^{\mu \rho}- \frac{p^{\mu}\, q_{2}^{\rho}}{p\cdot q}\right) p_\rho  = 0\,,
\phantom{\Delta_\perp^\mu +\frac{t}{2 p\cdot q} p}
&& \left( g^{\nu \rho}- \frac{p^{\nu}\, q_{1}^{\rho}}{p\cdot q}\right) p_\rho =0\,,
\\
\left(g^{\mu \rho}- \frac{p^{\mu}\, q_{2}^\rho}{p\cdot q}\right) \Delta_\rho= \Delta_\perp^\mu +\frac{t}{2 p\cdot q} p^\mu\,,
&&
\left( g^{\nu \rho}- \frac{p^{\nu}\, q_{1}^\rho}{p\cdot q}\right) \Delta_\rho=
\Delta_\perp^\nu -\frac{t}{2 p\cdot q} p^\nu\,,
\nonumber
\end{eqnarray}
where  $\Delta_\perp^\rho = \Delta^\rho -\eta p^\rho$, ensure the electromagnetic gauge invariance. Notice that in the above tensors the longitudinal
components were chosen in the form%
\footnote{In previous studies on the subject, we wrote the longitudinal pieces in terms of $q_{1\mu} - q_1\cdot q_2\, p_\mu/ {p\cdot q} $  and
$q_{2\nu} - q_1\cdot q_2\, p_\nu/ {p\cdot q} $, which to twist-three accuracy can be replaced by the vectors that are displayed above.},
$$
q_{2\mu} - \frac{q_2^2}{p\cdot q} p_\mu\quad\mbox{and}\quad
q_{1\nu} - \frac{q_1^2}{p\cdot q} p_\nu\, .
$$
Obviously, they do not contribute in the real photon limit. Last but not least,  the current conservation in the transverse helicity-flip amplitudes can be
implemented by utilizing the transverse projectors (\ref{P-trans}), yielding the substitution
\begin{eqnarray}
\tau^\perp_{\mu\nu;\rho\sigma}\, \frac{\Delta^\rho T^{\sigma}}{M^2}\;\; \rightarrow\;\;
\left(g_\mu^{\phantom{\mu} \alpha}- \frac{p_{\mu}\, q_{2}^\alpha}{p\cdot q}\right)
\left(g_\nu^{\phantom{\nu} \beta}- \frac{p_{\nu}\, q_{1}^\beta}{p\cdot q}\right) \tau^\perp_{\alpha\beta;\rho\sigma}\, \frac{\Delta^\rho T^{\sigma}}{M^2} \,.
\end{eqnarray}
It is important to realize that a different choice of dual light-cone vectors will result into a parameterization, where the kinematically suppressed
effects will be incorporated in a different fashion, see discussions in Ref.~\cite{Belitsky:2008bz}.

By analogy with the hadronic electromagnetic current, decomposed in terms of the Dirac bilinears accompanied by Dirac and Pauli form factors
(\ref{EMcurrentFFs}), we now introduce a similar representation for the VCS in CFFs, expanded in terms of the longitudinal and
transverse components. To generalize the first three terms in r.h.s.\ of Eq.\ (\ref{T_{munu}-twist2}) to the setup incorporating the exact kinematics, we
replace the light-cone vector $n$ projecting out the (axial-)vector CFFs by the average photon momentum $q$, whose leading component is indeed $n$
but it also encodes subleading twist effects as well. Then we use the following expansion
\begin{eqnarray}
\label{Def-V1}
V_i^{ \rho}\!\!\!&=&\!\!\!
\bar{u}_2
\left(
\frac{p^\rho}{p \cdot q} \left[\not\!q\,  {\cal H}+ i \sigma_{\alpha\beta}\, \frac{q^{\alpha}\Delta^\beta}{2 M}\, {\cal E}\right]
+ \frac{\Delta_{\perp}^\rho}{p \cdot q}  \left[
\not\!q\, {\cal H}_i + i \sigma_{\alpha\beta}\, \frac{q^\alpha \Delta^\beta}{2 M}\, {\cal E}_i
\right]
\right) u_1
\, , \\
\label{Def-A1}
A_i^{ \rho}\!\!\!&=&\!\!\!
\bar{u}_2\left(
\frac{p^\rho}{p \cdot q} \left[\not\!q \gamma_5\, \widetilde{\cal H}+ \frac{\Delta\cdot q}{2 M} \gamma_5\, \widetilde{\cal E} \right] +
\frac{\Delta_{\perp}^\rho}{p\cdot q} \left[ \not\!q \gamma_5\,  \widetilde{\cal H}_i + \frac{q \cdot \Delta}{2 M}\gamma_5\, \widetilde{\cal E}_i
\right]
\right) u_1
\, ,
\end{eqnarray}
with the subscript $i$ standing for $i \in  \{ {\rm T},{\rm L}\}$. The same expansion can be adopted for the $i \in  \{ {\rm LT}, {\rm TL}\}$ cases. However, for
these, the component proportional to $p^\rho$ will not contribute to the VCS and can be ignored, while the transverse part is
approximatively expressible in terms
of twist-three CFFs introduced earlier, namely,
\begin{align}
{\cal F}_{\rm LT}\stackrel{\rm tw-3}{=}{\cal F}^3_+ - {\cal F}^3_-
\, , \qquad
{\cal F}_{\rm TL}\stackrel{\rm tw-3}{=}{\cal F}^3_+ + {\cal F}^3_-
\, .
\end{align}
To write the transversity CFFs in the same form, let us recall the following facts about the amplitude $T_\rho$ entering the leading twist
DVCS tensor (\ref{T_{munu}-twist2}). It is parametrized by four transverse photon helicity-flip amplitudes \cite{Belitsky:2001ns} according to the
suggestion of \cite{Diehl:2001pm}  as follows
\begin{eqnarray}
\label{Tamp}
T_\sigma = \overline{u}_2\left[
i\sigma_{\alpha\sigma} \frac{q^\alpha}{p\cdot q}  {\cal H}_{T}^{\scriptsize\cite{Diehl:2001pm}}
+
 \frac{\Delta_{\perp\sigma}}{2 M^2} \widetilde{\cal H}_{T}^{\scriptsize\cite{Diehl:2001pm}}
 +
\frac{1}{2M} \left(\frac{\not\!q}{p\cdot q} \Delta_\sigma - \frac{\Delta\cdot q}{p\cdot q}\gamma_\sigma\right) {\cal E}_{T}^{\scriptsize\cite{Diehl:2001pm}}
+
\frac{\not\!q p_\sigma  -\gamma_\sigma p\cdot q}{2M p\cdot q}
\widetilde{\cal E}_{T}^{\scriptsize\cite{Diehl:2001pm}}
\right] u_1\, .
\nonumber\\
\end{eqnarray}
However, this  can be represented analogously to Eqs.\ (\ref{Def-V1}) and (\ref{Def-A1}) in terms of two parity-even and -odd Dirac bilinears
\begin{align}
\label{NewTV}
V_{\rm TT}^{ \rho}
&=
\frac{p^\rho}{p \cdot q} \bar{u}_2 \left[\not\!q\,  {\cal H}_T + i \sigma_{\alpha\beta}\, \frac{q^{\alpha}\Delta^\beta}{2 M}\, {\cal E}_T \right]
u_1
\, , \\
\label{NewTA}
A_{\rm TT}^{ \rho}
&=
\frac{p^\rho}{p \cdot q} \bar{u}_2 \left[\not\!q \gamma_5\, \widetilde{\cal H}_T + \frac{\Delta\cdot q}{2 M} \gamma_5\, \widetilde{\cal E}_T \right] u_1
\, ,
\end{align}
that  are proportional to $\Delta^\perp_\sigma$ and $\widetilde \Delta_\sigma^\perp=\varepsilon_{\sigma\Delta p q}/{p \cdot q}$, respectively,
\begin{eqnarray}
\label{TasAandV}
T_{\rho} \big({\cal F}_{T}^{\scriptsize\cite{Diehl:2001pm}}\big)
=
\Delta^\perp_{\rho}\, \frac{q\cdot V_{\rm TT}({\cal F}_{T})}{p\cdot q}
+
i \widetilde{\Delta}^\perp_{\rho}\, \frac{q\cdot A_{\rm TT}({\cal F}_{T})}{p\cdot q}
+
{\cal O}(1/Q^2)\,,
\end{eqnarray}
where $\widetilde{\Delta}_\mu^\perp$ is transverse with respect to both photon momenta $q_1$ and $q_2$.
The relation among the two sets of CFFs, introduced in Eq.\ (\ref{Tamp}) and Eqs.\ (\ref{NewTV}), (\ref{NewTA}),
respectively, is found to be
\begin{eqnarray}
\label{FT2FT-1}
{\cal H}_{T}  \!\!\!&=&\!\!\!
{\cal H}^{\scriptsize\cite{Diehl:2001pm}}_{T} +  {\cal E}^{\scriptsize\cite{Diehl:2001pm}}_{T}+ 2 \widetilde{\cal H}^{\scriptsize\cite{Diehl:2001pm}}_{T}
-
\frac{t}{ t+(4 M^2-t) \left(\eta^2+\xi^2 \frac{t}{Q^2}\right)}\left[(1-\eta^2){\cal H}^{\scriptsize\cite{Diehl:2001pm}}_{T}
-
\eta^2 {\cal E}^{\scriptsize\cite{Diehl:2001pm}}_{\rm T}-\eta \widetilde{\cal E}^{\cite{Diehl:2001pm}}_{T}
\right],
\nonumber\\
\\
{\cal E}_{T}  \!\!\!&=&\!\!\! -2\widetilde{\cal H}^{\scriptsize\cite{Diehl:2001pm}}_{T}
+
\frac{4M^2}{ t+(4 M^2-t) \left(\eta^2+\xi^2 \frac{t}{Q^2}\right)}\left[(1-\eta^2){\cal H}^{\scriptsize\cite{Diehl:2001pm}}_{T}
-
\eta^2 {\cal E}^{\scriptsize\cite{Diehl:2001pm}}_{T} - \eta \widetilde{\cal E}^{\scriptsize\cite{Diehl:2001pm}}_{T}
\right],
\\
\nonumber{\ }\\
\widetilde{\cal H}_{T}  \!\!\!&=&\!\!\!
\frac{4M^2}{ t+(4 M^2-t) \left(\eta^2+\xi^2 \frac{t}{Q^2}\right)}\left[
\eta {\cal H}^{\scriptsize\cite{Diehl:2001pm}}_{T}+\frac{\eta\, t}{4M^2} {\cal E}^{\scriptsize\cite{Diehl:2001pm}}_{T}
+
\frac{t}{4M^2}\widetilde{\cal E}^{\scriptsize\cite{Diehl:2001pm}}_{T}
\right],
\phantom{\Bigg|}
\\
\label{FT2FT-4}
\widetilde{\cal E}_{T}  \!\!\!&=&\!\!\! \frac{4 M^2}{\eta\,  t }
{\cal H}_{T}^{\scriptsize\cite{Diehl:2001pm}}
-
\frac{4M^2}{ t+(4 M^2-t) \left(\eta^2+\xi^2 \frac{t}{Q^2}\right)}
\left[
\eta\, \frac{4 M^2}{ t }  {\cal H}_{T}^{\scriptsize\cite{Diehl:2001pm}} +
\eta {\cal E}^{\scriptsize\cite{Diehl:2001pm}}_{T} + \widetilde{\cal E}^{\scriptsize\cite{Diehl:2001pm}}_{T}
\right],
\phantom{\Bigg|}
\end{eqnarray}
which reduce in the real photon  case $q_2^2=0$ to Eqs.~(\ref{FT2FT-1})--(\ref{FT2FT-4}).
Notice that the convenience of the representation (\ref{TasAandV}) had forced us to introduce kinematical singularities into the transverse CFFs. These
are exhibited in the VCS kinematics as poles in $t -t_{\rm min}$,
$$
 t+(4 M^2-t) \left(\eta^2+\xi^2 \frac{t}{Q^2}\right) = -\frac{4 \tK^2 }{(2-\xB+ \frac{\xB t}{\Q^2})^2} \propto (t-t_{\rm min}),
$$
Since in the helicity amplitudes  the transversity CFFs are multiplied with $\tK^2$, these poles  will be canceled and, moreover, if we change in the
expressions for the cross section back to the basis of Ref.~\cite{Diehl:2001pm}, we will find in the transversity sector the CFFs are appropriately
accompanied by a factor of $\tK^2$ so that Fourier harmonics possess the expected behavior in the $t\to t_{\rm min}$  limit, see discussion in Sect.~\ref{sect-TDVCS}.

Having defined the tensor structures and the corresponding CFFs, we are now in a position to write down the complete Compton scattering tensor:
\begin{eqnarray}
\label{Ten-BM}
T_{\mu \nu}
\!\!\!&=&\!\!\!
-\widetilde g_{\mu\nu} \frac{q \cdot V_{\rm T}}{p \cdot q} + i\widetilde \varepsilon_{\mu \nu } \frac{q\cdot A_{\rm T}}{p\cdot q} +
\left(\! q_{2\, \mu} -\frac{q_2^2}{p \cdot q} p_{\mu}\!\right) \left(\! q_{1\, \nu} -\frac{q_1^2}{p \cdot q} p_{\nu} \!\right)
\frac{q\cdot V_{\rm L}}{p\cdot q}
\nonumber\\
&& \!\!\! +
\left(\! q_{1\, \nu} -\frac{q_1^2}{p \cdot q} p_{\nu} \!\right) \left(\! g_{\mu \rho}- \frac{p_{\mu}\, q_{2\, \rho}}{p\cdot q} \!\right)
\left[ \frac{V_{\rm LT}^\rho}{p \cdot q}   + \frac{i \epsilon^\rho_{\phantom{\rho} q p  \sigma}}{p\cdot q }\, \frac{A^\sigma_{\rm LT}}{p \cdot q}  \right]
\\
&& \!\!\! +
\left(\! q_{2\, \mu} -\frac{q_2^2}{p \cdot q} p_{\mu} \!\right) \left(\! g_{\nu \rho}- \frac{p_{\nu}\, q_{1\, \rho}}{p\cdot q} \!\right)
\left[ \frac{V_{\rm TL}^\rho}{p \cdot q}   + \frac{i \epsilon^\rho_{\phantom{\rho} q p  \sigma}}{p\cdot q }\, \frac{A_{\rm TL}^\sigma}{p \cdot q}  \right]
\nonumber\\
&& \!\!\! +
\left(g_\mu^{\phantom{\mu} \rho}- \frac{p_{\mu}\, q_{2}^\rho}{p\cdot q}\right)
\left(g_\nu^{\phantom{\nu} \sigma}- \frac{p_{\nu}\, q_{1}^\sigma}{p\cdot q}\right)\left[
\frac{ \Delta_\rho  \Delta_\sigma + \widetilde{\Delta}^\perp_\rho  \widetilde{\Delta}^\perp_\sigma}{2M^2}\, \frac{q\cdot V_{\rm TT}}{p\cdot q}  +
\frac{\Delta_\rho  \widetilde{\Delta}^\perp_\sigma +  \widetilde{\Delta}^\perp_\rho  \Delta_\sigma }{2M^2}\,  \frac{q\cdot A_{\rm TT}}{p\cdot q}
\right] ,
\nonumber
\end{eqnarray}
where the electromagnetic gauge invariance is implemented exactly. From our exact parameterization of the tensor  (\ref{Ten-BM})
we can now find the helicity-dependent CFFs  $\mathcal{F}_{ab}$  used in Sect.~\ref{sect-TDVCS} and \ref{sect-INT} for the evaluation of the
cross sections in the target rest frame. Comparing Eq.~(\ref{cal-Tab}) with (\ref{Ten-BM}) projected onto the photon polarization vectors, we find
\begin{eqnarray}
\label{ExactHelicityFpb}
{\cal F}_{+b} &\!\!\!=\!\!\!& \left[
\frac{1+b\sqrt{1+\epsilon^2}}{2 \sqrt{1+\epsilon^2}}+
\frac{(1 - \xB) \xB^2 (4 M^2 - t) \left(1+\frac{t}{\Q^2}\right)}{\Q^2\,\sqrt{1+\epsilon^2}\left(2-\xB+\frac{\xB t}{\Q^2}\right)^2}
\right] {\cal F}_{\rm T}  \\
&&\!\! +\frac{1-b\sqrt{1+\epsilon^2}}{2 \sqrt{1+\epsilon^2}} \frac{2\widetilde{K}^2}{M^2\left(2-\xB+\frac{\xB t}{\Q^2}\right)^2} {\cal F}_{\rm TT}
+ \frac{4 \xB^2 \tK^2}{\Q^2 \sqrt{1+\epsilon^2}\left(2-\xB +\frac{\xB t}{\Q^2}\right)^3} {\cal F}_{\rm LT}
\,,
\nonumber\\
\label{ExactHelicityF0p}
{\cal F}_{0+}&\!\!\!=\!\!\!&  \frac{(-1)\sqrt{2}\, \widetilde{K}}{\sqrt{1+\epsilon^2} \Q \left(2-\xB+\frac{\xB t}{\Q^2}\right)} \Bigg\{
\frac{2\xB }{ 2-\xB+\frac{\xB t }{\Q^2}} \left[1+\frac{2 \xB^2 \left(4 M^2-t\right)}{\Q^2\left(2-\xB +\frac{\xB t}{\Q^2}\right)}\right]
{\cal F}_{\rm LT}
\\&&\!\!
+\xB \left[1 + \frac{2\xB (4 M^2-t)}{\Q^2\left(2-\xB +\frac{\xB t}{\Q^2}\right)}\right]  {\cal F}_{\rm T} +
\frac{4 \xB^2 M^2-\left(2\xB + \epsilon^2\right)t}{2M^2\left(2-\xB+\frac{\xB t}{\Q^2}\right)} {\cal F}_{\rm TT}
\Bigg\},
\nonumber
\end{eqnarray}
where $b=\pm 1$ is the helicity of the real photon. Notice that the longitudinal helicity-flip CFFs (\ref{ExactHelicityF0p}) are proportional to the kinematical
factor $\tK$. Moreover, the transverse photon helicity-flip CFFs (\ref{ExactHelicityFpb}) with $b=-1$ are mostly proportional to $\tK^2$, except for the term
\begin{eqnarray}
\frac{1-\sqrt{1+\epsilon^2}}{2 \sqrt{1+\epsilon^2}}+
\frac{(1 - \xB) \xB^2 (4 M^2 - t) \left(1+\frac{t}{\Q^2}\right)}{\Q^2\,\sqrt{1+\epsilon^2}\left(2-\xB+\frac{\xB t}{\Q^2}\right)^2}
 &\!\!\! = \!\!\! &
\frac{t_{\rm min}-t}{\Q^2}\, \frac{1+\sqrt{1+\epsilon^2}-2 \xB}{2\left(2-\xB+\frac{\xB t}{\Q^2}\right)^2}
\\
&&\times
\left[4(1-\xB) \frac{1-\sqrt{1+\epsilon^2}}{2 \sqrt{1+\epsilon^2}}  + \frac{\left(1+\frac{t}{\Q^2}\right) \xB^2}{\sqrt{1+\epsilon^2}} \right]
\, ,
\nonumber
\end{eqnarray}
which is proportional to $t_{\rm min}-t$. Here, the r.h.s.~of this relation may also be written as $\tK^2/(t-t_{\rm max})$, see Eq.~(\ref{K-tilde}).
Let us also point out that the transformations (\ref{ExactHelicityFpb},\ref{ExactHelicityF0p}) exist in the limit $s \to M^2$ or $u\to M^2$.
In the simultaneous limit, where $(2-\xB + \xB t/\Q^2) \propto (s-u)$ vanishes, we encounter  $1/(s-u)$ singularities that are
associated with the longitudinal helicity-flip CFFs. This artifact may appear as an obstacle only in the
low-energy expansion and can be overcome in a straightforward manner, either by removing this singularity by a simple reparametrization
of the VCS tensor (\ref{Ten-BM}), e.g., ${\cal F}_{\rm LT} \to (p\cdot q/M^2) {\cal F}_{\rm LT} $, or be regarded as a constraint for the low energy behavior of its CFFs.

For completeness we also quote in Appendix \ref{TarrachHelicityAmplitudes} the form of the helicity transitions for the hadronic tensor parametrization
introduced by Tarrach \cite{Tarrach_1975tu}. The latter does not suffer from kinematical singularities as well and it will be used in the next section in the
low-energy expansion. Thereby, we find that with the CFF basis (\ref{cff-electric})--(\ref{cal-TabA-1}) the map is singularity-free for any values of kinematical
variables, i.e., as for Born amplitudes the ``electric''  longitudinal helicity-flip CFFs are proportional to the kinematical factor $\tK$, as one can read off from
explicit formulae (\ref{R^{(1)}_{0+}})--(\ref{R^{(12)}_{0+}}), while the $1/\tK$- behavior of the remaining functions can be absorbed by their rescaling, as
suggested in the preceding section. An analogous structure holds for the map of the transverse helicity-flip amplitudes [not explicitly shown in
Eqs.~(\ref{R^{(1)}_{-+}})--(\ref{R^{(12)}_{-+}})], where again the ``electric'' combination of CFFs is always anticipated with a factor $t-t_{\rm min}$.

\section{Generalized polarizabilities and low energy expansion}
\label{GPs}

Having discussed at length the deeply virtual regime that gives access to GPDs and provides a set of observables that exactly account for
kinematically suppressed effects, let us turn to the opposite limit when the incoming photon becomes quasi-real or even real.  In fact the
formalism presented in the previous sections can also be utilized in these cases as well. In doing so, we set $\xB=\Q^2/(s+\Q^2-M^2)$ as we take
the limit $\Q^2\to 0$ where $s=(p_1+ q_1)^2$ is the center-of-mass energy in the real Compton scattering process.  Obviously, the Bjorken and
the $\epsilon$ variables vanish as $\Q^2$ tends to zero, while the $\widetilde{K}$-factor takes the following value
$$
\lim_{\Q^2\to 0} \tK = \sqrt{ -t \left(1+\frac{t\, s}{\left(s-M^2\right)^2}\right)}.
$$

To start with, it is easy to verify that for a point-like particle the longitudinal spin-flip CFFs  ${\cal H}_{0+}$, ${\cal E}_{0+}$, ${\cal \widetilde{H}}_{0+}$
and the combination ${\cal \overline{E}}_{0+} = (\Delta\cdot q/p\cdot q)\, {\cal \widetilde{E}}_{0+} \propto \Q^2 {\cal \widetilde{E}}_{0+} $ entering the $\cal C$-coefficients all vanish as
$\mathcal{Q}^2 \to 0$. The real-photon limit also exists for the remaining eight transverse-helicity CFFs, where  $\overline{\cal E}_{\pm+}$
vanish. Thus, as it is known, six amplitudes remain for real Compton scattering.
No further peculiarities arise in the squared DVCS amplitudes presented in Sect.~\ref{sect-TDVCS}, and thus they can be used in a
straightforward fashion to recover, e.g.,  the Klein-Nishina formula and its extension to a polarized point-like target from the helicity-dependent CFFs
(\ref{H_+b-pointlike})--(\ref{tE_0+-pointlike}),
\begin{align}
\label{WQ-real}
\frac{d^2 \sigma}{d\cos(\theta_{\gamma\gamma}) d\varphi} &=
\frac{ R^2 }{2}\left( \frac{\omega^\prime}{\omega} \right)^2
\Bigg[
\frac{\omega}{\omega^\prime}+\frac{\omega^\prime}{\omega}- \sin^2(\theta_{\gamma\gamma})
\\
&
+
\lambda\,\Lambda \left( \frac{\omega}{\omega^\prime}-\frac{\omega^\prime}{\omega}\right) \cos(\theta_{\gamma\gamma}) \cos(\theta)
-
\lambda\,\Lambda \left(1-\frac{\omega^\prime}{\omega}\right) \sin(\theta_{\gamma\gamma}) \sin(\theta) \cos(\varphi)
\Bigg]
\, .
\nonumber
\end{align}
Here $R = \alpha_{\rm em}/M$ is the classical radius of the point particle and $\omega$ is the energy of the initial-state photon,
$$
\omega^\prime= \frac{\omega}{1+ \frac{\omega}{M}\left[1-\cos(\theta_{\gamma\gamma})\right]}
$$
is the energy of the outgoing photon, while $\theta_{\gamma\gamma}$ is the photon scattering angle in the laboratory frame. This provides a further
consistency cross-check on our analytical results.

Since we have devised a parametrization of the Compton tensor in the previous section that can be used for any kinematical settings, we can relate
our CFFs to Polarizabilities and their generalizations used in the description of the deformation response of the nucleon to the external long-wave
electromagnetic probe. To define the generalized polarizabilities, let us recall that according to the Low's theorem \cite{Low:1954kd}, in the low-energy
expansion of the Compton amplitude in the energy of the outgoing photon $q_2^0 = \omega^\prime$, the pole $(\omega^\prime)^{-1}$ and the
constant $(\omega^\prime)^0$ terms are entirely determined by the elastic form factors of the nucleon $F_1$ and $F_2$. However, the linear in
$\omega^\prime$-term in the expansion has yet another component that cannot be solely expressed in terms of the form factors and is encoded
through generalized polarizabilities. These are functions of the incoming photon three-momentum. Depending on the polarization of the incoming
and outgoing photons and their multipolarity, one can introduce ten different functions. To do it in a consistent fashion without the contamination from
the form factor contributions, one conventionally splits the total Compton amplitude into the Born term and the rest,
\begin{align}
T^{\mu\nu} = T^{\mu\nu}_{\rm Born} + T^{\mu\nu}_{\rm non-Born}
\, ,
\end{align}
where the first contribution $T^{\mu\nu}_{\rm Born} $ stems from the nucleon exchange between the electromagnetic vertices,
\begin{align}
\label{BornAmplitude}
T^{\mu\nu}_{\rm Born}
= - 4\pi \alpha_{\rm em} \,
\bar{u}_2\, \Gamma^\mu (-q_2) \left( {\not\! p_2} + {\not\! q}_2 - M \right)^{-1} \Gamma^\nu (q_1)\,  u_1
+ \mbox{(cross term)}
\, ,
\end{align}
with the $\Gamma$'s defined in Eq.\ (\ref{EMcurrentFFs}). As we pointed out above, its low-energy expansion starts with the inverse power of
the photon energy, while the leading term in $T^{\mu\nu}_{\rm non-Born} $ is $O (\omega^\prime)$. In complete analogy with static multipoles
that yield electric and magnetic dipoles for the linear coordinate moments of charge densities, in order to generate a linear in $\omega^\prime$ effect,
the outgoing photon should be either electric or magnetic. Thus the polarizabilities are labelled by the type $\rho_1 (\rho_2)$ of the incoming
(outgoing) photon, with $\rho \in\{ 0,1,2\}$ corresponding to scalar, magnetic and electric multipoles, respectively, the initial (final) orbital momentum
$L_1(L_2)$ and spin-flip nature of the transitions with $S\in\{0,1\}$ standing for nonflip and flip, accordingly, $P^{(\rho_2 L_2, \rho_1 L_1)S}$. The final state
electric multipoles can be traded in terms of other charge multipoles via the Siegert's theorem \cite{Siegert:1937yt}, while the initial state electric sector
reduces to the charge ones only up to an additional contribution from the so-called mixed generalized polarizabilities $P^{(\rho_2 L_2, L_1)S}$
\cite{Guichon:1995pu}.

We use the center-of-mass frame as spelled out in the Appendix \ref{CMkinematics} for relating our helicity CFFs to generalized polarizabilities.
The low-energy
expansion is performed with respect to the energy of the outgoing photon $\omega^\prime$, with polarizabilities being functions of the momentum of the
incoming virtual quantum $\bar{q}$. Two of the generalized polarizabilities are $\bar{q}^2$-generalization of the electric $\alpha$ and magnetic $\beta$
polarizabilities measured in real Compton scattering,
\begin{align}
P^{(01,01)0} (\bar{q}^2) =  - \sqrt{\frac{2}{3}}  \frac{\alpha (\bar{q}^2)}{\alpha_{\rm em}}
\, , \qquad
P^{(11,11)0} (\bar{q}^2) =  - \sqrt{\frac{8}{3}} \frac{\beta (\bar{q}^2)}{\alpha_{\rm em}}
\, ,
\end{align}
with factored out dependence on the fine structure constant $\alpha_{\rm em}$. As an intermediate step, we constructed the low-energy expansion for CFFs
in terms of twelve Tarrach's structure functions $f_i$. The results are presented in the Appendix \ref{LowEnergyTarrach}. However, imposing the implications
of the charge conjugation symmetry and nucleon crossing, some of the $f$'s vanish at low energy, i.e., $f_3$, $f_4$, $f_8$ and $f_{10}$ are of order
$O (\omega')$ and thus vanish at leading order \cite{Drechsel:1997xv}. This yields a set of relations for the ten generalized polarizabilities resulting in just six
independent ones. We verified along the way the low-energy expansion for the $A$ amplitudes defined by Guichon et al. in terms of the Tarrach's structure
functions $f_i$ calculated in Ref.~\cite{Drechsel:1997xv}.

The  helicity CFFs then read in terms of generalized polarizabilities, where we neglect the Born contribution and suppress all higher order terms in $\omega^\prime$:
\begin{itemize}
\item ($+1$,$+1$) helicity CFFs
\end{itemize}
\begin{align}
{\cal H}_{++}&=\frac{\omega'}{2 \sqrt{2}}\sqrt{\frac{E_i}{E_i+M}}\bigg\{M \bigg[6 \bar{q} P^{(11,11)1}+\sqrt{6}(\bar{q}-\omega_0) P^{(11,11)0}(1+\cos\vartheta) \bigg]
\nonumber
\\
&\ \ \ \ \ \ +3 \bar{q} \bigg[ (\bar{q} \cos\vartheta- \omega_0) P^{(11,11)1}+\sqrt{2}\bar{q} (\bar{q}-\omega_0\cos\vartheta) P^{(01,12)1}
\bigg]\bigg\} \, ,
\\[3.75mm]
{\cal E}_{++}&=\frac{\omega'M}{2\sqrt{2} \bar{q}}
 \sqrt{\frac{E_i}{E_i+M}} \bigg\{6(2 M-\omega_0) (2 M-\omega_0+\bar{q} \cos\vartheta) P^{(11,11)1}
\\
&\ \ \ \ \ \  +\sqrt{2} \bar{q} \bigg[6 \omega_0  (\omega_0-2 M) P^{(01,12)1}\cos\vartheta+\sqrt{3} \omega_0  P^{(11,11)0}(1+\cos\vartheta)+\bar{q} \big[6 (2 M-\omega_0) P^{(01,12)1}
\nonumber
\\
&\ \ \ \ \ \ -\sqrt{3} P^{(11,11)0} (1+\cos\vartheta)\big]\bigg]\bigg\}
\, , \nonumber\\
\widetilde{\cal H}_{++}&=\frac{3\omega' \bar{q}^3(\bar{q}-\omega_0 \cos\vartheta) }{4 \omega_0^2} \sqrt{\frac{E_i}{E_i+M}} \bigg\{\sqrt{2} P^{(11,11)1}-2 \omega_0 P^{(01,12)1}\bigg\} \, ,
\\[3.75mm]
\widetilde{\cal E}_{++}&=\frac{3\omega' M \bar{q}^2(2 M-\omega_0+\bar{q} \cos\vartheta) }{2 \omega_0^2}\sqrt{\frac{E_i}{E_i+M}} \bigg\{\sqrt{2} P^{(11,11)1}-2 \omega_0 P^{(01,12)1}\bigg\}
\, ,
\end{align}
\pagebreak[3]

\begin{itemize}
\item ($0$,$+1$) helicity CFFs
\end{itemize}
\begin{align}
{\cal H}_{0+}&=-\frac{\omega'\sqrt{-M\omega_0}}{2 \sqrt{2} q \sin\theta} \sqrt{\frac{E_i}{E_i+M}}\bigg\{2 \sqrt{3} (q-\omega_0 \cos\theta) P^{(11,00)1}+q \bigg[3\omega_0 P^{(01,01)1}-6 q P^{(01,01)1}\cos\theta
\nonumber
\\
&\ \ \ \ \ \ +\sqrt{6} q(q-\omega_0 \cos\theta) P^{(11,02)1}+3\omega_0P^{(01,01)1} \cos2\theta +2 \sqrt{6} M P^{(01,01)0} (1-\cos2\theta) \bigg]\bigg\}
\, ,
\\
{\cal E}_{0+}&=-\frac{ \omega'M\sqrt{-M\omega_0} }{\sqrt{2}\omega_0 q\sin\theta} \sqrt{\frac{E_i}{E_i+M}}
\bigg\{\omega_0 \bigg[(2 M-\omega_0) \big(6 P^{(01,01)1}+\sqrt{6}\omega_0 P^{(11,02)1}\big)-2 \sqrt{3} P^{(11,00)1}\bigg]\cos\theta
\nonumber
\\
&\ \ \ \ \ \ +q \bigg[2 \sqrt{3} P^{(11,00)1}+\omega_0 \big[\sqrt{6} P^{(01,01)0}+3 P^{(01,01)1}+\sqrt{6}(\omega_0-2 M)P^{(11,02)1} \big]
\nonumber
\\
&\ \ \ \ \ \ +\omega_0\big[3 P^{(01,01)1}-\sqrt{6} P^{(01,01)0}\big]\cos2\theta\bigg]\bigg\}
\, ,
\\[3.75mm]
\widetilde{\cal H}_{0+}&=-\frac{\omega'M (q-\omega_0 \cos\theta)}{2 \sqrt{2} b\sin\theta \sqrt{-M \omega_0}} \sqrt{\frac{E_i}{E_i+M}} \bigg\{6 q P^{(01,01)1}-\sqrt{3} \big(2 P^{(11,00)1}+\sqrt{2}q^2 P^{(11,02)1}\big) \cos\theta) \bigg\}
\, ,
\\[3.75mm]
\widetilde{\cal E}_{0+}&=-\frac{\omega'M(q-\omega_0 \cos\theta)\sqrt{-M \omega_0}}{\sqrt{2} b \omega_0^2\sin\theta} \sqrt{\frac{E_i}{E_i+M}}  \bigg\{6 q P^{(01,01)1}-\sqrt{3} \big(2 P^{(11,00)1}+\sqrt{2}q^2P^{(11,02)1} \big) \cos\theta\bigg\}
\, .
\end{align}

\begin{itemize}
\item ($-1$,$+1$) helicity CFFs
\end{itemize}
\begin{align}
{\cal H}_{-+}&= \frac{\omega'}{2 \sqrt{2}} \sqrt{\frac{E_i}{E_i+M}}\bigg\{3 \sqrt{2} \bar{q}^3 P^{(01,12)1}-3\omega_0 \bar{q} P^{(11,11)1} +M \bigg[6 \bar{q}P^{(11,11)1}+\sqrt{6}(\omega_0+\bar{q}) P^{(11,11)0} \bigg]
\nonumber
\\
&\ \ \ \ \ \ -\bigg[\sqrt{6} M(\omega_0+\bar{q}) P^{(11,11)0} -3 \bar{q}^2 \big(P^{(11,11)1}-\sqrt{2}\omega_0P^{(01,12)1} \big)\bigg] \cos\vartheta\bigg\}
\, ,
\\[3.75mm]
{\cal E}_{-+}&=\frac{\omega'M}{2 \sqrt{2} \bar{q}} \sqrt{\frac{E_i}{E_i+M}} \bigg\{\sqrt{6} \bar{q} (\omega_0+\bar{q}) P^{(11,11)0}(\cos\vartheta-1)
\\
&
\ \ \ \ \ \ + 6(2 M-\omega_0) (2 M-\omega_0+\bar{q} \cos\vartheta)P^{(11,11)1}
- 6 \sqrt{2}\bigg[\omega_0(\omega_0-2 M)^2-\bar{q}^3 \cos\vartheta\bigg] P^{(01,12)1} \bigg\} \, ,
\nonumber\\[3.75mm]
\widetilde{\cal H}_{-+}&=\frac{3\omega' \bar{q}^3(\bar{q}-\omega_0 \cos\vartheta)}{4 \omega_0^2} \sqrt{\frac{E_i}{E_i+M}}\bigg\{2 \omega_0 P^{(01,12)1} -\sqrt{2}P^{(11,11)1}\bigg\}
\, ,\\[3.75mm]
\widetilde{\cal E}_{-+}&=\frac{3\omega' M \bar{q}^2(\omega_0-2 M-\bar{q} \cos\vartheta) }{2 \omega_0^2}
\sqrt{\frac{E_i}{E_i+M}} \bigg\{\sqrt{2}P^{(11,11)1}-2\omega_0 P^{(01,12)1} \bigg\}
\, ,
\end{align}
where we introduced the convention for the initial-state energy under the condition of the vanishing final-state one \cite{Drechsel:1997xv},
\begin{align}
\omega_0\equiv\omega|_{\omega'=0}=M-E_i=M-\sqrt{M^2+\bar{q}^2}
\, ,
\end{align}
where $E_i$ is the incoming nucleon's energy.

\section{Conclusions}

In this paper we developed a unified framework for virtual Compton scattering that uses helicity Compton form factors for the analysis of different
regimes of the processes, interpolating between deeply virtual and quasi-real. The main ingredients of our consideration include: a clear separation
between the leptonic and hadronic components via  computation of helicity amplitudes in the target rest frame and an exact reconstruction of the
kinematical tensor decomposition for the hadronic Compton amplitude. The target rest frame is special since the entire dependence on the azimuthal
angle of the photon-nucleon scattering plane is encoded in the leptonic part of the cross section that was calculated exactly, overcoming the limitation
of the scheme from Ref.~\cite{Belitsky:2001ns} adopted previously for the analysis of electroproduction. Since partial results for the hadronic helicity
amplitudes with unpolarized and longitudinally polarized targets were available before, we complemented them with the ones for the transversally
polarized nucleon as well. We also incorporated the double photon helicity-flip amplitudes, induced by the gluon transversity GPDs, into the analysis. To
relate the helicity CFFs with the conventional ones emerging in the OPE analysis of the Compton amplitude, we introduced an exact Lorentz
decomposition for the latter that is free from kinematical singularities and computed the relations exactly.

The formalism presented here will be implemented in existing CFF/GPD fitting codes for the deeply virtual kinematics
\cite{Kumericki:2007sa,Kumericki:2009uq}.  This is required for an unbiased random variable map of an (almost) complete DVCS
measurement by the HERMES collaboration \cite{Airapetian:2008aa,Airapetian:2010ab,Airapetian:2011uq,Airapetian:2012mq}, providing 34
asymmetries in 12 kinematical bins, into the space of CFFs. Another advantage of such a tool is that one can  easily switch between various
``parton-to-hadron'' conventions, which will allow one for a precise numerical cross check with other existing software packages, which
are adopted for GPD model predictions \cite{Vanderhaeghen:1998uc,Kroll:2012sm}.  In addition, the kinematical power-suppressed corrections
from Ref.~\cite{Braun:2011dgBraun:2011zr,Braun:2012bgBraun:2012hq} can be conveniently taken into account relying on our formalism, thereby,
avoiding a recalculation of the leptoproduction cross section.

The second part of the paper is dedicated to the consideration of the low-energy limit of Compton scattering and relation of the CFFs  to the generalized
polarizabilities in the center-of-mass frame. Along the way, we re-expressed our helicity form factors in terms of the structure functions of the VCS
amplitude introduced by Tarrach, as well as defined their low-energy expansion. Our formulae set allows one to provide in a rather straightforward manner
the low energy expansion of the cross section for all possible polarization options  in terms of generalized polarizabilities in an analytic form. This may, as
in the DVCS case, provide a useful guideline for further experimental measurements.  Our analysis suggests a complementary approach to low-energy
limit: instead of relying on a low energy expansion in order to extract generalized polarizabilities, where a subtraction procedure (yielding its own
ambiguities) is needed to extract genuine information about nucleon deformation, one may adopt the known scheme used in DVCS kinematics and
seek a complete measurement of  CFFs at low final state photon energies.

In summary, our work, in addition to providing new results for exact kinematical treatment of real photon leptoproduction, can be viewed as a formalism
that uses the same set of hadronic variables, i.e., CFFs, at large and low virtuality of the initial state photon and provides a concise dictionary for translating
the former to generalized polarizabilities at low energies. This could provide a unified framework for experimental studies of CFFs in the entire
range of virtualities.

\vspace{5mm}

We are grateful to B. Pasquini for useful correspondence and to B.L.G. Bakker,  H.~Spiesberger, and M.~Stratmann for discussions.
D.M.\ would like to thank the Cosmology Initiative at ASU for the hospitality at different stages of this work. The work of A.B.\ and Y.J.\ was
supported by the U.S. National Science Foundation under the grant PHY-1068286 and of D.M.\ by  the Joint Research Activity Study of
Strongly Interacting Matter (HadronPhysics3, Grant Agreement No.~283286) under the Seventh Framework Program of the European Community.

\appendix
\renewcommand{\theequation}{\Alph{section}.\arabic{equation}}%

\section{Kinematical decomposition in target rest frame}
\label{KinematicalDecomposition}
\setcounter{equation}{0}
Let us first quote particles' momenta involved in scattering in the rest frame of the target. The components of the corresponding four-vectors read
\be
p_1
= (M, 0, 0, 0)
\, , \quad
q_1
=
\frac{\mathcal{Q}}{\epsilon}
\left(1, 0, 0, - \sqrt{1 + \epsilon^2}
\right)
\, , \quad
k
=
\frac{\mathcal{Q}}{y \epsilon}
\left(
1 , \sin\theta_l, 0, \cos\theta_l
\right) \, ,
\ee
with the lepton scattering angle being
\be
\cos \theta_l = - \frac{1 + \frac{y \epsilon^2}{2}}{\sqrt{1 + \epsilon^2}}
\, , \qquad
\sin \theta_l = \frac{\epsilon \sqrt{ 1 - y -  \frac{y^2 \epsilon^2}{4}}}{\sqrt{1 + \epsilon^2}}
\, .
\ee
The outgoing momenta are parameterized in terms of the scattering angles in the hadronic plane,
see Fig.\ \ref{fig:frame},
\begin{align}
q_2
&=
\frac{\mathcal{Q}^2 + \xB t}{2 M \xB}
\left(
1 ,
\cos\varphi_\gamma \sin\theta_\gamma ,
\sin\varphi_\gamma \sin\theta_\gamma,
\cos\theta_\gamma
\right) \, , \\
p_2
&=
\left(
M - \frac{t}{2 M}, \sqrt{- t + \frac{t^2}{4 M^2}} \cos\phi \sin\theta_p ,
\sqrt{- t + \frac{t^2}{4 M^2}} \sin\phi \sin\theta_p, \sqrt{- t + \frac{t^2}{4 M^2}}
\cos\theta_p
\right) \, ,
\end{align}
where the polar angles read in terms of the kinematical variables of the phase space
\be
\cos \theta_\gamma
=
-
\frac{1 + \frac{\epsilon^2}{2} \frac{\mathcal{Q}^2 + t}{\mathcal{Q}^2 + \xB t}}{\sqrt{1 + \epsilon^2}}
\, , \qquad
\cos \theta_p
= -
\frac{\epsilon^2 (\mathcal{Q}^2 - t) - 2 \xB t}{4 \xB M \sqrt{1 + \epsilon^2} \sqrt{- t + \frac{t^2}{4 M^2}}}
\, .
\ee
The azimuthal angle of the photon $\varphi_\gamma$ is related to the one of the outgoing hadron $\phi$
via $\varphi_\gamma = \phi + \pi$.

The photon polarization vectors (\ref{Polarization1}), (\ref{Polarization2}) can be kinematically decomposed in term of those involved in the virtual
Compton scattering process as follows:
\begin{eqnarray}
\label{eps_1(0)}
\varepsilon_1^\mu(0)&\!\!\! =\!\!\!&
-\frac{1}{{\cal Q}\sqrt{1 + \epsilon^2}}\, q_1^\mu - \frac{2 \xB}{{\cal Q}\sqrt{1 + \epsilon^2}}\, p_1^\mu
\\
\label{eps_1(1)}
\varepsilon_1^\mu(\pm 1)&\!\!\! =\!\!\!& \frac{\sqrt{1 + \epsilon^2}}{\sqrt{2}\widetilde K } \left[ \Delta^\mu -
\frac{\epsilon^2\left({\cal Q}^2 -t\right) -2\xB t}{2{\cal Q}^2\left(1 + \epsilon^2\right)} q_1^\mu+
\xB \frac{{\cal Q}^2  - t + 2 \xB t}{{\cal Q}^2\left(1 + \epsilon^2\right)}
p_1^\mu \right]  \mp  \frac{\xB}{\sqrt{2} \widetilde K} \frac{i\epsilon_{pq\Delta}^{\phantom{pq\Delta}\mu}}{{\cal Q}^2}  \,,
\\
\label{eps_2(1)}
\varepsilon_2^\mu(\pm 1)&\!\!\! =\!\!\!&  \frac{1+\frac{ \epsilon^2}{2} \frac{{\Q}^2 + t}{\Q^2 + \xB t}}{\sqrt{2}\widetilde K } \left[ \Delta^\mu -
\frac{\epsilon^2\left({\cal Q}^2 -t\right) -2\xB t}{2{\cal Q}^2\left(1 + \epsilon^2\right)} q_1^\mu+
\xB \frac{{\cal Q}^2  - t + 2 \xB t}{{\cal Q}^2\left(1 + \epsilon^2\right)}
p_1^\mu \right]
\nonumber\\
&& +  \frac{\widetilde{K}}{\sqrt{2}\left(1+\epsilon^2\right)\left(\Q^2+\xB t\right)}\left[\epsilon^2\,q_1^\mu- 2\,\xB\,  p_1^\mu \right]
\mp  \frac{\xB}{\sqrt{2} \widetilde K} \frac{i\epsilon_{pq\Delta}^{\phantom{pq\Delta}\mu}}{{\cal Q}^2}\,,
\end{eqnarray}
where $\tK$ is defined in Eq.~(\ref{K-tilde}).
Analogously one finds for the nucleon polarization vector (\ref{S-vector})
\begin{eqnarray}
\label{S_L}
 S^\mu_{\rm L}&\!\!\! =\!\!\!&  \frac{1}{\sqrt{1 + \epsilon^2}}\left[ \frac{1}{M}\,  p_1^\mu - \frac{\epsilon}{\Q}\, q_1^\mu\right],
 \\
\label{S_T}
 S^\mu_{\rm T}&\!\!\! =\!\!\!& \frac{\sqrt{1 + \epsilon^2}}{\widetilde K } \left[ \Delta^\mu -
\frac{\epsilon^2\left({\cal Q}^2 -t\right) -2\xB t}{2{\cal Q}^2\left(1 + \epsilon^2\right)}\, q_1^\mu+
\xB \frac{{\cal Q}^2  - t + 2 \xB t}{{\cal Q}^2\left(1 + \epsilon^2\right)}\,
p_1^\mu \right] \cos(\varphi)  - \frac{\xB}{i \widetilde K} \frac{i\varepsilon_{pq\Delta}^{\phantom{pq\Delta}\mu}}{{\cal Q}^2}  \sin(\varphi)\,.
\nonumber\\
\end{eqnarray}
The photon polarization vectors (\ref{eps_1(0)})--(\ref{eps_2(1)}) remain well defined in
the whole physical region including the phase-space boundary $t=t_{\rm min}$
where $\widetilde{K}$ vanishes.

We add the following useful relation that was used multiple times in simplification of analytical results
\begin{eqnarray}
\frac{\tK^2}{\Q^2} =  -\left(\!1+\frac{t}{\Q^2}\!\right)^2 \frac{\epsilon^2}{4}  -(1-\xB) \left(\! 1+\frac{\xB t}{\Q^2}\!\right)\frac{t}{\Q^2}\,.
\end{eqnarray}

\section{Fourier harmonics of the leptonic tensor}
\label{FourierHarmonics}
\setcounter{equation}{0}

Let us present explicit expressions for the Fourier coefficients entering the leptonic part of the
interference term \re{Def-Sqa-Int-Hel}, see Sect.~\ref{sect-INT}.
 As in
Ref.~\cite{Belitsky:2001ns}, we use the following shorthand notation
$$K = \sqrt{1-y - \frac{\epsilon ^2}{4} y^2}\,\frac{{\widetilde K}}{\Q}\,,$$
where $t^\prime$ is
$$
t^\prime = t - t_{\rm min}
\, .
$$

\subsection{Unpolarized and transversally polarized ${\rm TP} -$ target}
\label{C&S-leptonic}

The angular coefficients for unpolarized target and the transversally polarized ${\rm TP} -$ part are
given by the expressions
\begin{eqnarray}
C_{ab} (n)\,,\; C^{V}_{ab} (n)\,, \;  C^{A}_{ab} (n)\,, && \mbox{for}\quad n\in \{0,1,2,3\}\,,
\nonumber\\
S_{ab} (n)\,,\; S^{V}_{ab} (n)\,,\; S^{A}_{ab} (n)\,, &&  \mbox{for}\quad n\in \{1,2\}\,.
\nonumber
\end{eqnarray}
Note that these coefficients are identical with $C^{{\rm unp},\cdots}_{ab} (n)$ and $S^{{\rm unp},,\cdots}_{ab} (n)/\lambda$  of
Ref.~\cite{Belitsky:2010jw} and that the third odd harmonics, i.e.,
$$
S_{ab} (n=3) = S^{V}_{ab} (n=3) = S^{A}_{ab} (n=3)=0 \, ,
$$
and the following third even harmonics in longitudinal helicity flip CFFs
$$
C_{0b} (n=3) = C^{V}_{0b} (n=3) = C^{A}_{0b} (n=3)=0 \,
$$
vanish and will be not listed.

\begin{itemize}
\item Conserved photon-helicity coefficients:
\end{itemize}
\begin{align}
C_{++} (n &=0)
=-\frac{4(2-y)\left(1+\sqrt{1+\epsilon^2}\right)}{(1+\epsilon^2)^{2}}
\Bigg\{
\frac{{\widetilde K}^2}{\Q^2}\frac{(2-y)^2}{\sqrt{1+\epsilon^2}}
 \\
&+
\frac{t}{\Q^2}\left(1-y-\frac{\epsilon^2}{4} y^2 \right)(2-\xB)
\Bigg(1+
\frac{2\xB\left(2-\xB+ \frac{\sqrt{1+\epsilon^2}-1}{2}+
\frac{\epsilon^2}{2\xB}\right)\frac{t}{\Q^2}+
\epsilon^2}{(2-\xB)(1+\sqrt{1+\epsilon^2})}\Bigg)\Bigg\}
\, ,\nonumber \\[3mm]
C^{V}_{++} (n &=0)=
\frac{8 (2-y)}{(1+\epsilon^2)^2}\frac{\xB t}{\Q^2}
\Bigg\{
\frac{(2-y)^2 {\widetilde K}^2}{\sqrt{1+\epsilon^2}\Q^2}+
\left(1-y-\frac{\epsilon^2}{4} y^2 \right)\frac{1+\sqrt{1+\epsilon^2}}{2}
\nonumber \\
&\hspace{5cm}\times\left(1+\frac{t}{\Q^2}\right)\left(1+
\frac{\sqrt{1+\epsilon^2}-1+2\xB}{1+\sqrt{1+\epsilon^2}}\frac{t}{\Q^2} \right)\Bigg\},
\nonumber \\[3mm]
C^{A}_{++} (n &=0)=
\frac{8(2-y)}{(1+\epsilon^2)^2}\frac{t}{\Q^2}
\Bigg\{
\frac{(2-y)^2 {\widetilde K}^2}{\sqrt{1+\epsilon^2}\Q^2}\frac{1+\sqrt{1+\epsilon^2}
-2\xB}{2}
-
\left(1-y-\frac{\epsilon^2}{4} y^2 \right)
\Bigg[
\frac{2{\widetilde K}^2}{\Q^2}
 \nonumber \\
&-
\frac{1+\sqrt{1+\epsilon^2}}{2}
\left(1+\sqrt{1+\epsilon^2}-\xB
+\left(
\sqrt{1+\epsilon^2}-1+\xB\frac{3+\sqrt{1+\epsilon^2}-2\xB}{1+\sqrt{1+\epsilon^2}}
 \right)\frac{t}{\Q^2} \right)
 \Bigg]\Bigg\},
\nonumber \\[3mm]
C_{++} (n &=1)
=
\frac{-16 K\left(1-y-\frac{\epsilon^2}{4} y^2 \right)}{(1+\epsilon^2)^{5/2}}
\left\{
\left(1+(1-\xB)\frac{\sqrt{\epsilon^2+1}-1}{2\xB}+
\frac{\epsilon^2}{4\xB} \right)\frac{\xB t}{\Q^2}-
\frac{3\epsilon^2}{4} \right\}
\nonumber \\
&-4 K\left(2-2 y+y^2+ \frac{\epsilon^2}{2}y^2\right)
\frac{1+\sqrt{1+\epsilon^2}-\epsilon^2}{(1+\epsilon^2)^{5/2}}
\Bigg\{\!1-(1-3\xB)\frac{t}{{\cal Q}^2} \nonumber
\\&\qquad\qquad\qquad\qquad\qquad\qquad\qquad\qquad\qquad\qquad+
\frac{1- \sqrt{1+ \epsilon^2}+3\epsilon^2}{1+ \sqrt{1+ \epsilon^2}-
\epsilon^2}\frac{\xB t}{\Q^2}\Bigg\},
\nonumber \\[3mm]
C^{V}_{++} (n &=1)
=
\frac{16 K}{(1+\epsilon^2)^{5/2}}\frac{ \xB t}{\Q^2}
\Bigg\{(2-y)^2\left(1-(1-2\xB)\frac{t}{\Q^2} \right) \nonumber \\
&\hspace{6cm}
+
\left(1-y-\frac{\epsilon^2}{4} y^2 \right)
\frac{1+\sqrt{1+\epsilon^2}-2\xB}{2}\frac{t^\prime}{\Q^2}\Bigg\},
\nonumber \\[3mm]
C^{A}_{++} (n &=1)=
\frac{-16 K}{(1+\epsilon^2)^2}\frac{t}{\Q^2}
\Bigg\{
\left(1-y-\frac{\epsilon^2}{4} y^2 \right)\left(1-(1-2\xB)\frac{t}{\Q^2}
+\frac{4\xB(1-\xB)+\epsilon^2}{4\sqrt{1+\epsilon^2}}\frac{t^\prime}{\Q^2}\right)
 \nonumber\\
&\quad
-(2-y)^2\left(1-\frac{\xB}{2}+\frac{1+\sqrt{1+\epsilon^2}-2\xB}{4}\left(1-\frac{t}{\Q^2}\right)
+\frac{4\xB(1-\xB)+ \epsilon^2}{2\sqrt{1+\epsilon^2}}\frac{t^\prime}{\Q^2} \right)\Bigg\},
\nonumber \\[3mm]
C_{++} (n &=2)
=
\frac{8(2-y)\left(1-y-\frac{\epsilon^2}{4} y^2 \right)}{(1+\epsilon^2)^{2}}
\Bigg\{
\frac{2\epsilon^2}{\sqrt{1+ \
\epsilon^2}(1+\sqrt{1+\epsilon^2})}\frac{{\widetilde K}^2}{\Q^2}
 \nonumber \\
&\qquad\qquad\qquad\qquad\qquad\qquad\qquad\qquad+
\frac{\xB t\,t^{\prime}}{\Q^4}\left(1-\xB-\frac{\sqrt{1+\epsilon^2}-1}{2}
+ \frac{\epsilon^2}{2\xB} \right)\Bigg\}\, ,
\nonumber \\[3mm]
C^{V}_{++} (n &=2)=
\frac{8(2-y)\left(1-y-\frac{\epsilon^2}{4} y^2 \right)}{(1+\epsilon^2)^{2}}
\frac{ \xB t}{\Q^2}
\Bigg\{
\frac{4{\widetilde K}^2}{\sqrt{1+\epsilon^2}\Q^2}+
\frac{1+\sqrt{1+\epsilon^2}-2\xB}{2}\left(1+ \frac{t}{\Q^2}\right)
\frac{t^\prime}{\Q^2}\Bigg\},
\nonumber \\[3mm]
C^{A}_{++} (n &=2)=
\frac{4(2-y)\left(1-y-\frac{\epsilon^2}{4} y^2 \right)}{(1+\epsilon^2)^{2}}
\frac{t}{\Q^2}
\Bigg\{
\frac{4(1-2\xB){\widetilde K}^2}{\sqrt{1+\epsilon^2}\Q^2}
-\left(3-\sqrt{1+\epsilon^2}-2\xB
+\frac{\epsilon^2}{\xB} \right)\frac{\xB t^\prime}{\Q^2}
\Bigg\},
\nonumber \\[3mm]
C_{++} (n &=3)
=-8 K\left(1-y-\frac{\epsilon^2}{4} y^2 \right)
\frac{\sqrt{1+\epsilon^2}-1}{(1+\epsilon^2)^{5/2}}
\left\{(1- \xB)\frac{t}{{\cal Q}^2}+
\frac{\sqrt{1+\epsilon^2}-1}{2}\left(1+ \frac{t}{{\cal Q}^2} \right)
\right\}
\, ,\nonumber \\[3mm]
C^{V}_{++} (n &=3)
=
-\frac{8 K\left(1-y-\frac{\epsilon^2}{4} y^2 \right)}{(1+\epsilon^2)^{5/2}}
\frac{ \xB t}{\Q^2}
\left\{
\sqrt{1+\epsilon^2}-1+ \left (1+\sqrt{1+\epsilon^2}-2\xB\right)
\frac{t}{\Q^2} \right\},
\nonumber \\[3mm]
C^{A}_{++} (n &=3)
=
\frac{16 K\left(1-y-\frac{\epsilon^2}{4} y^2 \right)}{(1+\epsilon^2)^{5/2}}
\frac{t\,t^\prime}{\Q^4}
\left\{
\xB(1-\xB)+\frac{\epsilon^2}{4} \right\},
\nonumber \\[3mm]
S_{++} (n &=1)
= \frac{8 K (2-y)y}{1+ \epsilon^2}
\Bigg\{1+
\frac{1-\xB+\frac{\sqrt{1+\epsilon^2}-1}{2}}{1+\epsilon^2}\,
\frac{t^{\prime}}{\Q^2}\Bigg\}
\, ,\nonumber \\[3mm]
S^{V}_{++} (n &=1)
=
-\frac{8 K (2-y) y}{(1+\epsilon^2)^{2}}\frac{ \xB t}{\Q^2}
\left\{
\sqrt{1+\epsilon^2}-1+ \left(1+\sqrt{1+\epsilon^2}-2\xB\right)
\frac{t}{\Q^2}
\right\},\nonumber \\[3mm]
S^{A}_{++} (n &=1)=
\frac{8 K (2-y)y}{(1+\epsilon^2)}\frac{t}{\Q^2}
\Big\{1-(1-2\xB)\frac{1+\sqrt{1+\epsilon^2}-2\xB}{2(1+\epsilon^2)}
\frac{t^\prime}{\Q^2}\Big\}
\nonumber \\[3mm]
S_{++} (n &=2)
=
-\frac{4\left(1-y-\frac{\epsilon^2}{4} y^2\right)y}{(1+\epsilon^2)^{3/2}}
\left(1+\sqrt{1+\epsilon^2}-2\xB\right)
\nonumber \\
&\qquad\times
\frac{t^{\prime}}{\Q^2}
\Bigg\{
\frac{\epsilon^2- \xB (\sqrt{1+\epsilon^2}-1)}{1+\sqrt{\epsilon^2+1}-2\xB}-
\frac{2\xB+\epsilon^2}{2\sqrt{1+\epsilon^2}}\,\frac{t^{\prime}}{\Q^2}
\Bigg\},\nonumber \\[3mm]
S^{V}_{++} (n &=2)
=
-\frac{
4\left(1-y-\frac{\epsilon^2}{4} y^2 \right) y
}{(1+\epsilon^2)^{2}}\frac{\xB t}{\Q^2}
\nonumber \\
&\times
\left(1-(1-2\xB)\frac{t}{\Q^2} \right)
\left\{
\sqrt{1+\epsilon^2}-1+ \left(1+\sqrt{1+\epsilon^2}-2\xB
\right)\frac{t}{\Q^2} \right\},
\nonumber \\[3mm]
S^{A}_{++} (n &=2)
=
-\frac{8\left(1-y-\frac{\epsilon^2}{4} y^2 \right)y}{(1+\epsilon^2)^{2}}
\frac{t\,t^\prime}{\Q^4}
\left(1-\frac{\xB}{2}+\frac{3 \epsilon^2}{4}\right)
\nonumber \\
&\quad\times
\left(1+\sqrt{1+\epsilon^2}-2\xB \right)
\Bigg(
1+ \frac{4 (1- \xB)\xB+ \epsilon^2}{4-2\xB+3\epsilon^2}\frac{t}{\Q^2}
\Bigg)
\, .
\nonumber
\end{align}

\begin{itemize}
\item Longitudinal-transverse coefficients:
\end{itemize}
\begin{align}
C_{0+} (n &=0)
=
\frac{12\sqrt{2} K (2-y)\sqrt{1-y-\frac{\epsilon^2}{4} y^2}}{\left(1+\epsilon^2\right)^{5/2}}
\left\{
\epsilon^2+ \frac{2-6\xB-\epsilon^2}{3}\frac{t}{\Q^2} \right\}
\, ,\\[3mm]
C^{V}_{0+} (n &=0)
=
\frac{
24\sqrt{2} K (2-y)\sqrt{1-y-\frac{\epsilon^2}{4} y^2}
}{
\left(1+\epsilon^2\right)^{5/2}
}
\frac{\xB t}{\Q^2}\left\{1-(1-2\xB)\frac{t}{\Q^2} \right\}
\, ,\nonumber \\[3mm]
C^{A}_{0+} (n &=0)
=
\frac{4\sqrt{2} K (2-y)\sqrt{1-y-\frac{\epsilon^2}{4}  y^2}}{\left(1+\epsilon^2\right)^{5/2}}
\frac{t}{\Q^2} (8-6\xB+5\epsilon^2)
\left\{1- \frac{t}{\Q^2}\frac{2-12\xB (1- \xB)- \epsilon^2}{8-6\xB+5\epsilon^2} \right\}
\, ,\nonumber \\[3mm]
C_{0+} (n &=1)
=
\frac{8\sqrt{2}\sqrt{1-y-\frac{\epsilon^2}{4} y^2}}{\left(1+\epsilon^2\right)^2}
\Bigg\{
(2-y)^2\frac{t^\prime}{\Q^2}
\Bigg(
1- \xB+
 \frac{(1-\xB)\xB+ \frac{\epsilon^2}{4}}{\sqrt{1+\epsilon^2}}
\frac{t^\prime}{\Q^2}
\Bigg)
\nonumber \\
&\qquad+\frac{1-y- \frac{\epsilon^2}{4} y^2}{\sqrt{1+\epsilon^2}}
\left(
1-(1-2\xB)\frac{t}{\Q^2}
\right)
\left(
\epsilon^2-2\left(1+\frac{\epsilon^2}{2\xB}
\right)\frac{\xB t}{\Q^2}
\right)
\Bigg\}
\, ,\nonumber \\[3mm]
C^{V}_{0+} (n &=1)
=
\frac{
16\sqrt{2}\sqrt{1-y-\frac{\epsilon^2}{4} y^2}
}{
\left(1+\epsilon^2\right)^{5/2}
}
\frac{\xB t}{\Q^2}
\left\{
\frac{\widetilde{K}^2 (2-y)^2}{\Q^2}+
\left(1-(1-2\xB)\frac{t}{\Q^2} \right)^2
\left(1-y-\frac{\epsilon^2}{4} y^2 \right) \right\}
\, ,\nonumber \\[3mm]
C^{A}_{0+} (n &=1)
=
\frac{8\sqrt{2}\sqrt{1-y-\frac{\epsilon^2}{4} y^2}}{\left(1+\epsilon^2\right)^{5/2}}
\frac{t}{\Q^2}
\Bigg\{
\frac{\widetilde{K}^2}{\Q^2} (1-2\xB)(2-y)^2
\nonumber \\
&+
\left(1-(1-2\xB)\frac{t}{\Q^2} \right)\left(1-y-\frac{\epsilon^2}{4} y^2\right)
\left(4-2\xB+3\epsilon^2+\frac{t}{\Q^2} (4\xB (1- \xB)+ \epsilon^2) \right)
\Bigg\}
\, ,\nonumber \\[3mm]
C_{0+} (n &=2)
=
-\frac{
8\sqrt{2} K (2-y)\sqrt{1-y-\frac{\epsilon^2}{4} y^2}
}{
\left(1+\epsilon^2\right)^{5/2}}
\left(1+\frac{\epsilon^2}{2}\right)
\left\{
1+ \frac{1+\frac{\epsilon^2}{2\xB}}{1+\frac{\epsilon^2}{2}}\frac{\xB t}{\Q^2}
 \right\}
\, ,\nonumber \\[3mm]
C^{V}_{0+} (n &=2)
=
\frac{
8\sqrt{2} K (2-y)\sqrt{1-y-\frac{\epsilon^2}{4} y^2}
}{
\left(1+\epsilon^2\right)^{5/2}
}
\frac{\xB t}{\Q^2}
\left(1-(1-2\xB)\frac{t}{\Q^2} \right)
\, ,\nonumber \\[3mm]
C^{A}_{0+} (n &=2)
=
\frac{
8\sqrt{2} K (2-y)\sqrt{1-y-\frac{\epsilon^2}{4} y^2}
}{\left(1+\epsilon^2\right)^{2}}
\frac{t}{\Q^2}
\Bigg\{
1-\xB+
\frac{t^\prime}{2\Q^2}\frac{4\xB (1- \xB)+ \epsilon^2}{\sqrt{1+ \epsilon^2}}
\Bigg\}
\, ,\nonumber \\[3mm]
S_{0+} (n &=1)
=
\frac{8\sqrt{2} (2-y) y\sqrt{1-y-\frac{\epsilon^2}{4} y^2}
}{
\left(1+\epsilon^2\right)^2
}
\frac{\widetilde{K}^2}{\Q^2}
\, ,\nonumber \\[3mm]
S^{V}_{0+} (n &=1)
=
\frac{4\sqrt{2} y (2-y)\sqrt{1-y-\frac{\epsilon^2}{4} y^2}}{\left(1+\epsilon^2\right)^{2}}
\frac{\xB t}{\Q^2}
\left\{4 (1-\xB)\frac{t}{\Q^2}\left(1+ \frac{\xB t}{\Q^2} \right)+
\epsilon^2\left(1+ \frac{t}{\Q^2} \right)^2 \right\}
\, ,\nonumber \\[3mm]
S^{A}_{0+} (n &=1)
=
- \frac{
8\sqrt{2} y (2-y) (1-2\xB)\sqrt{1-y-\frac{\epsilon^2}{4} y^2}
}{
\left(1+\epsilon^2\right)^{2}
}
\frac{t \widetilde{K}^2}{\Q^4}
\, ,\nonumber \\[3mm]
S_{0+} (n &=2)
=
\frac{
8\sqrt{2} K y\sqrt{1-y-\frac{\epsilon^2}{4} y^2}
}{
\left(1+\epsilon^2\right)^2
}
\left(1+\frac{\epsilon^2}{2}\right)
\left\{
1+ \frac{1+\frac{\epsilon^2}{2\xB}}{1+\frac{\epsilon^2}{2}}\frac{\xB t}{\Q^2}
 \right\}
\, ,\nonumber \\[3mm]
S^{V}_{0+} (n &=2)
=- \frac{8\sqrt{2} K y\sqrt{1-y-\frac{\epsilon^2}{4}
y^2}}{\left(1+\epsilon^2\right)^{2}}
\frac{\xB t}{\Q^2}
\left\{1-(1-2\xB)\frac{t}{\Q^2} \right\}
\, ,\nonumber \\[3mm]
S^{A}_{0+} (n &=2)
=
- \frac{
2\sqrt{2} K y\sqrt{1-y-\frac{\epsilon^2}{4} y^2}
}{
\left(1+\epsilon^2\right)^{2}
}
\frac{t}{\Q^2}
\left(4-4\xB+2\epsilon^2+\frac{2 t}{\Q^2} (4\xB (1- \xB)+ \epsilon^2) \right) ,
\nonumber
\end{align}

\begin{itemize}
\item Transverse-transverse helicity-flip coefficients:
\end{itemize}
\begin{align}
C_{-+} (n &=0)
= \frac{8 (2-y)}{\left(1+\epsilon^2\right)^{3/2}}
\Bigg\{(2-y)^2\frac{\sqrt{1+\epsilon^2}-1}{2(1+\epsilon^2)}
\frac{\widetilde{K}^2}{\Q^2}
\\&\quad
+\frac{1-y-\frac{\epsilon^2}{4} y^2}{\sqrt{1+\epsilon^2}}
\Bigg(1-\xB-\frac{\sqrt{1+\epsilon^2}-1}{2}+
\frac{\epsilon^2}{2\xB}\Bigg)\frac{\xB t\,t^\prime}{\Q^4}
\Bigg\}
\, ,\nonumber \\[3mm]
C^{V}_{-+} (n &=0)
=
\frac{4 (2-y)}{(1+ \epsilon^2)^{5/2}}\frac{\xB t}{\Q^2}
\Bigg\{
\frac{2\widetilde{K}^2}{\Q^2}
\left(2-2 y+y^2+ \frac{\epsilon^2}{2} y^2\right)
-\left(1-(1-2\xB)\frac{t}{\Q^2} \right)
\nonumber \\
&\qquad\times
\left(1-y-\frac{\epsilon^2}{4} y^2\right)\left(
\sqrt{1+ \epsilon^2}-1+ \left( \sqrt{1+ \epsilon^2}+1-2\xB \right)\frac{t}{\Q^2}
\right)
\Bigg\}
\, .\nonumber \\[3mm]
C^{A}_{-+} (n &=0)
=
\frac{4 (2-y)}{(1+ \epsilon^2)^{2}}\frac{t}{\Q^2}
\Bigg\{
\frac{t^\prime}{\Q^2}\left(1-y- \frac{\epsilon^2}{4} y^2 \right)
\left(2\xB^2- \epsilon^2-3\xB+ \xB\sqrt{1+ \epsilon^2} \right)
\nonumber
\\&+
\frac{\widetilde{K}^2}{\Q^2\sqrt{1+ \epsilon^2}}
\left(4-2\xB (2-y)^2-4 y+y^2-y^2 (1+ \epsilon^2)^{3/2} \right)
\Bigg\}
\, ,\nonumber \\[3mm]
C_{-+} (n &=1)
=
\frac{8 K}{\left(1+\epsilon^2\right)^{3/2}}
\Bigg\{
(2-y)^2\frac{2-\sqrt{1+\epsilon^2}}{1+\epsilon^2}
\Bigg(
\frac{\sqrt{1+\epsilon^2}-1+\epsilon^2}{2\left(2-\sqrt{1+\epsilon^2}
\right)}\left(1-\frac{t}{\Q^2}\right)-
\frac{\xB t}{\Q^2}\Bigg)
\nonumber \\
&+
2\frac{1-y-\frac{\epsilon^2}{4} y^2}{\sqrt{1+\epsilon^2}}
\Bigg(
\frac{1-\sqrt{1+\epsilon^2}+\frac{\epsilon^2}{2}}{2\sqrt{1+\epsilon^2}}
+\frac{t}{\Q^2}\left(
1-\frac{3\xB}{2}+\frac{\xB+\frac{\epsilon^2}{2}}{\ 2\sqrt{1+\epsilon^2}}
\right)\Bigg)
\Bigg\}\,  ,
\nonumber \\[3mm]
C^{V}_{-+} (n &=1)
=
\frac{8 K}{(1+ \epsilon^2)^{5/2}}\frac{\xB t}{\Q^2}
\Bigg\{2\left(1-(1-2\xB)\frac{t}{\Q^2} \right)
\left(2-2 y+y^2+\frac{\epsilon^2}{2} y^2\right) \nonumber \\
&+
\left(1-y-  \frac{\epsilon^2}{4} y^2\right)\left(3- \sqrt{1+ \
\epsilon^2}- \left(3 (1-2\xB)+ \sqrt{1+ \epsilon^2} \right)
\frac{t}{\Q^2} \right)
\Bigg\}
\, ,\nonumber \\[3mm]
C^{A}_{-+} (n &=1)
=
\frac{4 K}{(1+ \epsilon^2)^{5/2}}\frac{t}{\Q^2}
\Bigg\{
\left(2-2 y+y^2+ \frac{\epsilon^2}{2} y^2\right)
\nonumber \\
&\times
\left(5-4\xB+3\epsilon^2- \sqrt{1+\epsilon^2}-
\frac{t}{\Q^2}\left(1- \epsilon^2- \sqrt{1+ \epsilon^2}
-2\xB (4-4\xB- \sqrt{1+ \epsilon^2}) \right) \right)
\nonumber \\&+
\left(1-y- \frac{\epsilon^2}{4} y^2 \right)
\nonumber \\
&\times
\left(8+5\epsilon^2-6\xB+2\xB\sqrt{1+ \epsilon^2}-
\frac{t}{\Q^2}\left(2- \epsilon^2+2\sqrt{1+ \epsilon^2}
-4\xB (3-3\xB+ \sqrt{1+ \epsilon^2}) \right) \right)
\Bigg\}
\, ,\nonumber \\[3mm]
C_{-+} (n &=2)
=
4(2-y)\left(1-y- \frac{\epsilon^2}{4} y^2\right)
\frac{1+ \sqrt{1+\epsilon^2}}{\left(1+\epsilon^2\right)^{5/2}}
\Bigg\{(2-3\xB)\frac{t}{\Q^2}
\nonumber \\
&+
\left(1-2\xB+  \frac{2 (1-\xB)}{1+\sqrt{1+\epsilon^2}}
\right)\frac{\xB t^2}{\Q^4}+
\Bigg(1+ \frac{\sqrt{1+\epsilon^2}+
\xB+(1-\xB)\frac{t}{\Q^2}}{1+\sqrt{1+\epsilon^2}}\frac{t}{\Q^2}
\Bigg)\epsilon^2\Bigg\}
\, ,\nonumber \\[3mm]
C^{V}_{-+} (n &=2)
=
\frac{4 (2-y)\left(1-y- \frac{\epsilon^2}{4} y^2\right)
}{(1+\epsilon^2)^{5/2}}\frac{\xB t}{\Q^2}
\Bigg\{4\frac{\widetilde{K}^2}{\Q^2} +
1+\sqrt{1+ \epsilon^2}
\nonumber \\
&+
\frac{t}{\Q^2}\left((1-2\xB)\left(1-2\xB- \sqrt{1+ \epsilon^2} \right)
\frac{t}{\Q^2}-2+4\xB+2\xB\sqrt{1+ \epsilon^2} \right)\Bigg\}
\, ,\nonumber \\[3mm]
C^{A}_{-+} (n &=2)
=
\frac{16 (2-y)\left(1-y- \frac{\epsilon^2}{4} y^2\right)
}{(1+\epsilon^2)^{3/2}}\frac{t}{\Q^2}
\Bigg\{
\frac{\widetilde{K}^2}{\Q^2}\frac{1-2\xB}{1+ \epsilon^2}
\nonumber\\
&-
\frac{1- \xB}{4\xB (1- \xB)+ \epsilon^2}
\left(2\xB^2-\epsilon^2-3\xB- \xB\sqrt{1+ \epsilon^2} \right)-
\frac{t^\prime}{\Q^2}
\frac{2\xB^2- \epsilon^2-3\xB-\xB\sqrt{1+\epsilon^2}}{4\sqrt{1+ \epsilon^2}}
\Bigg\}
\, ,\nonumber \\[3mm]
C_{-+} (n &=3)
=
-8K\left(1-y-\frac{\epsilon^2}{4} y^2 \right)
\frac{1+ \sqrt{1+\epsilon^2}+
\frac{\epsilon^2}{2}}{\left(1+\epsilon^2 \right)^{5/2}}
\Bigg\{
1+\frac{1+\sqrt{1+\epsilon^2}+
\frac{\epsilon^2}{2\xB}}{1+\sqrt{1+\epsilon^2}
+
\frac{\epsilon^2}{2}}\frac{\xB t}{\Q^2}\Bigg\}
\, ,\nonumber \\[3mm]
C^{V}_{-+} (n &=3)
=
\frac{8 K\left(1-y- \frac{\epsilon^2}{4} y^2\right)
}{(1+\epsilon^2)^{5/2}}\frac{\xB t}{\Q^2}
\left(1+ \sqrt{1+ \epsilon^2} \right)
\Bigg\{
1-\frac{t}{\Q^2}\frac{1-2\xB- \sqrt{1+ \epsilon^2}}{1+ \sqrt{1+\epsilon^2}}
\Bigg\}
\, ,\nonumber \\[3mm]
C^{A}_{-+} (n &=3)
=
\frac{16K\left(1-y- \frac{\epsilon^2}{4} y^2\right)
}{(1+\epsilon^2)^{2}}\frac{t}{\Q^2}
\Bigg\{
1- \xB+ \frac{t^\prime}{\Q^2}
\frac{4\xB (1- \xB)+\epsilon^2}{4\sqrt{1+ \epsilon^2}}
\Bigg\}
\, ,\nonumber \\[3mm]
S_{-+} (n &=1)
=
\frac{4 K (2-y) y}{\left(1+\epsilon^2\right)^2}
\Bigg\{
1- \sqrt{1+\epsilon^2}+2\epsilon^2-2
\left(1+\frac{\sqrt{1+\epsilon^2}-1}{2\xB} \right)\frac{\xB t}{\Q^2}
\Bigg\}
\,,\nonumber \\[3mm]
S^{V}_{-+} (n &=1)
= \frac{8 K y (2-y)}{(1+ \epsilon^2)^{2}}\frac{\xB t}{\Q^2}
\left(1+ \sqrt{1+ \epsilon^2} \right)
\Bigg\{
1-\frac{t}{\Q^2}\frac{1-2\xB- \sqrt{1+ \epsilon^2}}{1+ \sqrt{1+\epsilon^2}}
\Bigg\}
\, ,\nonumber \\[3mm]
S^{A}_{-+} (n &=1)
=
\frac{4 K y (2-y)}{(1+ \epsilon^2)^{2}}\frac{t}{\Q^2}
\Bigg\{
3+2\epsilon^2 \nonumber \\
&+
\sqrt{1+ \epsilon^2}-2\xB-2\xB\sqrt{1+ \epsilon^2}-
\frac{t}{\Q^2} (1-2\xB)\left(1-2\xB- \sqrt{1+ \epsilon^2} \right)
\Bigg\}
\, ,\nonumber \\[3mm]
S_{-+} (n &=2)
=
2 y\left(1-y-\frac{\epsilon^2}{4} y^2 \right)
\frac{1+\sqrt{1+\epsilon^2}}{\left(1+\epsilon^2\right)^2}
\Bigg(\epsilon^2-2\Bigg(1+\frac{\epsilon^2}{2\xB}\Bigg)
\frac{\xB t}{\Q^2}\Bigg)
\nonumber \\
&\qquad
\times\Bigg\{
1+\frac{\sqrt{1+\epsilon^2}-1+2\xB}{1+\sqrt{1+\epsilon^2}}
\frac{t}{\Q^2}\Bigg\}
\, ,\nonumber \\[3mm]
S^{V}_{-+} (n &=2)
=
\frac{4 y\left(1-y- \frac{\epsilon^2}{4} y^2\right)
}{(1+\epsilon^2)^{2}}
\frac{\xB t}{\Q^2}
\nonumber \\
&\times
\left(1+ \sqrt{1+ \epsilon^2} \right)
\left(1-(1-2\xB)\frac{t}{\Q^2} \right)
\Bigg\{1-
\frac{t}{\Q^2}\frac{1-2\xB- \sqrt{1+ \epsilon^2}}{1+ \sqrt{1+\epsilon^2}}
\Bigg\}
\, ,\nonumber \\[3mm]
S^{A}_{-+} (n &=2)
=
\frac{2y\left(1-y- \frac{\epsilon^2}{4} y^2\right)
}{(1+\epsilon^2)^{2}}\frac{t}{\Q^2}
\left(4-2\xB+3\epsilon^2+ \frac{t}{\Q^2}
\left(4\xB (1- \xB)+\epsilon^2 \right) \right)
\nonumber \\
&\times
\left(1+ \sqrt{1+ \epsilon^2}-
\frac{t}{\Q^2}\left(1-2\xB- \sqrt{1+ \epsilon^2} \right) \right)
\, .\nonumber
\end{align}

\subsection{Longitudinally and transversally polarized ${\rm TP} +$ target}
\label{dC&dS-leptonic}

The angular coefficients for longitudinally and transversally polarized ${\rm TP} +$ parts are
determined by the expressions
\begin{eqnarray}
\delta C_{ab} (n)\,,\; \delta C^{V}_{ab} (n)\,, \; \delta C^{A}_{ab} (n)\,, && \mbox{for}\quad n\in \{0,1,2\}\,,
\nonumber\\
\delta S_{ab} (n)\,,\; \delta S^{V}_{ab} (n)\,,\; \delta S^{A}_{ab} (n)\,, &&  \mbox{for}\quad n\in \{1,2,3\}\,.
\nonumber
\end{eqnarray}
Note again, as in the previous section, these coefficients are identical with $C^{{\rm LP},\cdots}_{ab} (n)/\lambda \Lambda$ and
$S^{{\rm LP},,\cdots}_{ab} (n)/\Lambda$  of Ref.~\cite{Belitsky:2010jw} and that the third even harmonics, i.e.,
$$
\delta C_{ab} (n=3)=\delta C^V_{ab} (n=3)=\delta C^{A}_{ab} (n=3)=0,
$$
and the following third odd harmonics in longitudinal helicity flip CFFs
$$
\delta S_{0b} (n=3) = \delta S^{V}_{0b} (n=3) = \delta S^{A}_{0b} (n=3)=0 \,
$$
vanish and thus will not be presented below.\\

\begin{itemize}
\item Conserved photon-helicity coefficients:
\end{itemize}
\begin{align}
\delta C_{++} (n &=0)
=
-\frac{4 y\left(1+\sqrt{1+\epsilon^2}\right)}{(1+\epsilon^2)^{5/2}}
\Bigg\{(2-y)^2\frac{{\widetilde K}^2}{\Q^2}
\\
&+
\left(1-y-\frac{\epsilon^2}{4}y^2\right)
\left(\frac{\xB t}{\Q^2}-\left(1-\frac{t}{\Q^2}\right)
\frac{\epsilon^2}{2} \right)\left(1+\frac{\sqrt{1+\epsilon^2}-1+2\xB}{1+\sqrt{1+\epsilon^2}}
\frac{t}{\Q^2}\right)\Bigg\},
\nonumber \\[3mm]
\delta C^{V}_{++} (n &=0)
=
\frac{4 y\left(1+\sqrt{1+\epsilon^2}\right)}{\left(1+\epsilon^2\right)^{5/2}}\frac{t}{\Q^2}
\Bigg\{(2-y)^2\,
\frac{1+\sqrt{1+\epsilon^2}-2\xB}{1+\sqrt{1+\epsilon^2}}\,
\frac{{\widetilde K}^2}{\Q^2}+ \left(1-y-\frac{\epsilon^2}{4} y^2\right)
\nonumber\\
&\times\left(2-\xB+\frac{3\epsilon^2}{2}
\right)\left(1+\frac{4 (1-\xB)\xB+\epsilon^2}{4-2\xB+3\epsilon^2}
\frac{t}{\Q^2} \right)\left(1+\frac{\sqrt{1+\epsilon^2}-1+2\xB}{1+\sqrt{1+\epsilon^2}}
\frac{t}{\Q^2} \right)\Bigg\},
\nonumber \\[3mm]
\delta C^{A}_{++} (n &=0)
=
\frac{4 y}{\left(1+\epsilon^2\right)^{5/2}}
\frac{\xB t}{\Q^2}
\Bigg\{2 (2-y)^2\frac{{\widetilde K}^2}{\Q^2}+
\left(1-y-\frac{\epsilon^2}{4} y^2\right)
(1+\sqrt{1+\epsilon^2})
\nonumber \\
&\hspace{3.5cm}\times
\left(1-(1-2\xB)\frac{t}{\Q^2} \right)
\left(
1+\frac{\sqrt{1+\epsilon^2}-1+2\xB}{1+\sqrt{1+\epsilon^2}}\frac{t}{\Q^2}
\right)
\Bigg\},
\nonumber \\[3mm]
\delta C_{++} (n &=1)
=
-\frac{4 K y (2-y)}{\left(1+\epsilon^2\right)^{5/2}}
(1+\sqrt{1+\epsilon^2}-\epsilon^2)
\Bigg\{
1-\left(1-2\xB\frac{2+\sqrt{1+\epsilon^2}}{1+\sqrt{1+\epsilon^2}-\epsilon^2}\right)\frac{t}{\Q^2}
\Bigg\},
\nonumber \\[3mm]
\delta C^{V}_{++} (n &=1)
=
\frac{8 K (2-y) y}{\left(1+\epsilon^2\right)^{2}}
\left( \sqrt{1+\epsilon^2}+2(1- \xB) \right)\frac{t}{\Q^2}
\Bigg\{
1-\frac{1+\frac{1-\epsilon^2}{\sqrt{1+\epsilon^2}}-2\xB
\left(1+\frac{4 (1-\xB)}{\sqrt{1+\epsilon^2}}\right)}{2 \left(\sqrt{1+\epsilon^2}+2 (1-\xB)\right)}
\frac{t^\prime}{\Q^2}
\Bigg\},
\nonumber \\[3mm]
\delta C^{A}_{++} (n &=1)
=
\frac{16 K (2-y) y}{\left(1+\epsilon^2\right)^{5/2}}
\frac{\xB t}{\Q^2}\left(1-(1-2\xB)\frac{t}{\Q^2}\right),
\nonumber \\[3mm]
\delta C_{++} (n &=2)
=
-\frac{4 y\left(1-y-\frac{\epsilon^2}{4} y^2\right)}{\left(1+\epsilon^2\right)^{5/2}}
\left( \frac{\xB t}{\Q^2}-\left(1-\frac{t}{\Q^2}\right)\frac{\epsilon^2}{2} \right)
\nonumber \\
&\hspace{7cm}\times
\left\{1-\sqrt{1+\epsilon^2}- \left(1+\sqrt{1+\epsilon^2}-2\xB\right)\frac{t}{\Q^2} \right\},
\nonumber \\[3mm]
\delta C^{V}_{++} (n &=2)
=
-\frac{2 y\left(1-y-\frac{\epsilon^2}{4} y^2\right)}{\left(1+\epsilon^2\right)^{5/2}}
\left(4-2\xB+3\epsilon^2 \right)\frac{t}{\Q^2}\left(1+\frac{4 (1-\xB)\xB+\epsilon^2}{4-2\xB+3\epsilon^2}\frac{t}{\Q^2} \right)
\nonumber \\
&\hspace{7cm}\times
\left\{\sqrt{1+\epsilon^2}-1+\left(1+\sqrt{1+\epsilon^2}-2\xB\right)\frac{t}{\Q^2}\right\},
\nonumber \\[3mm]
\delta C^{A}_{++} (n &=2)
= \frac{4 y\left(1-y-\frac{\epsilon^2}{4} y^2\right)}{\
\left(1+\epsilon^2\right)^{5/2}}
\frac{\xB t}{\Q^2}\left(1-(1-2\xB)\frac{t}{\Q^2}\right)
\nonumber \\
&\hspace{7cm}\times
\Bigg\{1-\sqrt{1+\epsilon^2}-\left(1+\sqrt{1+\epsilon^2}-2\xB\right)\
\frac{t}{\Q^2}\Bigg\},
\nonumber \\[3mm]
\delta S_{++} (n &=1)
=
\frac{4 K\left(2-2 y+y^2+ \frac{\epsilon^2}{2} y^2 \right)}{\left(1+\epsilon^2\right)^3}
(1+\sqrt{1+\epsilon^2})
\left\{2\sqrt{1+\epsilon^2}-1+
\frac{1+\sqrt{1+\epsilon^2}-2\xB}{1+\sqrt{1+\epsilon^2}}\frac{t}{\Q^2} \right\}
\nonumber \\
&\quad+
\frac{8 K\left(1-y-\frac{\epsilon^2}{4} y^2 \right)}{\left(1+\epsilon^2\right)^3}
\Bigg\{
\frac{3\epsilon^2}{2}+
\left(1-\sqrt{1+\epsilon^2}-\frac{\epsilon^2}{2}- \xB\left(3-\sqrt{1+\epsilon^2}\right)\right)\frac{t}{\Q^2}
\Bigg\},
\nonumber \\[3mm]
\delta S^{V}_{++} (n &=1)
=
\frac{8 K\left(2-2 y+y^2+ \frac{\epsilon^2}{2} y^2 \right)}{\left(1+\epsilon^2\right)^2}\frac{t}{\Q^2}
\Bigg\{1-\frac{(1-2\xB)\left(1+\sqrt{1+\epsilon^2}-2\xB\right)}{2(1+\epsilon^2)}\frac{t^\prime}{\Q^2}
\Bigg\}
\nonumber \\
&\quad+ \frac{32 K\left(1-y-\frac{\epsilon^2}{4} y^2 \right)}{\left(1+\epsilon^2\right)^3}
\left(1- \frac{3+\sqrt{1+\epsilon^2}}{4}\xB+ \frac{5\epsilon^2}{8}\right)\frac{t}{\Q^2}
\nonumber \\
&\hspace{3cm}\times
\Bigg\{
1-\frac{1-\sqrt{1+\epsilon^2}-
\frac{\epsilon^2}{2}-2\xB\left(3 (1-\xB)-\sqrt{1+\epsilon^2}\right)
}{4-\xB\left(\sqrt{1+\epsilon^2}+3\right)+\frac{5\epsilon^2}{2}}\frac{t}{\Q^2}
\Bigg\},
\nonumber \\[3mm]
\delta S^{A}_{++} (n &=1)
=- \frac{8 K\left(2-2 y+y^2+ \frac{\epsilon^2}{2} y^2 \right)}{\left(1+\epsilon^2\right)^3}\frac{\xB t}{\Q^2}
\Bigg\{
\sqrt{1+\epsilon^2}-1+(1+\sqrt{1+\epsilon^2}-2\xB)\frac{t}{\Q^2}\Bigg\}
\nonumber \\
&\qquad+ \frac{8 K\left(1-y-\frac{\epsilon^2}{4} y^2 \right)}{\left(1+\epsilon^2\right)^3}
(3+\sqrt{1+\epsilon^2})\frac{\xB t}{\Q^2}
\Bigg\{1- \
\frac{3-\sqrt{1+\epsilon^2}-6\xB}{3+\sqrt{1+\epsilon^2}}\frac{t}{\Q^2}\Bigg\},
\nonumber \\[3mm]
\delta S_{++} (n &=2)
=
-\frac{4 (2-y)\left(1-y-\frac{\epsilon^2}{4} y^2 \right)}{\left(1+\epsilon^2\right)^{5/2}}
\nonumber \\
&\times
\Bigg\{
\frac{4 {\widetilde K}^2}{\sqrt{1+\epsilon^2}\Q^2} + \
(1+\sqrt{1+\epsilon^2}-2\xB)\left(1+\sqrt{1+\epsilon^2}+\frac{\xB t}{\Q^2}\right)\frac{t^\prime}{\Q^2}
\Bigg\},\nonumber \\[3mm]
\delta S^{V}_{++} (n &=2)
=
\frac{4 (2-y)\left(1-y-\frac{\epsilon^2}{4} y^2 \right)}{\left(1+\epsilon^2\right)^{5/2}}\frac{t}{\Q^2}
\Bigg\{
\frac{4 (1-2\xB) {\widetilde K}^2}{\sqrt{1+\epsilon^2}\Q^2}-
\left(3-\sqrt{1+\epsilon^2}-2\xB+ \frac{\epsilon^2}{\xB} \right)\frac{\xB t^\prime}{\Q^2}\Bigg\},
\nonumber \\[3mm]
\delta S^{A}_{++} (n &=2)
= \frac{4 (2-y)\left(1-y-\frac{\epsilon^2}{4} y^2 \right)}{\left(1+\epsilon^2\right)^{3}}\frac{\xB t}{\Q^2}
\Bigg\{
\frac{4 {\widetilde K}^2}{\Q^2}-
\left(1+\sqrt{1+\epsilon^2}-2\xB \right)\left(1- \frac{(1-2\xB) t}{\Q^2}\right)\frac{t^\prime}{\Q^2}\Bigg\},
\nonumber \\[3mm]
\delta S_{++} (n &=3)
=-\frac{4 K\left(1-y-\frac{\epsilon^2}{4} y^2 \right)}{\left(1+\epsilon^2\right)^{3}}
\frac{1+\sqrt{1+\epsilon^2}-2\xB}{1+\sqrt{1+\epsilon^2}}
\frac{\epsilon^2 t^\prime}{\Q^2},\nonumber \\[3mm]
\delta S^{V}_{++} (n &=3)
= \frac{4 K\left(1-y-\frac{\epsilon^2}{4} y^2 \right)}{\left(1+\epsilon^2\right)^{3}}
 \left(4 (1-\xB)\xB+ \epsilon^2 \right)\frac{t\,t^\prime}{\Q^4},
\nonumber \\[3mm]
\delta S^{A}_{++} (n &=3)
=- \frac{8 K\left(1-y-\frac{\epsilon^2}{4} y^2 \right)}{\left(1+\epsilon^2\right)^{3}}
\left(1+\sqrt{1+\epsilon^2}-2\xB \right)\frac{\xB t\,t^\prime}{\Q^4}
\,  . \nonumber
\end{align}

\begin{itemize}
\item Photon helicity-flip amplitudes by one unit:
\end{itemize}
\begin{align}
\delta C_{0+} (n &=0)
=  \frac{8\sqrt{2} K (1- \xB) y\sqrt{1-y-\frac{\epsilon^2}{4} y^2}}{(1+\epsilon^2)^2}
\frac{t}{\Q^2}
\, ,\\[3mm]
\delta C^{V}_{0+} (n &=0)
=  \frac{8\sqrt{2} K y\sqrt{1-y- \frac{\epsilon^2}{4} y^2}}{(1+\epsilon^2)^2}
\frac{t}{\Q^2}
\left(
\xB- \frac{t}{\Q^2} (1-2\xB) \right)
\, ,\nonumber \\[3mm]
\delta C^{A}_{0+} (n &=0)
=- \frac{8\sqrt{2} K y\sqrt{1-y- \frac{\epsilon^2}{4} \
y^2}}{(1+\epsilon^2)^2}
\frac{\xB t}{\Q^2}\left(1+ \frac{t}{\Q^2} \right)
\, ,\nonumber \\[3mm]
\delta C_{0+} (n &=1)
=- \frac{8\sqrt{2}  y (2-y)\sqrt{1-y-\frac{\epsilon^2}{4}y^2}}{(1+\epsilon^2)^2}
\frac{\widetilde{K}^2}{\Q^2}
\, ,\nonumber \\[3mm]
\delta C^{V}_{0+} (n &=1)
=  \frac{8\sqrt{2} y (2-y)\sqrt{1-y-\frac{\epsilon^2}{4} y^2 }}{(1+\epsilon^2)^2}
\frac{t\widetilde{K}^2}{\Q^4}
\, ,\nonumber \\[3mm]
\delta C^{A}_{0+} (n &=1)
=  0
\, ,\nonumber \\[3mm]
\delta C_{0+} (n &=2)
=- \frac{8\sqrt{2} K y\sqrt{1-y-\frac{\epsilon^2}{4}y^2}}{(1+\epsilon^2)^2}
\left(1+ \frac{\xB t}{\Q^2} \right)
\, ,\nonumber \\[3mm]
\delta C^{V}_{0+} (n &=2)
=  \frac{8\sqrt{2} K y (1- \xB)\sqrt{1-y-\frac{\epsilon^2}{4} y^2 }}{(1+\epsilon^2)^2}
\frac{t}{\Q^2}
\, ,\nonumber \\[3mm]
\delta C^{A}_{0+} (n &=2)
=  \frac{8\sqrt{2} K y\sqrt{1-y- \frac{\epsilon^2}{4} \
y^2}}{(1+\epsilon^2)^2}
\frac{\xB t}{\Q^2}\left(1+ \frac{t}{\Q^2} \right)
\, ,\nonumber \\[3mm]
\delta S_{0+} (n &=1)
=
\frac{8\sqrt{2}\sqrt{1-y-\frac{\epsilon^2}{4}y^2}}{(1+\epsilon^2)^{5/2}}
\Bigg\{
\frac{\widetilde{K}^2}{\Q^2} (2-y)^2 \nonumber
\\&+
\left(1+ \frac{t}{\Q^2} \right)\left(1-y- \frac{\epsilon^2}{4} y^2\right)\left(2\frac{\xB t}{\Q^2}- \left(1- \frac{t}{\Q^2} \right)
\epsilon^2 \right)\Bigg\}
\, ,\nonumber \\[3mm]
\delta S^{V}_{0+} (n &=1)
=- \frac{8\sqrt{2}\sqrt{1-y- \frac{\epsilon^2}{4} y^2}}{(1+\epsilon^2)^{5/2}}
\frac{t}{\Q^2}
\Bigg\{
\frac{\widetilde{K}^2}{\Q^2} (2-y)^2 \nonumber
\\&+
\left(1+ \frac{t}{\Q^2} \right)\left(1-y- \frac{\epsilon^2}{4} y^2\right)\left(4-2\xB+3\epsilon^2+ \frac{t}{\Q^2} (4\xB (1- \xB)+\epsilon^2) \right)\Bigg\}
\, ,\nonumber \\[3mm]
\delta S^{A}_{0+} (n &=1)
=- \frac{16\sqrt{2}\left(1-y- \frac{\epsilon^2}{4} y^2\right)^{3/2}}{(1+\epsilon^2)^{5/2}}
\frac{\xB t}{\Q^2}\left(1+ \frac{t}{\Q^2} \right)
\left(1-(1-2\xB)\frac{t}{\Q^2} \right)
\, ,\nonumber \\[3mm]
\delta S_{0+} (n &=2)
=  \frac{8\sqrt{2} K (2-y)\sqrt{1-y- \frac{\epsilon^2}{4} y^2}}{(1+\epsilon^2)^{5/2}}
\left(1+ \frac{\xB t}{\Q^2} \right)
\, ,\nonumber \\[3mm]
\delta S^{V}_{0+} (n &=2)
=- \frac{8\sqrt{2} K (2-y) (1- \xB)\sqrt{1-y- \
\frac{\epsilon^2}{4} y^2}}{(1+\epsilon^2)^{5/2}}
\frac{t}{\Q^2}
\, ,\nonumber \\[3mm]
\delta S^{A}_{0+} (n &=2)
=- \frac{8\sqrt{2} K (2-y)\sqrt{1-y- \frac{\epsilon^2}{4} y^2}}{(1+\epsilon^2)^{5/2}}
\frac{\xB t}{\Q^2}\left(1+ \frac{t}{\Q^2} \right)
\, .\nonumber
\end{align}

\begin{itemize}
\item Photon helicity-flip amplitudes by two units:
\end{itemize}
\begin{align}
\delta C_{-+} (n &=0)
= \frac{4 y}{(1+\epsilon^2)^{5/2}}
\Bigg\{
\frac{\widetilde{K}^2}{\Q^2} (2-y)^2\left(1- \sqrt{1+ \epsilon^2} \right)
+
\frac{1}{2}\left(1-y- \frac{\epsilon^2}{4} y^2\right)
 \\
&\times
\left(2\frac{\xB t}{\Q^2}-
\left(1- \frac{t}{\Q^2} \right)\epsilon^2 \right)\left(1- \sqrt{1+ \
\epsilon^2}-
\frac{t}{\Q^2}\left(1-2\xB+ \sqrt{1+ \epsilon^2} \right) \right)\Bigg\}
\, , \nonumber \\[3mm]
\delta C^{V}_{-+} (n &=0)
=
\frac{2 y}{(1+\epsilon^2)^{5/2}}\frac{t}{\Q^2}
\Bigg\{(4-2\xB+3\epsilon^2)
\left(1-y- \frac{\epsilon^2}{4} y^2\right)\left(1+ \frac{t}{\Q^2}\frac{4\xB (1- \xB)+\epsilon^2}{4-2\xB+3\epsilon^2}
\right)
\nonumber \\
&\times
\left(
\sqrt{1+ \epsilon^2}-1+ \frac{t}{\Q^2}\left(1-2\xB+ \sqrt{1+ \
\epsilon^2} \right) \right)+2 (2-y)^2 (\sqrt{1+ \epsilon^2}-1+2\xB)\
\frac{\widetilde{K}^2}{\Q^2}\Bigg\}
\, ,\nonumber \\[3mm]
\delta C^{A}_{-+} (n &=0)
=  \frac{4\xB y}{(1+\epsilon^2)^{5/2}}
\frac{t}{\Q^2}
\Bigg\{2 (2-y)^2\left((1- \xB)\frac{t}{\Q^2}\left(1+ \frac{\xB t}{\Q^2} \right)+
\left(1+ \frac{t}{\Q^2} \right)^2\frac{\epsilon^2}{4} \right)
\nonumber \\
&-
\left(1-y- \frac{\epsilon^2}{4} y^2\right)\left(1-(1-2\xB)\frac{t}{\Q^2} \right)\left(1- \sqrt{1+\epsilon^2}-
\frac{t}{\Q^2} (1+ \sqrt{1+ \epsilon^2}-2\xB) \right)\Bigg\}
\, ,\nonumber \\[3mm]
\delta C_{-+} (n &=1)
= \frac{4 K y (2-y)}{(1+\epsilon^2)^{5/2}}
\Bigg\{1- \epsilon^2- \sqrt{1+ \epsilon^2}-
\frac{t}{\Q^2}\left(1- \epsilon^2- \sqrt{1+ \epsilon^2}-2\xB\left(2-\sqrt{1+ \epsilon^2} \right) \right)\Bigg\}
\, ,\nonumber \\[3mm]
\delta C^{V}_{-+} (n &=1)
=
- \frac{4 K y (2-y)}{(1+\epsilon^2)^{5/2}}\frac{t}{\Q^2}
\Bigg\{5-4\xB+3\epsilon^2-
\sqrt{1+ \epsilon^2} \nonumber \\
&-
\frac{t}{\Q^2}\left(1- \epsilon^2- \sqrt{1+ \epsilon^2}-2\xB
(4-4\xB- \sqrt{1+ \epsilon^2}) \right)\Bigg\}
\, ,\nonumber \\[3mm]
\delta C^{A}_{-+} (n &=1)
=- \frac{16 K \xB y (2-y)}{(1+\epsilon^2)^{5/2}}
\frac{t}{\Q^2}
\left(1-(1-2\xB)\frac{t}{\Q^2} \right)
\, ,\nonumber \\[3mm]
\delta C_{-+} (n &=2)
=- \frac{2 y\left(1-y- \frac{\epsilon^2}{4} y^2\right)}{(1+\epsilon^2)^{5/2}}
\Bigg\{
\epsilon^2\left(1+ \sqrt{1+ \epsilon^2} \right)
\nonumber \\
&-
2\frac{t}{\Q^2}\left((1- \xB)\epsilon^2+ \xB\left(1+\sqrt{1+ \epsilon^2} \right) \right)+
\frac{t^2}{\Q^4} (2\xB+ \epsilon^2)\left(1-2\xB- \sqrt{1+\epsilon^2} \right)\Bigg\}
\, ,\nonumber \\[3mm]
\delta C^{V}_{-+} (n &=2)
=
-\frac{2 y\left(1-y- \frac{\epsilon^2}{4} y^2\right)}{(1+\epsilon^2)^{5/2}}\frac{t}{\Q^2}
\left(4-2\xB+3\epsilon^2+ \frac{t}{\Q^2} (4\xB (1- \xB)+ \epsilon^2)
\right)
\nonumber \\
&\times
\left(1+ \sqrt{1+ \epsilon^2}- \frac{t}{\Q^2} (1- \sqrt{1+\epsilon^2}-2\xB) \right)
\, ,\nonumber \\[3mm]
\delta C^{A}_{-+} (n &=2)
=
- \frac{4\xB y\left(1-y- \frac{\epsilon^2}{4} y^2 \right)}{(1+\epsilon^2)^{5/2}}
\nonumber \\
&\times
\frac{t}{\Q^2}
\left(1-(1-2\xB)\frac{t}{\Q^2} \right)
\left\{1+ \sqrt{1+ \epsilon^2}- \frac{t}{\Q^2}\left(1- \sqrt{1+\epsilon^2}-2\xB \right) \right\}
\, ,\nonumber \\[3mm]
\delta S_{-+} (n &=1)
=- \frac{4 K}{(1+\epsilon^2)^{3}}
\Bigg\{(2-y)^2\left(1+2\epsilon^2-
\sqrt{1+ \epsilon^2}+
\frac{t}{\Q^2}\left(1-2\xB- \sqrt{1+ \epsilon^2} \right) \right)
 \nonumber \\
 &-
\left(1-y- \frac{\epsilon^2}{4} y^2\right)\left(2+\epsilon^2-2\sqrt{1+ \epsilon^2}+
\frac{t}{\Q^2}\left(
\epsilon^2-4\sqrt{1+ \epsilon^2}+2\xB (1+ \sqrt{1+ \epsilon^2}) \right) \right)\Bigg\}
\, ,\nonumber \\[3mm]
\delta S^{V}_{-+} (n &=1)
=- \frac{4 K}{(1+\epsilon^2)^3}
\frac{t}{\Q^2}
\Bigg\{
\left(2-2 y+y^2+ \frac{\epsilon^2}{2} y^2\right)
\nonumber \\
&
\times
\left(3+2\epsilon^2+ \sqrt{1+ \epsilon^2}-2\xB (1+ \sqrt{1+ \
\epsilon^2})-
\frac{t}{\Q^2} (1-2\xB) (1-2\xB- \sqrt{1+ \epsilon^2}) \right)
\nonumber \\
&+
\left(1-y- \frac{\epsilon^2}{4} y^2\right)\bigg(8+5\epsilon^2-2\xB
(3- \sqrt{1+ \epsilon^2}) \nonumber \\
&\qquad\qquad\qquad-
\frac{t}{\Q^2}\left(2- \epsilon^2+2\sqrt{1+ \epsilon^2}-12\xB (1-\xB)-4\xB\sqrt{1+ \epsilon^2} \right)\bigg)\Bigg\}
\, ,\nonumber \\[3mm]
\delta S^{A}_{-+} (n &=1)
=
- \frac{8 K\left(2-2 y+y^2+ \frac{\epsilon^2}{2} y^2\right)}{(1+\epsilon^2)^3}
(1+ \sqrt{1+ \epsilon^2})
\frac{\xB t}{\Q^2}
\left(1-  \frac{t}{\Q^2}\frac{1- \sqrt{1+ \epsilon^2}-2\xB}{1+\sqrt{1+ \epsilon^2}} \right)
\nonumber \\
&-
\frac{8 K\left(1-y-\frac{\epsilon^2}{4} y^2\right)}{(1+\epsilon^2)^3}
\frac{\xB t}{\Q^2}
\left\{3- \sqrt{1+ \epsilon^2}-
\frac{t}{\Q^2}\left(3+ \sqrt{1+ \epsilon^2}-6\xB \right) \right\}
\, ,\nonumber \\[3mm]
\delta S_{-+} (n &=2)
=- \frac{4 (2-y)\left(1-y- \frac{\epsilon^2}{4} y^2\right)}{(1+\epsilon^2)^{3}}
\Bigg\{
\frac{t^2}{\Q^4}\left(
\epsilon^2-2\xB^2 (2+ \sqrt{1+ \epsilon^2})+ \xB (3- \epsilon^2+\sqrt{1+ \epsilon^2}) \right)
\nonumber \\&+
\epsilon^2\left(1+ \sqrt{1+ \epsilon^2} \right)
+
\frac{t}{\Q^2}\left(2+2\sqrt{1+ \epsilon^2}+ \epsilon^2\sqrt{1+ \epsilon^2}-
\xB\left(3- \epsilon^2+3\sqrt{1+ \epsilon^2} \right) \right)
\Bigg\}
\, ,\nonumber \\[3mm]
\delta S^{V}_{-+} (n &=2)
=
- \frac{4 (2-y) \left(1-y- \frac{\epsilon^2}{4} y^2\right)}{(1+\epsilon^2)^{5/2}}
\frac{t}{\Q^2}
\Bigg\{(2- \xB) (1+ \sqrt{1+ \epsilon^2}) \nonumber \\
&+
\epsilon^2+
\frac{4\widetilde{K}^2 (1-2\xB)}{\Q^2\sqrt{1+ \epsilon^2}}+
\frac{t}{\Q^2}\left(
\epsilon^2+ \xB (3-2\xB+ \sqrt{1+ \epsilon^2}) \right)\Bigg\}
\, ,\nonumber \\[3mm]
\delta S^{A}_{-+} (n &=2)
=
- \frac{4 (2-y)\left(1-y- \frac{\epsilon^2}{4} y^2 \right)}{(1+\epsilon^2)^3}
\frac{\xB t}{\Q^2}
\Bigg\{1+4\frac{\widetilde{K}^2}{\Q^2}
\nonumber\\
&+
\sqrt{1+ \epsilon^2}-2\frac{t}{\Q^2}\left(1-2\xB- \xB\sqrt{1+\epsilon^2} \right)+
\frac{t^2}{\Q^4} (1-2\xB)\left(1-2\xB- \sqrt{1+ \epsilon^2} \right)
\Bigg\}
\, ,\nonumber \\[3mm]
\delta S_{-+} (n &=3)
= \frac{4 K\left(1-y- \frac{\epsilon^2}{4} y^2\right)}{(1+\epsilon^2)^{3}}
\Bigg\{2+ \epsilon^2+2\sqrt{1+ \epsilon^2}+
\frac{t}{\Q^2}\left( \epsilon^2+2\xB (1+ \sqrt{1+ \epsilon^2}) \right)\Bigg\}
\, ,\nonumber \\[3mm]
\delta S^{V}_{-+} (n &=3)
=
- \frac{4 K\left(1-y- \frac{\epsilon^2}{4} y^2 \right)}{(1+\epsilon^2)^{5/2}}
\frac{t}{\Q^2}
\Bigg\{4-4\xB+ \frac{t^\prime}{\Q^2}\frac{4\xB (1- \xB)+\epsilon^2}{\sqrt{1+ \epsilon^2}}\Bigg\}
\, ,\nonumber \\[3mm]
\delta S^{A}_{-+} (n &=3)
=
- \frac{8 K\left(1-y- \frac{\epsilon^2}{4} y^2\right)}{(1+\epsilon^2)^3}
\frac{\xB t}{\Q^2}
\left\{1+ \sqrt{1+ \epsilon^2}- \frac{t}{\Q^2}\left(1-2\xB- \sqrt{1+ \epsilon^2} \right) \right\}
\, . \nonumber
\end{align}

\section{Helicity amplitudes from Tarrach tensor}
\label{TarrachHelicityAmplitudes}
\setcounter{equation}{0}

Let us establish a relation of the Compton tensor parametrization introduced in Eq.\ (\ref{Ten-BM}) in terms of CFFs and the one by
Tarrach \cite{Tarrach_1975tu} (also quoted in \cite{Drechsel:1997xv}) that is used as a starting point for the low-energy expansion relevant for
generalized polarizabilities. The Tarrach's tensor is written as a linear superposition of independent tensor structures $\rho_{\mu\nu}$
accompanied by $f$ functions encoding the structural information about the nucleon,
\begin{align}
\label{TarrachTensor}
{\varepsilon}_{\mu} (a) T_{\mu\nu} \varepsilon^{'*}_{\nu} (b)
=
\sum_{k = 1}^{12} f_k \, \bar{u}_2 R^{(k)}_{ab} u_1
\qquad\mbox{with}\qquad
R^{(k)}_{ab} = {\varepsilon}_{\mu}(a) \rho^{\mu\nu}_k \varepsilon^{'*}_{\nu}  (b)
\, .
\end{align}
Now we compute the helicity amplitudes for all polarization states of the photons and express the result in terms of the Dirac structures
used in the parametrization of helicity CFFs in Eqs.\ (\ref{cal-TabV}) and (\ref{cal-TabA}) multiplied by the functions of the kinematical invariants. Comparing
Eqs.\ (\ref{DVCS2helicity}) and (\ref{cal-Tab})--(\ref{cal-TabA}) with what we will find below, one can establish relation formulas of CFFs and $f$'s.
In the following we present an overcomplete set of $3\times 12$ relations from which an interesting reader can express  helicity dependent CFFs
in terms of $f$'s or reverse.

\begin{itemize}
\item (1,1) helicity amplitude:
\end{itemize}
\begin{align}
R^{(1)}_{++}
&=\frac{(\not\!q \Q \epsilon - i\sigma_{q\Delta}\xB) \left((1-2 \xB)t-\Q^2-\left(t+\Q^2\right) \sqrt{1+\epsilon ^2}\right)}{2 \left(t \xB+(2-\xB) \Q^2\right)\sqrt{1+\epsilon ^2}}
\\[3.75mm]
R^{(2)}_{++}
&=(\not\!q\ \Q \epsilon -i\sigma_{q\Delta} \xB) \Bigg\{ \frac{t
\xB+(2-\xB) \Q^2}{4 \xB^2}-\frac{t^2 (1-2 \xB)+2 t (2-\xB) \Q^2-\Q^4}{4 \left(t \xB+(2-\xB)
\Q^2\right) \sqrt{1+\epsilon ^2}}
\nonumber
\\
&+\frac{\Q^2 \left(2 t \xB+t (1-\xB) \epsilon ^2+(1-\xB) \Q^2 \left(2+\epsilon
^2\right)\right)}{2 \xB^2 \left(t \xB+(2-\xB) \Q^2\right) \sqrt{1+\epsilon ^2}}\Bigg\}
\\[3.75mm]
R^{(3)}_{++}
&=\frac{(\not\!q\Q\epsilon-i\sigma_{q\Delta}\xB)\Q^2}{2\xB}\bigg\{\frac{\Q^2 \left(2-\xB+\epsilon^2\right)+t\left((3-2\xB)\xB+\epsilon ^2\right)}{\left(t\xB+(2-\xB)\Q^2\right)\sqrt{1+\epsilon ^2}}+1\bigg\}
\\[3.75mm]
R^{(4)}_{++}
&=-\not\!q\ \frac{t (1-\xB)}{\sqrt{1+\epsilon ^2}}+i\sigma_{q\Delta}\frac{ (1-\xB) \Q \epsilon
}{\xB\sqrt{1+\epsilon ^2}}-\not\!q\
\gamma^5\frac{\left(t \xB+(2-\xB) \Q^2\right) \left(1+\sqrt{1+\epsilon
^2}\right)}{2 \xB \sqrt{1+\epsilon ^2}}
\nonumber
\\
&-\gamma^5 \frac{M \left(t \xB+(2-\xB) \Q^2\right)\left(1-2\xB-\sqrt{1+\epsilon^2}\right)}{2\xB\sqrt{1+\epsilon ^2}}
\\[3.75mm]
R^{(5)}_{++}
&=\not\!q\ \frac{t\xB}{4 \sqrt{1+\epsilon ^2}}-i\sigma_{q\Delta}\frac{
\Q \epsilon }{4 \sqrt{1+\epsilon ^2}}+\not\!q\
\gamma^5 \frac{t \xB+\Q^2 \left(1+\sqrt{1+\epsilon
^2}\right)}{4 \sqrt{1+\epsilon ^2}}
\nonumber
\\
&+\gamma^5\frac{M\Q^2
\left(1-\xB-\sqrt{1+\epsilon ^2}\right)}{4 \sqrt{1+\epsilon ^2}}
\, , \\[3.75mm]
R^{(6)}_{++}
&=-\not\!q\
 \frac{\Q \left(t^2 (2-3 \xB)+(2-\xB) \Q^4\right)
\epsilon }{2 \left(t \xB+(2-\xB) \Q^2\right) \sqrt{1+\epsilon ^2}}
\nonumber
\\
&-i\sigma_{q\Delta}\Q^2 \frac{(2-\xB)\Q^2 \left(2(1-\xB)+\epsilon ^2\right)+t \xB \left(4-2 (3-\xB) \xB+\epsilon ^2\right)}{\xB \left(t\xB+(2-\xB) \Q^2\right) \sqrt{1+\epsilon ^2}}
\nonumber
\\
-&\not\!q\ \gamma^5\bigg\{M\left(t-\Q^2\right)+\frac{M(1-\xB) \left(t+\Q^2\right)}{\sqrt{1+\epsilon ^2}}\bigg\}
\nonumber
\\
&-\gamma^5 \Bigg\{\frac{\left(t \xB+(2-\xB) \Q^2\right)^2+\Q^2 \left(t+3 \Q^2\right) \epsilon
^2}{4 \xB^2}-\frac{M^2 (1-\xB) \left(t+\Q^2\right)}{\sqrt{1+\epsilon ^2}}
\nonumber
\\
&\ \ \ \ \ \ -\frac{\left(t \xB+(2-\xB) \Q^2\right)
\left(t \xB (1-2 \xB)+(2-3 \xB) \Q^2\right)}{4 \xB^2 \sqrt{1+\epsilon ^2}}\Bigg\}
\\[3.75mm]
R^{(7)}_{++}
&=\not\!q\frac{t \xB}{4 \sqrt{1+\epsilon ^2}}- i\sigma_{q\Delta}\frac{
\Q \epsilon }{4 \sqrt{1+\epsilon ^2}}+\not\!q\gamma^5\frac{t}{4}\bigg\{ \frac{1-\xB}{\sqrt{1+\epsilon
^2}}-1\bigg\}+\gamma^5 \frac{M\Q^2}{4}\bigg\{\frac{1-\xB}{\sqrt{1+\epsilon ^2}}-1\bigg\}
\\[3.75mm]
R^{(8)}_{++}
&=\not\!q\frac{\xB \Q\left(t+\Q^2\right)^2\epsilon}{4\left(t\xB+(2-\xB)\Q^2\right)\sqrt{1+\epsilon ^2}}+i\sigma_{q\Delta} \frac{ (1-\xB) \Q^2 \left(t \xB+\Q^2\right)}{\left(t \xB+(2-\xB)\Q^2\right)\sqrt{1+\epsilon^2}}+\not\!q\gamma^5\frac{\Q \left(t+\Q^2\right)\epsilon}{4\sqrt{1+\epsilon^2}}
\nonumber
\\
&+\gamma^5\frac{\Q^2}{8\xB}\bigg\{t \xB+(2-\xB) \Q^2-\frac{t (1-2 \xB) \xB+(2-3 \xB) \Q^2}{ \sqrt{1+\epsilon ^2}}\bigg\}
\\[3.75mm]
R^{(9)}_{++}
&=\not\!q\Q\epsilon\frac{t^2 \xB+4 t (1-\xB) \Q^2-\xB\Q^4 }{4 \left(t \xB+(2-\xB) \Q^2\right) \sqrt{1+\epsilon ^2}}-i\sigma_{q\Delta} \Q^2\bigg\{\frac{(1-\xB) \left(t \xB+\Q^2\right)}{\left(t \xB+(2-\xB) \Q^2\right) \sqrt{1+\epsilon ^2}}+\frac{\epsilon^2}{2 \xB \sqrt{1+\epsilon ^2}}\bigg\}
\nonumber
\\
&- \not\!q\gamma^5 \bigg\{\frac{M\left(t-\Q^2\right)}{2}-\frac{M(1-\xB)\left(t+\Q^2\right) }{2\sqrt{1+\epsilon^2}}\bigg\}
\nonumber
\\
&-\gamma^5\Q^2\bigg\{\frac{(1-\xB)\left(4(1-\xB)\left(t\xB+\Q^2\right)+\left(t+\Q^2\right) \epsilon ^2\right)}{8 \xB^2 \sqrt{1+\epsilon ^2}}
 +\frac{(4 (1-\xB) \left(t \xB+\Q^2\right)+\left(t+3 \Q^2\right) \epsilon ^2}{8 \xB^2}\bigg\}
\\[3.75mm]
R^{(10)}_{++}
&=\not\!q\frac{\xB\Q\left(t+\Q^2\right)\epsilon}{\left(t\xB+(2-\xB)\Q^2\right)\sqrt{1+\epsilon^2}}+ i\sigma_{q\Delta}\bigg\{\frac{t (1-2 \xB)\xB+(2-3\xB)\Q^2}{\left(t\xB+(2-\xB)\Q^2\right) \sqrt{1+\epsilon ^2}}-1\bigg\}+\not\!q\gamma^5\frac{\Q\epsilon}{\sqrt{1+\epsilon^2}}
\nonumber
\\
&+\frac{\gamma^5}{2}\bigg\{t+\frac{(2-\xB)\Q^2}{\xB}-\frac{t (1-2 \xB) \xB+\Q^2 \left(2-3 \xB+\epsilon ^2\right)}{\xB\sqrt{1+\epsilon ^2}}\bigg\}
\\[3.75mm]
R^{(11)}_{++}
&=\not\!q\ \frac{t \xB}{\sqrt{1+\epsilon ^2}}- i\sigma_{q\Delta}\frac{
\Q \epsilon }{\sqrt{1+\epsilon ^2}}+\not\!q\
\gamma^5\frac{t \xB+\Q^2 \left(1+\sqrt{1+\epsilon ^2}\right)}{\sqrt{1+\epsilon ^2}}
\nonumber
\\
&+\gamma^5 \Q \epsilon \frac{t\left(1-2 \xB-\sqrt{1+\epsilon ^2}\right)-\Q^2 \left(\xB+2 \sqrt{1+\epsilon ^2}\right)}{2 \xB \sqrt{1+\epsilon^2}}
\\[3.75mm]
R^{(12)}_{++}
&=\not\!q\ \frac{
 \xB \Q \left(t+\Q^2\right)^2 \epsilon }{4 \left(t \xB+(2-\xB) \Q^2\right)
\sqrt{1+\epsilon ^2}}+ i\sigma_{q\Delta}\frac{
 (1-\xB) \Q^2 \left(t \xB+\Q^2\right)}{\left(t
\xB+(2-\xB) \Q^2\right) \sqrt{1+\epsilon ^2}}+\not\!q\
\gamma^5\frac{\Q \left(t+\Q^2\right) \epsilon }{4 \sqrt{1+\epsilon
^2}}
\nonumber
\\
&-\gamma^5 \frac{\Q^2}{8}  \bigg\{t+\Q^2+\frac{\Q^2 \left(4-3\xB+\epsilon ^2\right)+t \left((3-2 \xB) \xB+\epsilon ^2\right)}{\sqrt{1+\epsilon ^2}}\bigg\}
\, .
\end{align}

\begin{itemize}
\item (0,1) helicity amplitude:
\end{itemize}
\begin{align}
\label{R^{(1)}_{0+}}
R^{(1)}_{0+}
&=\frac{\sqrt{2}(\not\!q\Q \epsilon-i\sigma_{q\Delta} \xB) \widetilde{K} \Q }{\left(t \xB+(2-\xB) \Q^2\right) \sqrt{1+\epsilon ^2}}
\\[3.75mm]
R^{(2)}_{0+}
&=-\frac{(\not\!q\Q \epsilon-i\sigma_{q\Delta} \xB)\widetilde{K} \Q  \left(t \xB^2-\Q^2 \left((2-\xB) \xB+2 \epsilon ^2\right)\right)}{\sqrt{2} \xB^2 \left(t \xB+(2-\xB)
\Q^2\right) \sqrt{1+\epsilon ^2}}
\\[3.75mm]
R^{(3)}_{0+}
&=-\frac{\sqrt{2}(\not\!q\Q \epsilon-i\sigma_{q\Delta} \xB) \widetilde{K} (1-\xB) \Q^3 }{\xB \left(t \xB+(2-\xB) \Q^2\right) \sqrt{1+\epsilon ^2}}
\\[3.75mm]
R^{(4)}_{0+}
&=(\not\!q\ t \xB-i\sigma_{q\Delta} \Q \epsilon )\frac{\Q^2\left(2-\xB+\epsilon ^2\right)+t \left((3-2 \xB) \xB+\epsilon ^2\right)}{\sqrt{2} \widetilde{K} \xB \Q \sqrt{1+\epsilon
^2}}
\nonumber
\\
&
+\not\!q\ \gamma^5\frac{\sqrt{2} \widetilde{K} \left(t \xB+(2-\xB) \Q^2\right)}{\xB \Q \sqrt{1+\epsilon ^2}}
\nonumber
\\
&+(\not\!q\ \gamma^5 t+\gamma^5M\Q^2)\frac{\epsilon  \left(t \xB+(2-\xB) \Q^2\right)
\left(t \left(4 (1-\xB) \xB+\epsilon ^2\right)+\Q^2 \left(4-2 \xB+3 \epsilon ^2\right)\right)}{4 \sqrt{2}M\widetilde{K} \xB^2\Q^2 \sqrt{1+\epsilon ^2}}
\\[3.75mm]
R^{(5)}_{0+}
&=
(\not\!q\ t \xB-i\sigma_{q\Delta} \Q \epsilon )\frac{t (1-2 \xB)-\Q^2 }{4 \sqrt{2} \widetilde{K} \Q \sqrt{1+\epsilon ^2}}-\not\!q\ \gamma^5\frac{\widetilde{K} \Q}{\sqrt{2} \sqrt{1+\epsilon^2}}
\nonumber
\\
&-(\not\!q\ \gamma^5 t+\gamma^5M\Q^2)\frac{\Q^2 \left(2-\xB+\epsilon ^2\right)+t \left((3-2 \xB) \xB+\epsilon ^2\right)}{4 \sqrt{2}\widetilde{K}\Q\sqrt{1+\epsilon ^2}}
\\[3.75mm]
R^{(6)}_{0+}
&=-\not\!q\ \frac{\sqrt{2} \widetilde{K} \Q^2 \left((2-\xB) \Q^2+t
\left(\xB (5-2 \xB)+2 \epsilon ^2\right)\right)}{\left(t \xB+(2-\xB) \Q^2\right) \epsilon \sqrt{1+\epsilon ^2}}
\nonumber
\\
&-(\not\!q\ t \xB-i\sigma_{q\Delta} \Q \epsilon )
\bigg\{\frac{\sqrt{2} t (1-\xB) \left(2 (2-\xB) \left(t \xB+\Q^2\right)+\left(t+\Q^2\right) \epsilon ^2\right)}{\widetilde{K}\epsilon \left(t \xB+(2-\xB)
\Q^2\right) \sqrt{1+\epsilon ^2}}
\nonumber\\
&\hspace{8cm}
+
\frac{2 (1-\xB) \left(t \xB+\Q^2\right)+\left(t+\Q^2\right) \epsilon ^2}{\sqrt{2} \widetilde{K} \xB \epsilon\sqrt{1+\epsilon ^2}}\bigg\}
\nonumber\\
&-\not\!q\ \gamma^5
\frac{\sqrt{2}
\widetilde{K} \left(t^2 \xB^2+\Q^4 \left(2-\xB+\epsilon ^2\right)+t \Q^2 \left((3-\xB) \xB+\epsilon ^2\right)\right)}{\xB
\Q^2 \epsilon \sqrt{1+\epsilon ^2}}
\nonumber
\\
&+(\not\!q\ \gamma^5 t+\gamma^5M\Q^2)\bigg\{
\frac{\widetilde{K} \Q \epsilon ^2}{\sqrt{2}M \xB^2 \sqrt{1+\epsilon^2}}
-
\left( t \xB+(2-\xB) \Q\right)\bigg[\frac{(1-\xB) \left(t \xB+\Q^2\right)^2}{\sqrt{2}M \widetilde{K}
\xB^2 \Q^3 \sqrt{1+\epsilon ^2}}
\nonumber\\
&\hspace{7cm}
+\frac{ \left(t^2 \xB+t (1+2 \xB) \Q^2+(3-\xB) \Q^4\right) \epsilon^2}{4 \sqrt{2}M \widetilde{K} \xB^2 \Q^3 \sqrt{1+\epsilon ^2}}\bigg]
\bigg\}
\\[3.75mm]
R^{(7)}_{0+}
&= (\not\!q\ t \xB-i\sigma_{q\Delta} \Q \epsilon )\frac{t (1-2 \xB)-\Q^2}{4 \sqrt{2} \widetilde{K} \Q \sqrt{1+\epsilon ^2}}
\nonumber
\\
&-\left(\not\!q\ \gamma^5 t+\gamma^5 M \Q^2\right) \frac{\Q^2 \left(2-\xB+\epsilon ^2\right)+t \left((3-2 \xB) \xB+\epsilon ^2\right)}{4 \sqrt{2}\widetilde{K} \Q \sqrt{1+\epsilon ^2}}
\\[3.75mm]
R^{(8)}_{0+}
&=-\not\!q\ \frac{\widetilde{K} \xB \Q^2 \left(t (1-2 \xB)-\Q^2\right)}{\sqrt{2}
\left(t \xB+(2-\xB) \Q^2\right) \epsilon \sqrt{1+\epsilon ^2}}
\nonumber
\\
&-(\not\!q\ t \xB-i\sigma_{q\Delta} \Q \epsilon )\frac{ (1-\xB) \left(t \xB+\Q^2\right)\left(t (1-2\xB)-\Q^2\right)}{\sqrt{2} \widetilde{K}\epsilon \left(t \xB+(2-\xB) \Q^2\right) \sqrt{1+\epsilon ^2}}
+\not\!q\ \gamma^5\frac{\widetilde{K} \left(t \xB+\Q^2\right)}{\sqrt{2} \epsilon \sqrt{1+\epsilon^2}}
\nonumber
\\
&+(\not\!q\ \gamma^5 t+\gamma^5M\Q^2)\frac{ 4 (1-\xB) \left(t \xB+\Q^2\right)^2+\left(t+\Q^2\right)\left(t \xB+(2-\xB) \Q^2\right) \epsilon ^2}{8 \sqrt{2}M \widetilde{K} \xB \Q\sqrt{1+\epsilon ^2}}
\\[3.75mm]
R^{(9)}_{0+}
&=\not\!q\ \widetilde{K} \Q^2 \frac{t \xB (1-2 \xB)-\Q^2
\left(\xB-2 \left(1+\epsilon ^2\right)\right)}{\sqrt{2} \left(t \xB+(2-\xB) \Q^2\right) \epsilon \sqrt{1+\epsilon
^2}}
\nonumber
\\
&+(\not\!q\ t \xB-i\sigma_{q\Delta} \Q \epsilon )\bigg\{\frac{2 (1-\xB) \left(t \xB+\Q^2\right)+\left(t+\Q^2\right)\epsilon ^2}{2 \sqrt{2} \widetilde{K} \xB\epsilon\sqrt{1+\epsilon ^2}}
\nonumber
\\
&\hspace{7cm}
-\frac{\left(t \xB+\Q^2\right)\left(2 t (1-\xB) \xB+\left(t+\Q^2\right) \epsilon ^2\right)}{\sqrt{2} \widetilde{K} \epsilon\left(t \xB+(2-\xB) \Q^2\right)\sqrt{1+\epsilon ^2}}\bigg\}
\nonumber
\\
&+\not\!q\ \gamma^5\frac{\widetilde{K}
\left(\Q^2 \left(2-\xB+\epsilon ^2\right)+t \left((3-2 \xB) \xB+\epsilon ^2\right)\right)}{\sqrt{2} \xB \epsilon
\sqrt{1+\epsilon ^2}}
\nonumber
\\
&+(\not\!q\ \gamma^5 t+\gamma^5M\Q^2)\frac{4 (1-\xB) \left(t \xB+\Q^2\right)+\left(t+\Q^2\right)\epsilon ^2}{8 \sqrt{2}M
\widetilde{K} \xB^2 \Q\sqrt{1+\epsilon ^2}} \bigg\{\Q^2 \left(2-\xB+\epsilon ^2\right)
\nonumber
\\
&\hspace{10cm}
+t \left(\xB (3-2 \xB)+\epsilon ^2\right)\bigg\}
\\[3.75mm]
R^{(10)}_{0+}
&=\not\!q\ \frac{2 \sqrt{2} \widetilde{K} \xB \left(t \xB+\Q^2\right)}{\left(t
\xB+(2-\xB) \Q^2\right) \epsilon \sqrt{1+\epsilon ^2}}
\nonumber
\\
&+(\not\!q\ t \xB-i\sigma_{q\Delta} \Q \epsilon ) \bigg\{\frac{2 \sqrt{2} (1-\xB) \left(t \xB+\Q^2\right)^2}{\widetilde{K}
\Q^2\epsilon \left(t \xB+(2-\xB) \Q^2\right) \sqrt{1+\epsilon ^2}}+\frac{\left(t+\Q^2\right) \epsilon}{\sqrt{2}
\widetilde{K} \Q^2 \sqrt{1+\epsilon ^2}}\bigg\}
\nonumber
\\
&+\not\!q\ \gamma^5\frac{2 \sqrt{2} \widetilde{K}
\left(t \xB+\Q^2\right)}{\Q^2 \epsilon \sqrt{1+\epsilon ^2}}
\nonumber
\\
&+(\not\!q\ \gamma^5 t+\gamma^5 M\Q^2)\frac{ 4 (1-\xB) \left(t \xB+\Q^2\right)^2+\left(t^2
\xB+t (1+2 \xB) \Q^2+(3-\xB) \Q^4\right) \epsilon ^2}{2 \sqrt{2}M \widetilde{K} \xB \Q^3 \sqrt{1+\epsilon^2}}
\\[3.75mm]
R^{(11)}_{0+}
&=(\not\!q\ t\xB-i\sigma_{q\Delta} \Q\epsilon)  \frac{t(1-2 \xB)-\Q^2}{\sqrt{2}
\widetilde{K} \Q \sqrt{1+\epsilon ^2}}
-\not\!q\ \gamma^5\frac{2\sqrt{2}\widetilde{K}\left(t+\Q^2\right)}{\Q\sqrt{1+\epsilon ^2}}
\nonumber
\\
&-(\not\!q\ \gamma^5t+\gamma^5M\Q^2)\epsilon \bigg\{\frac{2 \Q^2 \left(2 t+\Q^2\right)
+\left(t+\Q^2\right) \left(t+2 \Q^2\right) \epsilon ^2}{2 \sqrt{2}M \widetilde{K} \xB\Q^2 \sqrt{1+\epsilon ^2}}
\nonumber
\\
&\hspace{5cm}
-\frac{t \xB\left(2 t+\Q^2\right) }{\sqrt{2}M \widetilde{K}\Q^2 \sqrt{1+\epsilon ^2}}+\frac{4 t^2-t \Q^2-\Q^4 }{2 \sqrt{2}M
\widetilde{K}\Q^2 \sqrt{1+\epsilon ^2}}\bigg\}
\\[3.75mm]
\label{R^{(12)}_{0+}}
R^{(12)}_{0+}
&=-\not\!q\ \frac{\widetilde{K} \Q^2 \left(t (1-2 \xB) \xB+\Q^2
\left(2-\xB+2 \epsilon ^2\right)\right)}{\sqrt{2} \left(t \xB+(2-\xB) \Q^2\right) \epsilon \sqrt{1+\epsilon ^2}}
\
\nonumber
\\
&-(\not\!q\ t \xB-i\sigma_{q\Delta} \Q \epsilon ) \frac{ (1-\xB) \left(t \xB+\Q^2\right) \left(t (1-2 \xB)
\xB+\Q^2 \left(2-\xB+2 \epsilon ^2\right)\right)}{\sqrt{2} \widetilde{K} \xB\epsilon \left(t \xB+(2-\xB) \Q^2\right)
\sqrt{1+\epsilon ^2}}
\nonumber
\\
&-\not\!q\ \gamma^5\widetilde{K}\frac{t (1-\xB) \xB-\Q^2 \left(\xB-2 \left(1+\epsilon ^2\right)\right)}{\sqrt{2}\xB \epsilon \sqrt{1+\epsilon ^2}}
\nonumber
\\
&-(\not\!q\ \gamma^5t+\gamma^5M\Q^2) \bigg\{\frac{t^2 \xB^2}{2 \sqrt{2} M \widetilde{K} \Q \sqrt{1+\epsilon^2}}+\frac{t^2 \left(4-\epsilon^2\right)+\Q^4 \left(4+\epsilon ^2\right)-4 t \Q^2 \left(5+2 \epsilon ^2\right)}{8 \sqrt{2}M\widetilde{K}\Q \sqrt{1+\epsilon ^2}}
\nonumber
\\
&\ \ \ \ \ \ \ \ \ \ \ \ +\frac{\Q \sqrt{1+\epsilon ^2} \left(t \epsilon ^2+\Q^2 \left(4+\epsilon ^2\right)\right)}{4 \sqrt{2}M \widetilde{K} \xB^2}+\frac{ \left(t-\Q^2\right) \left(t \left(\epsilon ^2-8 \xB^2\right)+3 \Q^2\left(4+3 \epsilon ^2\right)\right)}{8 \sqrt{2}M \widetilde{K} \xB \Q \sqrt{1+\epsilon ^2}}\bigg\}
\end{align}

\begin{itemize}
\item (-1,1) helicity amplitude:
\end{itemize}
\begin{align}
\label{R^{(1)}_{-+}}
R^{(1)}_{-+}
&=(\not\!q\ \epsilon \Q-i\sigma_{q\Delta} \xB) \frac{ t-2
t \xB-\Q^2+\left(t+\Q^2\right) \sqrt{1+\epsilon ^2}}{2 (t \xB+(2-\xB) \Q^2) \sqrt{1+\epsilon
^2}}
\\[3.75mm]
R^{(2)}_{-+}
&= \frac{\not\!q\ \Q \epsilon-i\sigma_{q\Delta} \xB }{2\xB^2} \bigg\{\frac{(1-\xB) \Q^2 \left(t+\Q^2\right) \epsilon ^2}{\left(t \xB+(2-\xB) \Q^2\right) \sqrt{1+\epsilon ^2}}-\frac{(t \xB+(2-\xB) \Q^2)(1+\sqrt{1+\epsilon ^2})}{2\sqrt{1+\epsilon ^2}}
\nonumber
\\
&+\frac{\left(t \xB+\Q^2\right) \left(t \xB^2+(2-\xB)^2 \Q^2\right)}{\left(t\xB+(2-\xB) \Q^2\right) \sqrt{1+\epsilon ^2}}\bigg\}
\\[3.75mm]
R^{(3)}_{-+}
&=\frac{(\not\!q\ \epsilon \Q-i\sigma_{q\Delta}\xB)\Q^2}{2\xB}\Bigg\{\frac{
\Q^2 \left(2- \xB+\epsilon ^2\right)+t \left((3-2  \xB)  \xB+\epsilon ^2\right)}{\left(t  \xB+(2- \xB)
\Q^2\right) \sqrt{1+\epsilon ^2}}-1\Bigg\}
\\[3.75mm]
R^{(4)}_{-+}
&=-\not\!q\ \frac{t (1-\xB)}{\sqrt{1+\epsilon ^2}}+ i\sigma_{q\Delta}\frac{2M(1-\xB)}{\sqrt{1+\epsilon ^2}}-\not\!q\ \gamma^5\frac{\left(t \xB+(2-\xB) \Q^2\right) \left(1-\sqrt{1+\epsilon^2}\right)}{2 \xB \sqrt{1+\epsilon ^2}}
\nonumber
\\
&-\gamma^5 \frac{M \left(t \xB+(2-\xB) \Q^2\right) \left(1-2\xB+\sqrt{1+\epsilon^2}\right)}{2\xB\sqrt{1+\epsilon^2}}
\\[3.75mm]
R^{(5)}_{-+}
&=\not\!q\ \frac{
t \xB}{4 \sqrt{1+\epsilon ^2}}-i\sigma_{q\Delta}\frac{ \Q \epsilon }{4 \sqrt{1+\epsilon ^2}}+\not\!q\  \gamma^5\frac{ t \xB+\Q^2 \left(1-\sqrt{1+\epsilon^2}\right)}{4 \sqrt{1+\epsilon ^2}}
\nonumber
\\
&+\gamma^5\frac{M\Q^2\left(1-\xB+\sqrt{1+\epsilon ^2}\right)}{4\sqrt{1+\epsilon ^2}}
\\[3.75mm]
R^{(6)}_{-+}
&=-\not\!q\ M\xB\frac{t^2 (2-3 \xB)+(2-\xB) \Q^4}{\left(t \xB+(2-\xB) \Q^2\right) \sqrt{1+\epsilon ^2}}
\nonumber
\\
&-\frac{i\sigma_{q\Delta}}{\sqrt{1+\epsilon ^2}}\Bigg\{4 M^2 \xB+\frac{2 (2-\xB) (1-\xB) \Q^2 \left(t \xB+\Q^2\right)}{\xB \left(t \xB+(2-\xB) \Q^2\right)}\Bigg\}
\nonumber
\\
&-\not\!q\  \gamma^5 M \Bigg\{\frac{(1-\xB)\left(t+\Q^2\right) }{ \sqrt{1+\epsilon ^2}}- \left(t-\Q^2\right)
\Bigg\}+\gamma^5 \Bigg\{\frac{\left(t \xB+(2-\xB) \Q^2\right)^2}{4 \xB^2}+M^2\left(t+3 \Q^2\right)
\nonumber
\\
&\ \ \ \ \ \  +\frac{M^2(1-\xB) \left(t+\Q^2\right)}{\sqrt{1+\epsilon ^2}}+\frac{\left(t \xB+(2-\xB) \Q^2\right) \left(t \xB
(1-2 \xB)+(2-3 \xB) \Q^2\right)}{4 \xB^2 \sqrt{1+\epsilon ^2}}\Bigg\}
\\[3.75mm]
R^{(7)}_{-+}
&=\frac{\not\!q\ t \xB-i\sigma_{q\Delta} \Q \epsilon }{4 \sqrt{1+\epsilon
^2}}+\left(\not\!q\  \gamma^5 t+\gamma^5 M \Q^2\right)\frac{1-\xB+\sqrt{1+\epsilon ^2}}{4 \sqrt{1+\epsilon ^2}}
\\[3.75mm]
R^{(8)}_{-+}
&= \not\!q\ \frac{ \xB \Q \left(t+\Q^2\right)^2 \epsilon }{4 \left(t\xB+(2-\xB) \Q^2\right) \sqrt{1+\epsilon ^2}}+i\sigma_{q\Delta}\frac{ (1-\xB) \Q^2 \left(t \xB+\Q^2\right)}{\left(t\xB+(2-\xB) \Q^2\right) \sqrt{1+\epsilon ^2}}
\nonumber
\\
&+\not\!q\ \gamma^5\frac{  \Q \left(t+\Q^2\right) \epsilon }{4 \sqrt{1+\epsilon^2}} -\gamma^5\frac{\Q^2}{8\xB}\bigg\{t \xB+(2-\xB) \Q^2+\frac{t \xB (1-2 \xB)+(2-3
\xB) \Q^2}{\sqrt{1+\epsilon ^2}}\bigg\}
\\[3.75mm]
R^{(9)}_{-+}
&=\not\!q\ \frac{\epsilon  \Q \left(t^2 \xB+4 t (1-\xB) \Q^2-\xB \Q^4\right)}{4 \left(t \xB+(2-\xB) \Q^2\right) \sqrt{1+\epsilon ^2}}
\nonumber
\\
&-i\sigma_{q\Delta}\Q^2\frac{2 (1-\xB) \xB \left(t \xB+\Q^2\right)+\left(t \xB+(2-\xB)\Q^2\right) \epsilon ^2}{2 \xB \left(t \xB+(2-\xB) \Q^2\right) \sqrt{1+\epsilon ^2}}
\nonumber
\\
&+\not\!q\ \gamma^5\frac{M}{2}\bigg\{ t-\Q^2+\frac{(1-\xB)(t+\Q^2)}{\sqrt{1+\epsilon^2}}\bigg\}
\nonumber
\\
&-\gamma^5\frac{\Q^2}{8\xB^2}
\bigg\{\frac{(1-\xB)\left(4 (1-\xB) \left(t \xB+\Q^2\right)+\left(t+\Q^2\right) \epsilon ^2\right)}{\sqrt{1+\epsilon ^2}}
\nonumber
\\
&\ \ \ \ \ \ -\left(4 (1-\xB) \left(t \xB+\Q^2\right)+\left(t+3 \Q^2\right)
\epsilon ^2\right)\bigg\}
\\[3.75mm]
R^{(10)}_{-+}
&=\not\!q\ \frac{ \xB
\Q \left(t+\Q^2\right) \epsilon }{\left(t \xB+(2-\xB) \Q^2\right) \sqrt{1+\epsilon ^2}}+i\sigma_{q\Delta}
\bigg\{1+\frac{t (1-2 \xB) \xB+(2-3 \xB) \Q^2}{\left(t \xB+(2-\xB)
\Q^2\right) \sqrt{1+\epsilon ^2}}\bigg\}
\nonumber
\\
&+\not\!q\ \gamma^5 \frac{\Q \epsilon }{\sqrt{1+\epsilon ^2}}-\frac{\gamma^5}{2}
\bigg\{t+\frac{(2-\xB) \Q^2}{\xB}+\frac{t (1-2 \xB) \xB+\Q^2 \left(2-3 \xB+\epsilon ^2\right)}{\xB
\sqrt{1+\epsilon ^2}}\bigg\}
\\[3.75mm]
R^{(11)}_{-+}
&=
\not\!q\ \frac{ t\xB}{\sqrt{1+\epsilon ^2}}-i\sigma_{q\Delta}\frac{ \Q \epsilon }{\sqrt{1+\epsilon ^2}}+\not\!q\  \gamma^5\frac{t\xB+\Q^2\left(1-\sqrt{1+\epsilon^2}\right)}{\sqrt{1+\epsilon^2}}
\nonumber
\\
&+\gamma^5M\bigg\{ t+2\Q^2+\frac{t-2t\xB-\xB\Q^2}{\sqrt{1+\epsilon^2}}\bigg \}
\\[3.75mm]
\label{R^{(12)}_{-+}}
R^{(12)}_{-+}
&=\not\!q\ \frac{
 \xB \Q \left(t+\Q^2\right)^2 \epsilon }{4 \left(t\xB+(2-\xB) \Q^2\right) \sqrt{1+\epsilon ^2}}+i\sigma_{q\Delta}\frac{
 (1-\xB) \Q^2 \left(t \xB+\Q^2\right)}{\left(t\xB+(2-\xB) \Q^2\right) \sqrt{1+\epsilon ^2}}
\nonumber
\\
&+\not\!q\ \gamma^5 \frac{\Q \left(t+\Q^2\right) \epsilon }{4 \sqrt{1+\epsilon^2}}+\gamma^5 \frac{ \Q^2 }{8}\bigg\{t+\Q^2-\frac{\Q^2 \left(4-3 \xB+\epsilon^2\right)+t \left((3-2 \xB) \xB+\epsilon ^2\right)}{ \xB \sqrt{1+\epsilon ^2}}\bigg\}
 \, .
\end{align}

\section{Low Energy Expansion: CFFs and Tarrach $f$s}
\label{CMkinematics}
\setcounter{equation}{0}

For the low energy expansion  we adopt the momenta in  the center-of-mass frame  as defined in  Ref.~\cite{Drechsel:1997xv}:
\begin{align}
q_1 &=(\sqrt{\omega'^2+M^2}+\omega'-\sqrt{\bar{q}^2+M^2},0,0,\bar{q})
\\
q_2&=(\omega',\omega'\sin\vartheta,0,\omega'\cos\vartheta)
\\
p_1&=\big(\sqrt{\bar{q}^2+M^2},0,0,-\bar{q}\big)
\, ,  \nonumber\\
p_2&=(\sqrt{\omega'^2+M^2},-\omega'\sin\vartheta,0,-\omega'\cos\vartheta)
\, .
\end{align}

\subsection{Low energy expansions as functions of $f_i$}
\label{LowEnergyTarrach}

Here we quote our results for the leading term in the low-energy expansion for the helicity CFFs in terms of Tarrach's structure functions
(\ref{TarrachTensor}). We will keep only the leading Non-Born contributions, i.e., linear in $\omega^\prime$, and neglect all subleading $O (\omega^\prime)$
effects.

\begin{itemize}
\item ($+1$,$+1$) helicity amplitude:
\end{itemize}
\begin{align}
{\cal H}_{++}&=\frac{\omega' M}{2\bar{q}} \bigg\{4 \bar{q}^2 (f_{10}+ M \bar{q} f_{3})+ \omega_0\bigg[8 M f_{10}+\bar{q}^2\big(4 f_{11}+f_{5}+f_{7}-4M (f_{3}-2 f_{6}-f_{9})\big)\bigg]
\nonumber
\\
&\ \ \ \ \ \ -2\bar{q} (\bar{q}-\omega_0) f_{1} -\bar{q} \bigg[
\omega_0 \big(4 f_{10}+  \omega_0(4 f_{11}+f_{5}+f_{7}+8 M f_{6}+4 Mf_{9})\big)
\nonumber
\\
&\ \ \ \ \ \ -2(\bar{q}-\omega_0)(2M \bar{q} f_{3}  - f_{1})\bigg] \cos\vartheta \bigg\} \, ,
\\[3.75mm]
{\cal E}_{++}&=\frac{\omega' M} {\bar{q}}\bigg\{\bar{q} \bigg[ (\bar{q}-\omega_0)f_{1}-M \bar{q} (4 f_{11}+f_{5}+f_{7}+8
 M f_{6}+4 M f_{9}+2 \bar{q} f_{3}-2 \omega_0 f_{3})\bigg]
\nonumber
\\
&\ \ \ \ \ \ -2\bigg[\bar{q} (\bar{q}-\omega_0)+2 M (\bar{q}+\omega_0)\bigg] f_{10}+\bar{q} \bigg[(\bar{q}-\omega_0)f_{1}-2 \bar{q} ( M \bar{q} f_{3}+f_{10})
\nonumber
\\
&\ \ \ \ \ \ + \omega_0\big(2 f_{10} +M(4 f_{11}+f_{5}+f_{7}+8 M f_{6}+4 M f_{9}+2 \bar{q} f_{3})\big)\bigg] \cos\vartheta\bigg\} \, ,
\\[3.75mm]
\widetilde{\cal H}_{++}&=\frac{\omega' M (\omega_0 \cos\vartheta-\bar{q})}{2}\bigg\{4 f_{10}+ \bar{q}\big(4 f_{11}+f_{5}+f_{7}+8M f_{6} +4 M f_{9}\big)\bigg\} \, ,
\\[3.75mm]
\widetilde{\cal E}_{++}&=\frac{\omega' M^2 (\omega_0-2 M-\bar{q} \cos\vartheta)}{\bar{q}} \bigg\{4 f_{10}+\bar{q} \big(4 f_{11}+f_{5}+f_{7}+8 M f_{6}+4 M f_{9}\big)\bigg\}
\, .
\end{align}

\begin{itemize}
\item ($0$,$+1$) helicity amplitude:
\end{itemize}
\begin{align}
{\cal H}_{0+}&=-\frac{\omega'\sqrt{-M \omega_0}}{2 q}\bigg\{\bigg[q \big[4 M q^2 f_{12}+\omega_0 \big(4 f_{10}+\omega_0(4 f_{11}+f_{5}+f_{7})\big)\big] \cos\theta-2 q^2 \big[4  (M-\omega_0)f_{4}
\nonumber
\\
&\ \ \ \ \ \ +\omega_0(4 f_{11}+f_{5}+f_{7}+4M f_{12}) \big]-4\left(q^2+\omega_0^2\right) f_{10} \bigg] \cot\theta+\frac{q}{\sin\theta} \bigg[q^2(4 f_{11}-8 f_{4}+f_{5}+f_{7})
\nonumber
\\
&\ \ \ \ \ \ +4 \omega_0 (f_{10}+ M \omega_0f_{12})\bigg]+4 q\sin\theta \bigg[ q^2f_{4}+M \big(f_{1}+4 M^2 f_{2}-2 M \omega_0 (f_{2}+2 f_6+f_{9})\big)\bigg]\bigg\}
\, , \\[3.75mm]
{\cal E}_{0+}&=-\frac{\omega' M\sqrt{-M \omega_0}}{q}\bigg\{\bigg[2 q^2 \left(4 f_{11}+f_{5}+f_{7}\right)-8 M \left(f_{10}+2 M f_{4}-q^2f_{12}\right)+8 \omega_0 (f_{10}+3 M f_{4})
\nonumber
\\
&\ \ \ \ \ \ -8 \omega_0^2 f_{4}- q \omega_0 \cos\theta\left(f_{5}+f_{7}\right)\bigg] \cot\theta-\frac{q}{\sin\theta} \bigg[8 f_{10}-2 M \big(4 f_{11}-8 f_{4}+f_{5}+f_{7}+4 M f_{12}\big)
\nonumber
\\
&\ \ \ \ \ \ + \omega_0\big(8 f_{11}-8 f_{4}+f_{5}+f_{7}+8 M f_{12}\big)\bigg]+2q \sin\theta \bigg[ 2 \omega_0 \big(f_{11}-f_{4}+ M(f_{2}+2 f_{6}+f_{9})\big)
\nonumber
\\
&\ \ \ \ \ \ -f_{1}+2 f_{10}+4M f_{4}  -4 M^2 f_{2} \bigg]\bigg\} \, ,
\\[3.75mm]
 \widetilde{\cal H}_{0+}&=-\frac{\omega' M}{2 \sqrt{-M \omega_0}}\bigg\{\frac{q^2}{\sin\theta} \bigg[4 f_{10}+(8 M-6\omega_0) f_{4}+ \omega_0(4 f_{11}+f_{5}+f_{7}+4 M f_{12})-2 \omega_0 \cos2\theta f_{4}\bigg]
\nonumber
\\
&\ \ \ \ \ \  -\bigg[ q^3\big(4 f_{11}-8 f_{4}+f_{5}+f_{7}+4M f_{12}\big)+8q \omega_0 f_{10} + q \omega_0^2\big(4 f_{11}+f_{5}+f_{7}+4 M f_{12}\big)
\nonumber
\\
&\ \ \ \ \ \  -\omega_0\big( q^2 (4 f_{11}+f_{5}+f_{7}+4M f_{12})+4 \omega_0 f_{10}\big) \cos\theta\bigg]\cot\theta\bigg\} \, ,
\\[3.75mm]
\widetilde{\cal E}_{0+}&=-\frac{\omega' M^2(q-\omega_0 \cos\theta)}{\sin\theta \sqrt{-M \omega_0}}
\bigg\{\bigg[ q^2\big(4 f_{11}-4 f_{4}+f_{5}+f_{7}+4 M f_{12}\big)+4
 \omega_0f_{10}\bigg] \cos\theta
\nonumber
\\
&\ \ \ \ \ \ -q \bigg[4 f_{10}+\big(8 M-4 \omega_0\big)f_{4} + \omega_0\big(4 f_{11}+f_{5}+f_{7}+4 M f_{12}\big)\bigg] \bigg\} \, .
\end{align}

\begin{itemize}
\item ($-1$,$+1$) helicity amplitude:
\end{itemize}
\begin{align}
{\cal H}_{-+}&=\frac{\omega' M}{2 \bar{q}}\bigg\{4 \left(\bar{q}^2+2 M\omega_0\right)f_{10}+\bar{q}^2 \bigg[\omega_0(4 f_{11}+f_5+f_7+8Mf_6 +4 Mf_9 )-4 M (\bar{q}+\omega_0) f_3\bigg]
\nonumber
\\
&-2\bar{q}(\bar{q}+\omega_0)f_1-\bar{q}  \cos\vartheta\bigg[\bar{q}^2\big(4 f_{11}+f_5+f_7-4M(f_3-2 f_6-f_9)\big) +4 \omega_0f_{10}
\nonumber
\\
&\ \ \ \ \ \ +2 M \omega_0(4 f_{11}+f_5+f_7+8M f_6 +4 M f_9-2\bar{q} f_3)-2(\bar{q}+\omega_0) f_1\bigg]\bigg\}
\, ,
\\[3.75mm]
{\cal E}_{-+}&=\frac{\omega' M}{\bar{q}} \bigg\{2\big (2 M (\bar{q}-\omega_0)-\bar{q} (\bar{q}+\omega_0)\big)f_{10} +\bar{q} \bigg [(\bar{q}+\omega_0)f_1 -M
\bar{q} \big(4 f_{11}+f_7+4  M(2 f_6+f_9)
\nonumber
\\
&\ \ \ \ \ \ +f_5-2(\bar{q}+\omega_0) f_3 \big)\bigg]+\bar{q}\bigg[M \omega_0 (4 f_{11}+f_5+f_7+8M
f_6 +4M f_9 -2 \bar{q} f_3) -2M \bar{q}^2 f_3
\nonumber
\\
&\ \ \ \ \ \ + (\bar{q}+\omega_0)(2f_{10}-f_1)\bigg]\cos \vartheta \bigg\}
\, ,
\\[3.75mm]
 \widetilde{\cal H}_{-+}&=\frac{\omega' M}{2} \bigg\{ \bar{q}\big(4 f_{11}+f_{5}+f_7+8M f_{6} +4 M f_{9}\big)-4f_{10}\bigg\}\big(\bar{q}-\omega_0 \cos\vartheta\big)
 \, ,
\\[3.75mm]
\widetilde{\cal E}_{-+}&=\frac{\omega' M^2}{\bar{q}} \bigg\{\bar{q}\big(4 f_{11}+f_{5}+f_{7}+8 M f_{6}+4 M f_{9}\big)-4 f_{10} \bigg\}\big(2 M-\omega_0+\bar{q}\cos\vartheta\big)
\, .
\end{align}

\section{Born term for Compton scattering off nucleon}
\label{App:Born}
\setcounter{equation}{0}

In this Appendix, we list our results for the computation of the helicity CFFs in the Born approximation using the target rest frame. These will be extracted from
Eq.\ (\ref{DVCS2helicity}) where the covariant Compton amplitude is replaced by its Born approximation,
\begin{align}
{\cal T}^{\rm Born}_{ab} = (-1)^{a-1} \varepsilon_{1}^\mu (a)\, T_{\mu\nu}^{\rm Born}\,  \varepsilon_{2}^{\nu\,\ast} (b)
\, ,
\end{align}
making use of the definitions given in Eqs.\ (\ref{cal-Tab})--(\ref{cal-TabA}) and (\ref{BornAmplitude}). The r.h.s.\ can be decomposed into a sum of four terms
that differ by the form factor products accompanying them, i.e.,
$$
e_N F_1 (-q^2_1)\,,\quad  e_N F_2 (-q^2_1)\,,\quad  \kappa_N F_1 (-q_1^2)\,,\quad  \mbox{and} \quad \kappa_N F_2 (-q^2_1)\,,
$$
where we set $e_N=F_1(0)$ and $\kappa_N=F_2(0)$ and use in the following also the nucleon magnetic moment $\mu_N=e_N + \kappa_N$.  We find
(suppressing superscript Born)
\begin{align}
\label{H_+a-Born}
{\cal H}_{+b} &=
-\frac{\left(1+b \sqrt{1+\epsilon^2}\right) \left( 2-\xB+\frac{\xB t}{\Q^2} \right)^2 }{4\sqrt{1+\epsilon^2}(1-\xB)\left(\!1+\frac{\xB t}{\Q^2}\!\right)}\,
e_N F_1
-b\,\frac{\xB^2 \left(1+\frac{t}{\Q^2}\right)^2 }{4 (1-\xB) \left(1+\frac{\xB t}{\Q^2}\right)} \left[\kappa_N F_1 + \mu_N F_2\right]
\\&
-\frac{e_N\,\xB^2\left( 1+\frac{t}{\Q^2} \right)}{2 \sqrt{1+\epsilon^2} \left(1+\frac{\xB t}{\Q^2} \right) }\left\{
\frac{4M^2}{\Q^2}  \,F_1
- \frac{\left(2-\xB+\frac{\xB t}{\Q^2} \right) \frac{t}{\Q^2}\,\left[F_1+ F_2\right]  }{\left(\!1+\frac{t}{\Q^2}\!\right)(1-\xB)}
+\frac{\xB\left(\!1+\frac{t}{\Q^2}\!\right) \,F_2}{2  (1-\xB)}
\right\},
\nonumber
\\
{\cal E}_{+b}  &=
-b \left[\kappa_N F_1 + \mu_N F_2\right] - \frac{e_N\, \epsilon^2}{2\sqrt{1+\epsilon^2}}
\left\{ \frac{2\xB\,F_2}{\epsilon^2}
-  \frac{ \left( 1+\frac{t}{\Q^2} \right) F_1}{1+\frac{\xB t}{\Q^2} }
+ \frac{\left( 2-\xB+\frac{\xB t}{\Q^2} \right)
\left[F_1+F_2\right]}{(1-\xB)\left(\! 1+\frac{\xB t}{\Q^2} \!\right)}
\right\},
\nonumber\\
\\
\widetilde{\cal H}_{+b} &=
-\frac{\left(1+b \sqrt{1+\epsilon^2}\right)\xB \left(2-\xB+\frac{\xB t}{\Q^2}\right) }{2\sqrt{1+\epsilon^2}}
\left[\frac{\kappa_N F_2}{\epsilon^2}+  \frac{
\left(1-\frac{t}{\Q^2}\right)e_N\left[F_1+F_2\right]}{2  (1-\xB) \left(\!1+\frac{\xB t}{\Q^2}\!\right)}
\right]
-\frac{\xB}{4\sqrt{1+\epsilon^2}}
\nonumber\\
& \times \frac{2-\xB+\frac{\xB t}{\Q^2}}{(1-\xB) \left(\!1+\frac{\xB t}{\Q^2}\!\right)}
\left\{
 \mu_N \left[
 (2-\xB)  F_2+\frac{\xB t}{\Q^2} \left[2 F_1+ F_2\right]
\right]
+\left(1-\frac{t}{\Q^2}\right)\left[\kappa_N F_1-e_N F_2\right]
\right\}\,,
\nonumber\\
\\
\widetilde{\cal E}_{+b} &=
\frac{\left(1+b \sqrt{1+\epsilon^2}\right) \left( 2-\xB+\frac{\xB t}{\Q^2}\right)}{2 \sqrt{1+\epsilon^2}}
\left[
\frac{\kappa_N \left[ 2 F_1  + F_2\right]}{\xB}-
\frac{\epsilon^2 \left( 3+\frac{t}{\Q^2} \right)  e_N \left[F_1+F_2\right]}{2 \xB (1-\xB) \left(\!1+\frac{\xB t}{\Q^2}\!\right)}
\right]
\nonumber\\
&
-\frac{2-\xB+\frac{\xB t}{\Q^2}}{\xB \sqrt{1+\epsilon^2}} \left\{e_N (1-\xB) F_2+\kappa_N \left[F_1+ F_2\right]\right\}+
\frac{\epsilon^2 \left(2-\xB+\frac{\xB t}{\Q^2}\right) }{4 \xB \sqrt{1+\epsilon ^2}(1-\xB)\left(1+\frac{\xB  t}{\Q^2}\right)}
\nonumber\\
&\times\Bigg\{
e_N\left(3+ \frac{t}{\Q^2}\right)\left[F_1+F_2\right]-\mu_N\left(1+\frac{\xB t}{\Q^2}\right) \left[F_1+F_2\right]
+\mu_N (1-\xB) \left[\frac{t F_1}{\Q^2}-F_2\right]
\Bigg\},
\nonumber
\\
\\
{\cal H}_{0+}  &=
\frac{\sqrt{2}\widetilde{K} \xB}{\Q \sqrt{1+\epsilon^2}}\, e_N\!\left\{
\frac{2-\xB -\frac{\xB t}{\Q^2}+ \frac{8\xB M^2}{\Q^2}}{2(1-\xB)\left(1+\frac{\xB  t}{\Q^2}\right)}
\left[F_1-\frac{\Q^2}{4\xB M^2} F_2\right]
+
\frac{1-\frac{t}{4 M^2}}{1+\frac{\xB  t}{\Q^2}} F_2 \right\} - \frac{t\, {\cal E}_{0+} }{4 M^2}\,,
\\
{\cal E}_{0+}  &=
\frac{(-1)\sqrt{2}\widetilde{K}}{\Q \sqrt{1+\epsilon^2}}\,
\frac{e_N \left[\epsilon^2 F_1 -\xB^2 F_2\right]}{(1-\xB)\left(\!1+\frac{\xB  t}{\Q^2}\!\right)}
+ \frac{\left( 1+\frac{(1-\xB) \Q^2}{\Q^2+ \xB t } \right)\Q}{\sqrt{2} \tK \sqrt{1+\epsilon^2}}
\left[1+\frac{\xB t}{\Q^2}+\frac{\epsilon^2\left(1+\frac{t}{\Q^2}\right)}{2} \right]e_N F_2\,,
\nonumber\\
\\
\widetilde{\cal H}_{0+}  &
=
\frac{\sqrt{2}\widetilde{K}\xB\left(2-\xB+\frac{\xB t}{\Q^2}\right)}{\Q \sqrt{1+\epsilon^2}}
\left\{
\frac{e_N F_2}{\epsilon^2 }-
\frac{\left(1+\frac{t}{\Q^2}\right)\mu_N\left[F_1-\frac{\Q^2}{4\xB M^2}  F_2\right] }{2 (1-\xB)\left(1+\frac{\xB t}{\Q^2}\right)}
 \right\}
- \frac{t\, \widetilde{\cal E}_{0+} }{4 M^2} \,,
\\
\label{tE_+0-Born}
\widetilde{\cal E}_{0+}  &
=
\frac{(-1)\sqrt{2}\widetilde{K}}{\Q \sqrt{1+\epsilon^2}}\,
\frac{\left(2-\xB+\frac{\xB t}{\Q^2}\right) \frac{4\xB M^2}{\Q^2}}{2 (1-\xB)\left(1+\frac{\xB t}{\Q^2}\right)}
\mu_N \left[F_1-  \frac{\Q^2}{4\xB M^2} F_2\right]
\nonumber\\
& - \frac{\left(2-\xB + \frac{\xB t}{\Q^2}\right)\Q}{2\sqrt{2} \tK\xB \sqrt{1+\epsilon^2}}
 \left[
 4-2 \xB+3 \epsilon^2 +
\left(4 (1-\xB) \xB+\epsilon^2\right)  \frac{t}{\Q^2}
 \right]  e_N F_2
\,.
\end{align}
Notice that in the longitudinal helicity-flip CFFs a  spurious kinematical $1/\tK$ singularity appears, which cancels, however, in electric-like combinations
introduced in Eq.\ (\ref{cff-electric}). Hence, the Born result is well defined for any value of kinematical variables, except for the elastic poles
at $s= M^2$ ($x_B = 1$) and $u= M^2$ ($x_B = -\Q^2/t$).



\begin{thebibliography}{100}

\bibitem{Belitsky:2001ns}
A.V. Belitsky, D.~M{\"u}ller, and A.~Kirchner,
\newblock Nucl. Phys. {\bf B 629}, 323 (2002), hep-ph/0112108.

\bibitem{Drechsel:2002ar}
D.~Drechsel, B.~Pasquini, and M.~Vanderhaeghen,
\newblock Phys. Rept. {\bf 378}, 99 (2003), hep-ph/0212124.

\bibitem{Guichon:1995pu}
P.A. Guichon, G.~Liu, and A.W. Thomas,
\newblock Nucl. Phys. {\bf A 591}, 606 (1995), nucl-th/9605031.

\bibitem{Drechsel:1997xv}
D.~Drechsel, G.~Knochlein, A.Y. Korchin, A.~Metz, and S.~Scherer,
\newblock Phys. Rev. {\bf C 57}, 941 (1998), nucl-th/9704064.

\bibitem{Lepage:1980fj}
G.P. Lepage and S.J. Brodsky,
\newblock Phys. Rev. {\bf D 22}, 2157 (1980).

\bibitem{Radyushkin:1998rt}
A.V. Radyushkin,
\newblock Phys. Rev. {\bf D 58}, 114008 (1998), hep-ph/9803316.

\bibitem{Diehl:1998kh}
M.~Diehl, T.~Feldmann, R.~Jakob, and P.~Kroll,
\newblock Eur. Phys. J. {\bf C 8}, 409 (1999), hep-ph/9811253.

\bibitem{Mueller:1998fv}
D.~M{\"u}ller, D.~Robaschik, B.~Geyer, F.-M. Dittes, and J.~Ho\v{r}ej{\v s}i,
\newblock Fortschr. Phys. {\bf 42}, 101 (1994), hep-ph/9812448.

\bibitem{Radyushkin:1996nd}
A.V. Radyushkin,
\newblock Phys. Lett. {\bf B 380}, 417 (1996), hep-ph/9604317.

\bibitem{Ji:1996nm}
X.~Ji,
\newblock Phys. Rev. {\bf D 55}, 7114 (1997), hep-ph/9609381.

\bibitem{Belitsky:2010jw}
A.V. Belitsky and D.~M\"uller,
\newblock Phys. Rev. {\bf D 82}, 074010 (2010), arXiv:1005.5209 [hep-ph].

\bibitem{Tarrach_1975tu}
R.~Tarrach,
\newblock Nuovo Cim. {\bf A 28}, 409 (1975).

\bibitem{Belitsky:2008bz}
A.V. Belitsky and D.~M{\"u}ller,
\newblock Phys. Rev. {\bf D 79}, 014017 (2009), arXiv:0809.2890 [hep-ph].

\bibitem{DieGouPirRal97}
M.~Diehl, T.~Gousset, B.~Pire, and J.P. Ralston,
\newblock Phys. Lett. {\bf B 411}, 193 (1997), hep-ph/9706344.

\bibitem{ItzZub80}
C.~Itzykson and J.~Zuber,
\newblock {\em Quantum Field Theory} (McGraw-Hill, New York, 1980).

\bibitem{Belitsky:2005qn}
A.V. Belitsky and A.V. Radyushkin,
\newblock Phys. Rept. {\bf 418}, 1 (2005), hep-ph/0504030.

\bibitem{Braun:2011dgBraun:2011zr}
V.M. Braun and A.N. Manashov,
\newblock JHEP {\bf 01}, 085 (2012), arXiv:1111.6765 [hep-ph];
\newblock Phys. Rev. Lett. {\bf 107}, 202001 (2011), arXiv:1108.2394 [hep-ph].

\bibitem{Braun:2012bgBraun:2012hq}
V.M.~Braun, A.N.~Manashov, and B.~Pirnay,
\newblock Phys. Rev. {\bf D 86}, 014003 (2012), arXiv:1205.3332 [hep-ph];
\newblock (2012), arXiv:1209.2559 [hep-ph].

\bibitem{Diehl:2001pm}
M.~Diehl,
\newblock Eur. Phys. J. {\bf C 19}, 485 (2001), hep-ph/0101335.

\bibitem{PhysRev.110.240}
R.E. Prange,
\newblock Phys. Rev. {\bf 110}, 240 (1958).

\bibitem{PhysRev.126.789}
A.C. Hearn and E.~Leader,
\newblock Phys. Rev. {\bf 126}, 789 (1962).

\bibitem{Mankiewicz:1997bk}
L.~Mankiewicz, G.~Piller, E.~Stein, M.~V{\"a}nttinen, and T.~Weigl,
\newblock Phys. Lett. {\bf B 425}, 186 (1998), hep-ph/9712251.

\bibitem{BelMue00}
A.V.~Belitsky and D.~M{\"u}ller,
\newblock Phys. Lett. {\bf B 486}, 369 (2000), hep-ph/0005028.

\bibitem{Low:1954kd}
F.~Low,
\newblock Phys. Rev. {\bf 96}, 1428 (1954).

\bibitem{Siegert:1937yt}
A.~Siegert,
\newblock Phys. Rev. {\bf 52}, 787 (1937).

\bibitem{Kumericki:2007sa}
K.~Kumeri{\v c}ki, D.~M{\"u}ller, and K.~Passek-Kumeri{\v c}ki,
\newblock Nucl. Phys. {\bf B 794}, 244 (2008), hep-ph/0703179.

\bibitem{Kumericki:2009uq}
K.~Kumeri{\v c}ki and D.~M{\"u}ller,
\newblock Nucl. Phys. {\bf B 841}, 1 (2010), arXiv:0904.0458 [hep-ph].

\bibitem{Airapetian:2008aa}
HERMES, A.~Airapetian {\em et~al.},
\newblock JHEP {\bf 06}, 066 (2008), arXiv:0802.2499 [hep-ex].

\bibitem{Airapetian:2010ab}
HERMES, A.~Airapetian {\em et~al.},
\newblock JHEP {\bf 06}, 019 (2010), arXiv:1004.0177 [hep-ex].

\bibitem{Airapetian:2011uq}
The HERMES, A.~Airapetian {\em et~al.},
\newblock Phys. Lett. {\bf B 704}, 15 (2011), arXiv:1106.2990 [hep-ex].

\bibitem{Airapetian:2012mq}
HERMES, A.~Airapetian {\em et~al.},
\newblock JHEP {\bf 1207}, 032 (2012), arXiv:1203.6287 [hep-ex].

\bibitem{Vanderhaeghen:1998uc}
M.~Vanderhaeghen, P.A. Guichon, and M.~Guidal,
\newblock Phys. Rev. Lett. {\bf 80}, 5064 (1998).

\bibitem{Kroll:2012sm}
P.~Kroll, H.~Moutarde, and F.~Sabatie,
\newblock Eur. Phys. J. {\bf C 73}, 2278 (2013), arXiv:1210.6975 [hep-ph].

\end{thebibliography}
\end{document}